\documentclass[mnsc, nonblindrev]{arxiv} 

\OneAndAHalfSpacedXI 

\usepackage{natbib}
 \bibpunct[, ]{(}{)}{,}{a}{}{,}%
 \def\bibfont{\small}%

\usepackage{multibib}
\newcites{AP}{References for the Appendices}

\usepackage{multirow}
\usepackage{scrextend}
\usepackage{tabularx}
\newcolumntype{Y}{>{\hsize=1.33\hsize}X}
\newcolumntype{Z}{>{\hsize=0.67\hsize}X}
\newcolumntype{W}{>{\hsize=1.167\hsize}X}

\usepackage{booktabs}
\TheoremsNumberedThrough     
\ECRepeatTheorems

\EquationsNumberedThrough    

\usepackage{xcolor}
\usepackage{caption, subcaption}
\usepackage{color-edits}

\usepackage{xspace}
\usepackage{bbm}
\usepackage{cancel}
\usepackage{hyperref}

\definecolor{burgundy}{rgb}{0.5, 0.0, 0.13}

\addauthor{pf}{red}
\addauthor{yz}{magenta}
\addauthor{jw}{cyan}

\newcommand{\EE}{\mathbb{E}}
\newcommand{\PP}{\mathbb{P}}
\newcommand{\NN}{\mathbb{N}}

\newcommand{\Var}{\text{Var}}
\newcommand{\Cov}{\text{Cov}}
\newcommand{\A}{\mathcal{A}}
\newcommand{\diff}{\mathop{}\!\mathrm{d}}
\newcommand{\CP}{correlated pooling\xspace}
\newcommand{\NP}{naive pooling\xspace}
\newcommand{\CCP}{community-correlated pooling\xspace}
\newcommand{\HCP}{household-correlated pooling\xspace}
\newcommand{\cp}{\texttt{CP}\xspace}
\newcommand{\np}{\texttt{NP}\xspace}
\newcommand{\pool}{\texttt{POOL}\xspace}
\newcommand{\muL}{\textmu L}

\newcommand{\Dt}{\tilde{D}}

\usepackage{etoc}

\begin{document}


\RUNAUTHOR{Wan, Zhang, and Frazier}

\RUNTITLE{Correlation Improves Group Testing}

\TITLE{Correlation Improves Group Testing: \\ Modeling Concentration-Dependent Test Errors}
\newcommand*\samethanks[1][\value{footnote}]{\footnotemark[#1]}

\ARTICLEAUTHORS{%
\AUTHOR{Jiayue Wan\thanks{Equal contribution.}}
\AFF{School of Operations Research \& Information Engineering, Cornell University, NY 14850, \EMAIL{jw2529@cornell.edu}} 
\AUTHOR{Yujia Zhang\samethanks 
}
\AFF{Center for Applied Mathematics, Cornell University, NY 14850, \EMAIL{yz685@cornell.edu}}
\AUTHOR{Peter I. Frazier}
\AFF{School of Operations Research \& Information Engineering, Cornell University, NY 14850, \EMAIL{pf98@cornell.edu}}
} 

\ABSTRACT{Population-wide screening is a powerful tool for controlling infectious diseases. 
Group testing can enable such screening despite limited resources.
Viral concentration of pooled samples are often positively correlated, either because prevalence and sample collection are influenced by location, or through intentional enhancement via pooling samples according to risk or household.
Such correlation is known to improve efficiency when test sensitivity is fixed.
However, in reality, a test's sensitivity depends on the concentration of the analyte (e.g., viral RNA), as in the so-called dilution effect, where sensitivity decreases for larger pools.
We show that concentration-dependent test error alters correlation's effect under the most widely-used group testing procedure, the two-stage Dorfman procedure.
We prove that when test sensitivity increases with concentration,
pooling correlated samples together (\textit{\CP}) achieves asymptotically higher sensitivity than independently pooling the samples (\textit{\NP}). 
In contrast, in the concentration-independent case, correlation does not affect sensitivity. 
Moreover, with concentration-dependent errors, correlation can degrade test efficiency compared to \NP, whereas under concentration-independent errors, correlation always improves efficiency.
We propose an alternative measure of test resource usage, the number of positives found per test consumed, which we argue is better aligned with infection control, and show that \CP outperforms \NP on this measure.
In simulation,
we show that the effect of correlation under realistic concentration-dependent test error is meaningfully different from correlation's effect assuming fixed sensitivity.
Our findings underscore the importance for policy-makers of using models that incorporate naturally-occurring correlation and of considering ways of strengthening this correlation.
}


\maketitle

\pagebreak[4]
\section{Introduction}\label{sec:intro}
The severe acute respiratory syndrome coronavirus 2 (SARS-CoV-2) has claimed millions of lives while causing enormous economic losses. Large-scale screening using polymerase chain reaction (PCR) tests can curb the virus's spread \citep{mercer2021testing,xing2020rapid,barak2021lessons} through promptly identifying and isolating infected individuals and their contacts \citep{cleary2021using, brault2021group}, but it requires a massive amount of chemical reagent and access to many diagnostic testing machines. 

A promising solution is \textit{group testing}.\footnote{In this manuscript, we use \textit{pooled testing} to refer to pooling multiple samples together and testing them with a single test, and use group testing to refer to a testing protocol that utilizes pooled testing to improve testing efficiency.} The \textit{Dorfman procedure}, the first group testing protocol proposed in 1943 to screen soldiers for syphilis \citep{dorfman1944}, pools multiple samples and tests each pool using a single test. Especially in low-prevalence settings, group testing can save significant test resources compared to individual testing \citep{kim2007comparison}.
Group testing has proven effective in large-scale community screening worldwide and in controlling the spread of coronavirus disease 2019 (COVID-19). In May 2020, Wuhan screened nine million people over ten days using pools of five to ten \citep{wuhan2020}. Many K-12 schools and universities, including Cornell University, Duke University, and the University of Cambridge, used pools of 5 to 24 to conduct campus-wide screenings \citep{mendoza2021implementation,cornell2020,denny2020implementation, mahase2020covid}.

Mathematical analysis of group testing's improved resource utilization has largely assumed independence of pooled samples' infection status (e.g., \citealt{kim2007comparison,westreich2008optimizing}). However, several researchers \citep{barak2021lessons,basso2021effect,augenblick2020group,lin2020positively,comess2021statistical} have recently observed that human behavior and the logistics of sample collection naturally lead to correlations. Specifically, when one person is infected, others in their immediate social circles are likely also infected \citep{vang2021participation,rader2020crowding,lan2020work}. 
The literature observes that correlation can significantly affect the performance of pooled testing. Correlation tends to reduce the number of pools with virus-containing samples \citep{lendle2012group, deckert2020simulation}.
Mathematical analyses show that when tests are error-free, this correlation improves {\it test efficiency} (i.e., the number of people screened per test) \citep{augenblick2020group, lin2020positively}.
This remains true in the presence of test errors, where the \textit{test sensitivity} (i.e., the probability that testing a positive sample provides a correct result) is fixed regardless of the virus concentration (called the \textit{viral load}) in the sample
\citep{basso2021effect,aprahamian2019optimal,bilder2010informative,bilder2012pooled,mcmahan2012informative,mcmahan2012two}.

However, these mathematical analyses of correlation's effect on pooled testing ignore an important practical aspect: Test sensitivity depends on the concentration of the analyte of interest (e.g., viral RNA) in the sample.
Whether testing individually or in pools, the sensitivity of a PCR test is \textit{lower} for samples with a lower viral load due to its inherent detection limit \citep{van2020comparison}.
Existing studies have argued that modeling concentration-dependent test errors has important implications for designing pooled testing strategies, e.g., \citet{westreich2008optimizing} and \citet{brault2021group}. However, they do not consider correlation.

Our work bridges the gap in the literature by studying correlated group testing under concentration-dependent test errors (see Table~\ref{tab:literature_table} and Section~\ref{sec:litreview}). We argue that concentration-dependent test errors alter the way correlation impacts pooled testing. This is because correlation tends to increase the number of positive samples in positive-containing pools, thereby increasing the viral load in the pooled sample and elevating the likelihood of such pools testing positive. Moreover, this increase in sensitivity can {\it decrease} test efficiency because more pools test positive and require follow-up tests.
Neither effect is present in models assuming fixed test sensitivity considered in \citet{ basso2021effect,augenblick2020group,lin2020positively,aprahamian2019optimal}. 
\citet{comess2021statistical} studies the joint effect of increasing correlation and prevalence on test sensitivity under concentration-dependent test errors. They find that the combination of these two changes increases sensitivity, but do not elucidate whether the effect is due to correlation, increased prevalence, or both.
Recently, \citet{chatterjee2022capturing}\footnote{This work appeared after a draft of this manuscript was first posted in 2021.} extends \citet{aprahamian2019optimal} by considering a test error model that depends on the number of positive samples in the pool, without modeling the viral load of the positive samples.
In addition, we argue that modeling test errors realistically in pooled testing has significant implications for policy decisions, including the choice between \textit{repeated screening} versus shutdown and, in the case of screening, decisions of pool size and screening frequency. 

\begin{table}[t]
\centering
\TABLE
{Existing theoretical studies on group testing with different correlation structures and test error models. \label{tab:literature_table}}
{\begin{tabularx}{\textwidth}{ZWW}
\toprule
Correlation structure & \multicolumn{2}{c}{Test error model}\\ \cmidrule{2-3}
& Concentration-independent  & Concentration-dependent  \\ \midrule
Independent samples & \citet{dorfman1944, graff1972group, kim2007comparison} & \citet{hwang1976, hung1999robustness, wein1996pooled, westreich2008optimizing, mutesa2020, brault2021group} \\[4em]
Correlated samples & \citet{mcmahan2012informative, mcmahan2012two, aprahamian2019optimal, augenblick2020group, lin2020positively, basso2021effect}  & \citet{comess2021statistical}, \citet{chatterjee2022capturing}, \textbf{Our work} \\  \bottomrule
\end{tabularx}}
{\textit{Note}. See Section~\ref{sec:litreview} for a detailed discussion of the literature.}
\end{table}

We make these arguments through both theoretical analysis and simulation study under concentration-dependent test errors. 
We prove that, under a general correlation structure in the population
and when test sensitivity is monotone increasing in the concentrations of the samples in the pool, pooling correlated samples together (called \textit{\CP}) in the two-stage Dorfman procedure yields asymptotically higher sensitivity compared to independently pooling the samples (called \textit{\NP}) using the same pool size. In contrast, correlation has no impact on sensitivity in the concentration-independent case. 
Moreover, correlation can degrade test efficiency compared to \NP with the same pool size, which we demonstrate in an example. In contrast, under concentration-independent errors, correlation always improves test efficiency \citep{augenblick2020group, lin2020positively, basso2021effect}. 
Furthermore, we argue that test efficiency, the key performance metric in studies assuming concentration-independent errors, may not adequately capture a group testing procedure's effectiveness for repeated screening. This is because, in the concentration-dependent case, a protocol can exhibit high efficiency but low sensitivity, which is unfavorable for epidemic control. 
Instead, we propose an alternative measure of test resource usage, the \textit{effective efficiency}, which measures the number of positive cases identified per test consumed. It better captures a procedure’s effectiveness for repeated screening, complementing the conventional efficiency metric and other metrics balancing accuracy and test consumption \citep{aprahamian2019optimal}. We prove that \CP achieves asymptotically higher effective efficiency. 

These insights have significant implications for policy-makers, as we demonstrate in a realistic agent-based simulation.
We simulate the correlation in viral loads arising naturally from interactions in communities and households, which induces correlation within pools.
We adopt the perspective of a policy-maker using our simulation to assess screening policies, i.e., screening frequencies and pool sizes, in response to an emerging pandemic.
We draw three conclusions.
First, modeling concentration-dependent test errors realistically is essential for accurately quantifying the benefit of correlation, while using fixed test errors obscures correlation's benefit.
Second, policy-makers should consider correlation when choosing a policy that fully utilizes the test capacity.
Failure to do so risks underestimating screening policies' true effectiveness and making overly cautious policy decisions.
Third, enhancing correlation within pools can substantially improve epidemic outcomes. We recommend that policy-makers implement explicit measures to promote household pooling, such as encouraging families or roommates to get tested together and mailing sample collection kits to households \citep{stanfordSelfSwab_2020}. 
For example, given a 100-day test supply of $4\times 10^4$ for a population of $1\times 10^5$, a \textit{correlation-oblivious} policy-maker deems screening impractical and imposes a lockdown. 
A \textit{correlation-aware} policy-maker, who 
includes naturally-occurring correlation into the model they use to make decisions but does not take further measures to pool households together, 
opts for screening every five days with a pool size of ten, incurring $3.2\times 10^3$ infections on average. If, in addition, the policy-maker is able to {\it enhance correlation} by pooling households together, they would choose to screen every four days with a pool size of ten, incurring $2.6\times 10^3$ infections on average, a 20\% reduction compared to the best policy when not enhancing correlation.

To summarize, our contributions in this paper are:
\begin{itemize}
    \item We establish an analytical framework for modeling pooling methods in large populations, formulating a model of correlation in pools derived from an asymptotic analysis of a more general population-level model of infection spread and pool formation.

    \item We prove that under the general population-level model and in the presence of concentration-dependent test errors, using \CP in the Dorfman procedure achieves asymptotically higher sensitivity and effective efficiency than \NP.  Our work is the first to study sensitivity or efficiency theoretically under a general correlation model and realistic test errors.

    \item We propose an alternative metric for test usage called effective efficiency, defined as the number of positives identified per test consumed, which we argue captures a procedure’s effectiveness for repeated screening in epidemic control.

    \item 
    We develop a realistic agent-based simulation incorporating viral load progression and PCR tests to validate our theoretical results in the non-asymptotic regime. We show that modeling within-pool correlation under concentration-dependent test errors is crucial for decision-making, and that the effect of correlation is misrepresented under simplified test error models. Moreover, intentionally enhancing the correlation can further improve epidemic control.
\end{itemize}

The rest of this paper is organized as follows: Section~\ref{sec:litreview} reviews related work in more detail. 
Section~\ref{sec:theory} formulates the correlation model in pools derived from an asymptotic analysis of a more general population-level model and proves our main theoretical results.
Section~\ref{sec:exp} performs a case study highlighting the importance of correlation for policy-making. 
Section~\ref{sec:discussion} concludes and discusses future research.

\section{Related Work}\label{sec:litreview}
\subsection{Group Testing and Test Error Models}\label{subsec:group_testing_lit}
Group testing was proposed by \cite{dorfman1944} to screen enlisted soldiers for syphilis during World War II. The Dorfman procedure combines multiple samples and tests the pooled samples; only samples in a pool testing positive are tested individually.
This enables screening multiple individuals with a single test. Since then, many group testing protocols have been developed and studied theoretically. 
Group testing is also widely applied in the surveillance and control of various infectious diseases \citep{aprahamian2019optimal, kim2007comparison}, including COVID-19 \citep{mercer2021testing}.

The modeling of test sensitivity is a key component in understanding the performance of a group testing protocol. 
The model in \citet{dorfman1944} assumes perfect test sensitivity, which was later extended by \citet{graff1972group} to incorporate a fixed test error.
Many subsequent analyses have adopted the assumption of fixed test sensitivity, such as evaluations of test efficiency improvements in different pooling designs \citep{eberhardt2020multi}, development of more sophisticated test protocols \citep{kim2007comparison}, and estimation of disease prevalence \citep{tebbs2013two}.

However, modeling the sensitivity as a fixed constant fails to capture the concentration-dependent nature of test errors: the sensitivity of an assay, whether for pooled or individual samples, usually depends on the concentration of the analyte (e.g., virus or antigen).
Moreover, if a positive sample is combined with negative ones, the analyte concentration gets diluted \citep{wein1996pooled}. 
As a result, a pool dominated by negative samples may test negative, causing its positive members to be missed. 
This is called the \textit{dilution effect}. 
Such concentration-dependent test errors in pooled tests were first modeled by \cite{hwang1976} and incorporated in subsequent theoretical studies \citep{hung1999robustness, westreich2008optimizing, mutesa2020}.
Practically, the dilution effect has been observed in pooled testing for various diseases, including HIV \citep{kemper1998effects}, malaria \citep{bharti2009malaria, hsiang2010pcr}, and hepatitis B \citep{chatterjee2014sensitivity}.

Many studies have assessed the dilution effect in SARS-CoV-2 tests from both mathematical and empirical perspectives. \citet{pilcher2020group} assumes a temporal viral load progression in infected individuals, which, together with the detection limit of PCR tests, defines a ``window of detection"; under this setting, pooling is equivalent to raising the detection limit of the test and shortening the effective window of detection. \citet{brault2021group} proposes a similar quantification of decrease in sensitivity due to dilution based on a mathematical model for PCR. Some experimental studies \citep{Yelin2020,Lohse2020} evidence that pooling up to around 30 samples does not result in a loss of sensitivity, while \citet{bateman2020assessing} observes an increasing deterioration of sensitivity in pooling 5, 10, and 50 samples. 

\subsection{Correlation in Group Testing}\label{subsec:group_testing_correlation}
Most of the aforementioned literature assumes that the infection statuses of the samples within a pool, whether binary or not, are independent from each other. 
However, as we described in the introduction, correlation between samples is often present in reality and can potentially be leveraged for our advantage to combat the dilution effect.

One important cause of correlation is transmission within households. 
The \textit{secondary attack rate} (SAR), i.e., the probability that an infectious person in a household infects another given household member, is significant for many infectious diseases \citep{carcione2011secondary, whalen2011secondary,odaira2009assessment,meningococcal1976meningococcal,glynn2018variability}. For SARS-CoV-2, a meta-analysis \citep{madewell2020household} of 40 studies finds an average SAR of 16.6\% and a 95\% confidence interval of 14.0\%-19.3\%. 
Beyond household transmission, correlation in infection statuses among members of the same social group has also been observed among college students belonging to the same fraternity or sorority \citep{vang2021participation}, people living in the same neighborhood \citep{rader2020crowding}, and co-workers \citep{lan2020work}.

Group testing of correlated samples has not been fully explored. 
\citet{aprahamian2019optimal, bilder2012pooled, bilder2010informative, deckert2020simulation, mcmahan2012informative,mcmahan2012two} investigate group testing of a heterogeneous population with varying risk levels. 
In particular, \citet{aprahamian2019optimal} proposes risk-based Dorfman pooling designs to jointly optimize false negatives, false positives, and test consumption, with the option to consider equity across different risk groups. 
The performance measures in \citet{aprahamian2019optimal} are flexible and comprehensive, and their pooling algorithm is suitable if public health officials have detailed individual-level risk information and can dynamically implement different pool sizes. 
\citet{lendle2012group} uses simulation to show that correlation improves the efficiency of hierarchical and matrix-based group testing. 
\citet{lin2020positively} uses a regenerative process to model samples arriving at a testing site and computes the cost efficiency of group testing assuming perfect test accuracy. 
\citet{basso2021effect} models a constant pairwise correlation in infections using a Beta-Binomial distribution for the number of positives in a pool, and shows that such correlation improves efficiency.
These papers mostly focus on correlation's impact on efficiency while assuming a fixed, if not perfect, test sensitivity. As a result, the presence of correlation does not affect sensitivity.

However, correlation's impact on sensitivity has been observed empirically.
In large-scale screening conducted in Israel in 2020, \citet{barak2021lessons} finds that weakly positive samples (i.e., those with low viral load that would have likely been missed if all other samples in the pool were negative) were identified with higher probability when pooled together with strongly positive samples, which they call the \textit{hitchhiker effect}. They also observed that the sensitivity of group testing was higher than independent sampling would suggest, implying that the distribution of positive samples was not random. 

The closest paper in the literature to ours is \cite{comess2021statistical}, which is qualitatively motivated by similar considerations but makes theoretical contributions that are different in nature. 
There are two major distinctions between our work and \cite{comess2021statistical}. 

First, \cite{comess2021statistical} considers a specific model of correlation in which all participants in a pool are close contacts of each other and infections are acquired in a community infection stage followed by homogeneous secondary infections within the pool. As a result, the prevalence in the correlated pool is \textit{higher} than that in a naive pool (which only assumes community infection).
Hence, the model in \cite{comess2021statistical} is best suited for understanding the joint effect of increasing secondary transmission while pooling related samples together. 
We argue, however, that the choice of pooling strategy should be based on a comparison of their properties while holding the population's prevalence steady.
This is the approach we take in our paper.

Second, \cite{comess2021statistical} theoretically studies a different but related metric of test consumption. A theoretical result therein (Observation 5) defines an efficiency metric (the number of tests per sample) that assumes 100\% sensitivity of the pooled test and shows that the metric is identical for both pooling methods. This metric, though theoretically tractable, does not fully capture the difference in test consumption in practice, as is reported in their simulation results. Nevertheless, much of the intuition described in \cite{comess2021statistical} is consistent with our results. In particular, Observation 5 claims that the sensitivity is no worse under correlated pooling than under naive pooling. We prove a similar result in our Theorem~\ref{thm:fnr}. In addition, the simulated efficiencies in Figures 6 and 8 of \cite{comess2021statistical}, though not discussed by the authors, indicate that correlated pooling can have lower efficiency than naive pooling, which we demonstrate is possible in Section~\ref{subsec:revisit_eff}.

Beyond viral testing, group testing with correlation has been studied in the signal processing community. For example, graph structures may induce correlation among nodes and edges \citep{ganesan2017learning} or impose constraints on pool formulation \citep{cheraghchi2012graph}.

\section{Theoretical Results}\label{sec:theory}
As outlined in Section~\ref{sec:intro}, despite the recognized significance of correlation in analyzing group testing methods in the literature, our current theoretical understanding remains limited. 
Specifically, existing literature investigating correlation in sample infection status ignores a crucial factor, concentration-dependent test error, thereby yielding inaccurate conclusions. 
In this section, we aim to bridge this critical knowledge gap by studying how correlation impacts pooled testing where test error depends on the sample viral load. 
We focus on two central metrics, \textit{sensitivity} and \textit{effective efficiency}, which are crucial for evaluating the efficacy of pooling methods. We argue that effective efficiency, a novel metric we introduce, better captures a procedure's \textit{effectiveness for repeated screening} compared to the ordinary efficiency metric studied in the literature. 

\subsection{Model Setup}\label{subsec:model_setup}
We consider using pooled testing to test a large population of $N$ individuals whose viral loads are described by random variables $\{U_i : i = 1,\ldots,N\}$. \textit{Infected} individuals $i$ are those with $U_i > 0$. 

We study the two-stage Dorfman procedure \citep{dorfman1944}, in which samples are placed into non-overlapping, uniformly-sized pools. In the first stage of the Dorfman procedure, each pool is tested. In the second stage, samples from pools testing positive in the first stage are tested individually. A positive sample is correctly declared positive if and only if its pool tests positive \textit{and} it tests positive in the follow-up test.
We model the assignments of individuals to pools for testing by $\mathcal{A} := \{A_j : j=1,\ldots,N/n \}$, a partition of $\{1,\cdots,N\}$ into $N/n$ groups of size $n$.\footnote{For simplicity, we assume that $N$ is a multiple of $n$.}

We consider $A_j$ to be a random partition whose distribution depends on the pooling method used.
Specifically, under \textit{\NP} (\np), each pool is formed by picking $n$ individuals uniformly at random from the population without replacement. 
This is not a realistic model for how pooling is done in practice, but we introduce it because it is how pooling is studied in most of the academic literature.
In contrast, \textit{\CP} (\cp) is a general and more realistic structure that occurs naturally (e.g., because samples are collected from people who live in the same household or neighborhood and are tested together) and can be enhanced by explicit measures.
To support the analysis of the Dorfman procedure, which depends on the viral loads in one pool, we let $( V_{j,i} : i = 1,...,n ) = (U_i : i \in A_j) = \BFU_{A_j}$ indicate the viral loads in pool $j$.
For both correlated and naive pooling, once the pools are formed, we reorder the samples in each pool by applying independent random permutations of 1 through $n$. This simplifies analysis. 

To support varying the population size $N$ and the pooling method, we define a collection of probability measures,\footnote{
From a measure-theoretic perspective, the random quantities $\mathcal{A}$, $U_i$, and others defined below are (measurable) mappings from the event space to the outcome space. The mappings themselves do not depend on $N$, but the distributions of these random quantities under $\PP_{\np,\alpha}^{(N)}$ or $\PP_{\cp,\alpha}^{(N)}$ do because the measure itself depends on $N$.} 
$\mathbb{P}^{(N)}_{\np, \alpha}$ and
$\mathbb{P}^{(N)}_{\cp, \alpha}$, for each population size $N\in\mathbb{N}$. Under both $\mathbb{P}^{(N)}_{\np,\alpha}$ and $\mathbb{P}^{(N)}_{\cp,\alpha}$, the expectation of $\frac{1}{N} \sum_{i=1}^N 1\{U_i >0\}$ is $\alpha$ and so $\alpha$ indicates the prevalence, i.e., the probability that a person chosen uniformly at random has a positive viral load.  
Let $\EE_{\np,\alpha}^{(N)}[\cdot]$ and $\EE_{\cp,\alpha}^{(N)}[\cdot]$ denote the expectations taken under $\PP_{\np,\alpha}^{(N)}$ and $\PP_{\cp,\alpha}^{(N)}$, respectively.

\paragraph{Test outcomes.}
We model test outcomes as dependent on the sample viral loads. We first model the result of an individual test. Given an input sample with viral load $v$, we assume a test returns a positive result with probability $p(v):\mathbb{R}_{\geq 0}\to [0,1]$ and a negative result with probability $1-p(v)$. We refer to $p(v)$ as the \textit{test sensitivity function}. Here we assume $p(0)=0$, i.e., no false positives; later in Appendix~\ref{subsec:specificity}, we argue that a low individual test \textit{false positive rate} (FPR), e.g., 0.01\% \citep{ontario2020}, implies an FPR of \CP that is sufficiently low (i.e., a \textit{specificity} high enough, as specificity is 1 minus the FPR) for its deployment in repeated screening.  
We further assume that $p(v)>0$ for $v>0$, $p$ is monotone increasing in $v$, and that the result of a test, whether individual or pooled, when given its viral load, is conditionally independent from any other test. 

We denote the viral load in the pooled sample as $h(\BFv)$, where $\BFv = (v_1, \ldots, v_n)$ represents the viral loads of individual samples in the pool. We assume that the function $h(\cdot)$ is monotone increasing, meaning that for any two non-negative vectors $\BFu$ and $\BFv$ such that $\BFu \leq \BFv$ (i.e., $u_i \leq v_i$ for $i=1,\ldots, n$), $h(\BFu) \leq h(\BFv)$. 
This notation for $h(\cdot)$ is general. It allows us to model the dilution effect (see Section~\ref{subsec:group_testing_lit}) and accommodates alternative ways in which the viral load in the pooled sample depends on the individual viral loads.
When we model the dilution effect with a dilution factor equal to the pool size, as is both assumed by classical analyses \citep{zenios1998pooled} and used in empirical studies \citep{laverack2023cornell}, $h(\BFv)$ takes the average of the individual viral loads, i.e., $h(\BFv) = \frac{1}{n}\sum_{i=1}^{n}v_i = \bar{v}_n$. 
 
Consequently, a pooled test with viral loads $v_1,\cdots, v_n$ yields a positive result with probability $p(h(\BFv))$. Let $Y_j=\text{Ber}\left(p\left(h(\BFV_{j})\right)\right)$, where $\BFV_{j}$ represents the vector of $(V_{j, 1},\ldots, V_{j, n})$, denote the outcome of the pooled test of pool $j$ in the first stage. Let $W_{j,i}=\text{Ber}\left(p(V_{j,i})\right)$ denote what the outcome of the individual test for sample $i$ with viral load $V_{j,i}$ will be, if it is performed. 
Let $S_j =\sum_{i=1}^{n}1\{V_{j,i} > 0\}$ denote the number of infected individuals
and $D_j =\sum_{i=1}^{n}Y_{j}W_{j,i}$ the number of positives identified in  pool $j$.  
The conditional independence assumption stated above implies that the pooled and individual tests are conditionally independent given the viral loads of the participating samples.

\paragraph{Population-level outcomes.}
We define three population-level averages of pooling outcomes:
$\overline{S} = \frac1{|\mathcal{A}|} \sum_{j=1}^{|\mathcal{A}|} S_j$ and
$\overline{D} = \frac1{|\mathcal{A}|} \sum_{j=1}^{|\mathcal{A}|} D_j$,
which are the average number of infected individuals present and detected
per pool;
and $\overline{Y} = \frac1{|\mathcal{A}|} \sum_{j=1}^{|\mathcal{A}|} Y_j$,
which is the fraction of pools testing positive.

\subsection{Metrics of Interest}\label{subsec:metric_of_interest}
We will study two metrics, sensitivity\footnote{The sensitivity metric here is used to describe the \textit{overall} accuracy of a testing protocol and should not be confused with test sensitivity, which refers to the accuracy of a single test.} and effective efficiency, characterizing the performance of a pooling method on a population. They are central to summarizing a pooling method's utility for controlling the spread of infections.

\begin{definition}[Sensitivity]\label{def:sn}
Let $\beta$ denote the \textit{overall false negative rate}, or the fraction of positive samples falsely declared negative under the Dorfman procedure. That is, $\beta = 1 - \dfrac{\overline{D}}{\overline{S}}$. Sensitivity is defined as $1-\beta$.

\end{definition}
\begin{definition}[Effective Efficiency]\label{def:eff_eff}
Effective efficiency, denoted by $\gamma$, is defined as the number of positive cases identified per test consumed (including pooled and follow-up tests).
That is, $\gamma=\dfrac{\overline{D}}{1+n\overline{Y}}$.
\end{definition}

We propose the effective efficiency metric as an \textit{alternative} to the efficiency metric used in the literature to study test consumption. 
It also complements existing metrics in the literature balancing test accuracy and test consumption \citep{aprahamian2019optimal}.
The metric studied in \citet{aprahamian2019optimal} is suitable when the policy-maker knows the relative importance of different objectives, such as test accuracy and test consumption. In comparison, our metric is both interpretable and of significance in epidemic control, which we contextualize using a susceptible-infected-removed (SIR) model \citep{kermack1927contribution} in Appendix~\ref{subsec:SIR}.

To see the significance of the sensitivity and effective efficiency metrics described above, consider problem settings of selecting a repeated screening strategy in a large population, such as the ones we later study in Section~\ref{sec:exp} and Appendix~\ref{subsec:SIR}.
The ability of a testing method to control infections is largely determined by the rate at which it can identify positive individuals in a population and reduce the number of positives missed in screening: when an infected individual is identified, they can be isolated, preventing them from infecting other individuals and reducing the number of future infections.

Suppose we have a budget $bN$ ($b>0$) for the number of tests available that scales with the population size $N$. 
When $b$ is large, the rate at which we can test a person is not constrained by the testing budget. Thus, the number of positives found is determined by the sensitivity. 
When $b$ is smaller, the rate at which we can test a person is proportional to the testing budget. Therefore, the number of positives found is determined by the product of the testing budget and the effective efficiency. We include an in-depth discussion of these metrics and the ordinary efficiency metric in Section~\ref{subsec:revisit_eff}.

We will show in Section~\ref{subsec:theoretical_results} that \CP achieves asymptotically higher sensitivity and, under a mild condition, has an asymptotically higher effective efficiency. We illustrate these findings in the context of a realistic epidemic simulation in Section~\ref{sec:exp}.

\subsection{From Population-Level Model to Single-Pool Model}\label{subsec:convergence}
To support \textit{tractable} analysis of these metrics, we introduce an analytical framework for modeling pooling methods in large populations. This framework can be adapted for assessing various test procedures, including but not limited to the two-stage Dorfman procedure. We focus on the limit as the population size $N$ becomes large, enabling us to characterize the population-level outcomes using a simpler-to-analyze \textit{single-pool model}.

The key idea in this analysis is to let $J$ be a pool chosen uniformly at random from $\{1,\ldots,N/n\}$.  We then define quantities for this single pool that are analogous to the population-level quantities: let $V_i = V_{J,i}, i=1,\ldots,n$; $S=S_J$; $W_i=W_{J,i}, i=1,\ldots,n$; $Y=Y_J; D=D_J$.

Let $\PP_{\pool,\alpha}$ be the limiting joint distribution of the single-pool quantities $(V_i : i=1,\ldots,n)$, $S$, $(W_i : i=1,\ldots,n)$, $Y$, and $D$ as $N\to\infty$, for $\pool\in\{\np,\cp\}$.
Such convergence is consistent with the idea that adding one more person to the population should not radically change what happens to a single pool chosen at random from all pools.
We refer to this as the {\it single-pool model}. We show that, with the assumptions described below, the population-level quantities ($\overline{D}$, $\overline{S}$, and $\overline{Y}$) converge to constants as $N\to\infty$.

\paragraph{A model of association.}
We define a measure of association between the viral loads in $A_k$ (i.e., the $k$th pool by pooling assignment $\A$), and a random variable $X$ (e.g., a pool-level quantity in a different pool). This measure quantifies the extent to which the viral load in $A_k$ affects the distribution of random variable $X$:
\begin{align*}
    \Delta_{\pool,\alpha}^{(N)}(X, k) = \sup_{\BFu\in\mathbb{R}_{\geq 0}^{n}}\left|\EE^{(N)}_{\pool,\alpha}[X \mid \BFU_{A_k} = \BFu] - \EE^{(N)}_{\pool,\alpha}[X]\right|.
\end{align*}

For a fixed $k$, and a specific pool-level quantity $Z_j$ (i.e. $Z_j$ is one of $S_j$, $Y_j$, or $D_j$), we can define the set of indices $j$ ($j\neq k$) such that $Z_j$ has an association with $\BFU_{A_k}$ stronger to $\epsilon$:
$$
\left\{j: j\neq k,\ \Delta_{\pool,\alpha}^{(N)}(Z_j,k)>\epsilon\right\}.
$$

We denote by $m_{\pool,\alpha}^{(N)}(\epsilon, Z_{1:|\A|})$ the maximum size of such sets, across $k\in \{1,\cdots, |\A|\}$:
$$
m_{\pool, \alpha}^{(N)}(\epsilon, Z_{1:|\A|}) = \max_{k\in\{1,\cdots, |\A|\}} \left|\left\{j: j\neq k,\ \Delta_{\pool,\alpha}^{(N)}(Z_j,k)>\epsilon\right\}\right|.
$$

Now, we take $N$ to the asymptotic regime and make the following assumption: 
\begin{assumption}
   For $Z_{1:|\A|}\in\{S_{1:|\A|}, Y_{1:|\A|}, D_{1:|\A|}\}$ and $\pool\in\{\np,\cp\}$, there exists a sequence $\epsilon_N\downarrow 0$ such that $\lim_{N\rightarrow \infty}\frac{1}{N}m_{\pool, \alpha}^{(N)}(\epsilon_N, Z_{1:|\A|})=0$.
\label{assu:corr_decay}
\end{assumption} 

Assumption~\ref{assu:corr_decay} prescribes that as population size $N$ goes to infinity, the set of pool-level quantities that have association stronger than $\epsilon_N$ with the viral loads in pool $k$ grows slower than linearly in population size. In an epidemic like COVID-19, transmission typically takes place between close contacts \citepAP{world2020modes}. It is reasonable to assume that for two individuals to be associated in infection statuses, they have to be within a few degrees of contact with each other. Since the duration of the infectious period is finite, and a person's contact rate is typically bounded above by some constant \citepAP{hu2013scaling} even as population size grows large, the number of people connected to any individual in pool $k$ via within a few degrees of contact grows slower than linearly in population size. Because the pool-level quantities are conditionally independent from viral loads in other pools (given viral loads in the pool), the sub-linearity should be inherited. Hence, this assumption is well-justified.

It follows that as $N\to\infty$, the metrics of interest outlined in Section~\ref{subsec:metric_of_interest} converge in probability to their corresponding single-pool values, as guaranteed by the continuous mapping theorem.

\begin{proposition}\label{prop:convergence_of_metrics}
Under Assumption~\ref{assu:corr_decay}, the metrics $\beta$ and $\gamma$ converge in probability to their corresponding single-pool values, i.e., $1 - \frac{\EE_{\pool,\alpha}[D]}{\EE_{\pool,\alpha}[S]}$, $\frac{\EE_{\pool,\alpha}[D]}{1 + n\EE_{\pool,\alpha}[Y]}$ (denoted $\beta_{\pool,\alpha}$ and $\gamma_{\pool,\alpha}$ thereafter), respectively, as $N\to\infty$, for $\alpha>0$ and $\pool\in \{\np,\cp\}$.
\end{proposition}

Proposition~\ref{prop:convergence_of_metrics} justifies the analysis of metrics within the single-pool model because they characterize the asymptotic behavior of the population-level metrics of interest. 

\subsection{Properties of the Single-Pool Model}\label{subsec:single_pool_model}

Having justified the use of a single pool selected uniformly at random (i.e., pool $J$) for analyzing the asymptotic performance of a pooling method, we proceed to examine the single-pool values corresponding to the population-level metrics of interest, under probability measure $\PP_{\pool,\alpha}$, for $\pool\in\{\np, \cp\}$. 

To achieve this, it is essential to understand the fundamental distinction between naive and correlated pooling. In this section, we introduce Propositions~\ref{proposition:asymp_indep} and \ref{proposition:corr_pool}, which collectively highlight the primary feature differentiating the two pooling methods: as prevalence approaches zero, the probability that a positive-containing pool contains more than one positive sample diminishes for a randomly selected naive pool (Proposition~\ref{proposition:asymp_indep}) but
persists for a correlated pool (Proposition~\ref{proposition:corr_pool}). 

Under \np, pools are created by selecting $n$ individuals uniformly at random without replacement. This pooling method intuitively reduces correlation within pool $J$ as $N$ approaches infinity, as we demonstrate below.

\paragraph{A second model of association.} Analogous to the association model introduced in Section~\ref{subsec:metric_of_interest}, we further define a second measure of association, between the viral loads of one individual $i$ and a group of individuals $\BFj$ whose population indices are denoted $\{j_1,\cdots,j_{|\BFj|}\}$: 
\begin{align*}
    \Lambda_{\alpha}^{(N)}(i,\BFj) = \sup_{\substack{u\in\mathbb{R}_{\geq 0}\\\BFu\in\mathbb{R}_{\geq 0}^{|\BFj|}}} \left|\PP^{(N)}_\alpha(U_i\leq u\mid U_{\BFj} = \BFu) - \PP^{(N)}_\alpha(U_i\leq u)\right|.
\end{align*}
where $U_{\BFj} = (U_{j_1},\cdots, U_{j_{|\BFj|}})$. This measure quantifies the maximum change in the cumulative distribution function of $i$'s viral load with respect to the viral loads of $\BFj$. It reflects the degree to which conditioning on the viral loads of $\BFj$ affects the viral load of $i$. A larger $\Lambda_{\alpha}^{(N)}(i,\BFj)$ indicates a stronger association between $i$ and $\BFj$. The collection of individuals having association with $\BFj$ stronger than $\epsilon$ is 
\begin{align*}
    \{i:i\notin {\BFj}, \Lambda_{\alpha}^{(N)}(i,\BFj)>\epsilon\}.
\end{align*}

We denote by $d_{\alpha}^{(N)}(\epsilon)$ the maximum size of such sets, across any collection $\BFj$ of at most $n-1$ individuals:
\begin{align}
    d_{\alpha}^{(N)}(\epsilon) = \max_{\substack{\BFj\subset \{1,\cdots,N\}\\|\BFj|<n}} \left|\{i: i\notin {\BFj}, \Lambda_{\alpha}^{(N)}(i,\BFj)>\epsilon\}\right|.\label{eq:pop_model_d_def}
\end{align}
If $d_{\alpha}^{(N)}(\epsilon)$ is small relative to $N$, when we add an individual $i$ to the pool who is chosen uniformly from the larger population, they are unlikely to be in a set with high association $\Lambda_{\alpha}^{(N)}(i,\BFj)>\epsilon$ with the individuals already in the pool. This makes the viral loads in the pool unlikely to be strongly correlated. 

Now, we take $N$ to the asymptotic regime and make the following assumption. 
\begin{assumption}
There exists a sequence $\epsilon_N\downarrow 0$ such that $\lim_{N\rightarrow \infty}\frac{1}{N}d_{\alpha}^{(N)}(\epsilon_N)=0$.
\label{assu:corr_decay_indep}
\end{assumption}

Assumption~\ref{assu:corr_decay_indep} prescribes that as population size $N$ goes to infinity, for any collection $\BFj$ of individuals of size less than $n$, the set of individuals having association stronger than $\epsilon_N$ with $\BFj$ grows sublinearly in population size. The same arguments for Assumption~\ref{assu:corr_decay} apply when justifying this assumption. 

We show that, under Assumption~\ref{assu:corr_decay_indep}, samples in pool $J$ are independent, consistent with the conventional assumption commonly made in pooled testing analyses. This independence result aligns with what a policy-maker would assume for a finite population if they do not account for correlation. 

\begin{proposition}
\label{proposition:asymp_indep}
Under Assumption~\ref{assu:corr_decay_indep}, the viral loads in pool $J$ are independent under $\PP_{\np,\alpha}$. 
\end{proposition} 

Unlike in \np, samples within the correlated pool are expected to display distinct behavior due to their inherent correlation. To quantify such behavior, we characterize the correlation between viral loads in a correlated pool based on the notion of ``close contacts". Specifically, we assume that infected individuals and their close contacts are correlated in infection statuses and are likely to be placed into the same pool under \cp.  These assumptions are formalized mathematically in Assumption~\ref{assu:justify_CP}, and we leverage them to derive Proposition~\ref{proposition:corr_pool}.

\begin{assumption}
For each individual $i$ in the population, let $C_i$ denote the set of their close contacts. We model $C_i$ as deterministic. The following conditions hold: 
\begin{enumerate}
    \item (Bounded infection risk) For any $\alpha$, $\PP_{\alpha}^{(N)}(U_i > 0)\in \{0\}\cup [\epsilon_0\alpha,\Pi_0\alpha]$ where $0 <\epsilon_0 \leq 1\leq \Pi_0$.

    \item (Existence of close contacts for non-isolated individuals) $C_i\neq \emptyset$ if $\PP_{\alpha}^{(N)}(U_i>0) > 0$.

    \item (Correlation in infection status) There exists $c_1>0$ such that $\PP^{(N)}_{\alpha}(U_j>0\mid U_i>0)\geq c_1$\ $\forall j\in C_i$. This holds for any $\alpha$ and any $N$. 

    \item (Correlated pooling) There exists $c_2>0$ such that $\PP^{(N)}_{\cp,\alpha}(j \text{ is in the same pool as }i)\geq c_2$ $\forall j\in C_i$. This holds for any $\alpha$ and any $N$.
    \end{enumerate}
\label{assu:justify_CP}
\end{assumption}

Assumption~\ref{assu:justify_CP} captures important features of the spread of infectious diseases and the correlated pooling method.
The first sub-assumption prescribes that each individual in the population either (i) cannot be infected due to social isolation; or (ii) may be infected but the bounds of infection risk fall within the same order as the population-level prevalence. 
The second sub-assumption is justified because individuals with non-zero infection risk must have some degree of human-to-human contact. 
The third sub-assumption finds ample support in the literature regarding transmission between infected individuals and their close contacts \citep{world2020modes, madewell2020household}. 
The fourth sub-assumption describes the key feature assumed for correlated pools, namely that individuals who are close contacts of each other are placed into the same pool with a non-vanishing probability, even as $N$ goes to infinity. This is justified because in large-scale screening using group testing, correlation either arises naturally or can be enhanced through explicit measures, as discussed later in Section~\ref{subsec:motivation}. Together, they allow us to derive the following property of pool $J$ under $\PP_{\cp,\alpha}$.
\begin{proposition}
\label{proposition:corr_pool}
Under Assumption~\ref{assu:justify_CP}, $\PP_{\cp,\alpha}(S>1\mid S>0)>0$ for any $\alpha$.
\end{proposition}

Propositions~\ref{proposition:asymp_indep}~and~\ref{proposition:corr_pool} lay the foundation for our main theoretical results discussed in Section~\ref{subsec:theoretical_results}.

\subsection{Main Theoretical Results}\label{subsec:theoretical_results}
Building upon the properties of the single-pool model outlined above, we establish our primary theoretical findings. Specifically, we prove that \CP attains asymptotically higher sensitivity and, under a mild condition, achieves asymptotically higher effective efficiency. To the best of our knowledge, we are the first to (1) theoretically show that \CP has better sensitivity, and (2) theoretically characterize the effect of \CP on test usage while modeling concentration-dependent test errors.

First, we show that under a general class of test sensitivity functions, \cp achieves asymptotically higher sensitivity than \np in low-prevalence settings. We approach this by setting $\alpha\to 0^+$, as it facilitates tractable analysis. In addition, during the early stage of an epidemic when group testing protocols are considered for repeated screening, prevalence tends to be low.
\begin{theorem}\label{thm:fnr}
If $p(v)$ is monotone increasing in $v$, $\lim_{\alpha\rightarrow 0^+}\beta_{\np,\alpha} \geq \lim_{\alpha\rightarrow 0^+}\beta_{\cp,\alpha}$. If, in addition, $p(\cdot)$ and $h(\cdot)$ are both strictly monotone increasing, then the inequality is strict. \footnote{Here, $h(\cdot)$ is strictly monotone increasing if $h(\BFu)<h(\BFv)$ whenever $\BFu\geq \BFv$ but $\BFu\neq \BFv$.}
\end{theorem}

\proof{Proof sketch of Theorem~\ref{thm:fnr}.}

For $\pool\in\{\np,\cp\}$, we can show that the overall false negative rate is given by
$\beta_{\pool,\alpha} = 1 - \EE_{\pool,\alpha}\left[p\left(h(\BFV)\right) p(V_1)\mid V_1 > 0\right]$

For \NP, the $V_i$'s are i.i.d. As $\alpha\to 0^+$, the probability that a positive pool contains multiple positive samples vanishes, and we can show that $\lim_{\alpha\to 0^+}\beta_{\np,\alpha} = 1 - \EE\left[p\left(h(V_1,0,\ldots,0)\right) p(V_1)\mid V_1 > 0\right]$.

For \CP, a positive pool contains multiple positives with non-negligible probability, so we can write $\beta_{\cp,\alpha} = 1 - \sum_{\ell = 1}^n A_{\ell}\cdot \PP_{\cp,\alpha}(S=\ell\mid S > 0)$, where 
$A_\ell\overset{\Delta}{=}\EE_{\cp,\alpha}\left[p\left(h(\BFV)\right) p(V_1)\mid V_1 > 0, S=\ell\right]$.
When $\ell = 1$, 
$A_1 = \EE[p(h(V_1,0,\ldots,0))p(V_1)\mid V_1 > 0]=1-\lim_{\alpha\to 0^+}\beta_{\np,\alpha}$.
When $\ell \geq 2$, we have $h(\BFV) \geq h(V_1,0,\ldots, 0)$ because there exists at least one $i\neq 1$ such that $V_i > 0$ and $h(\cdot)$ is monotone increasing as described in Section~\ref{subsec:model_setup}. Assuming $p(v)$ is a monotone increasing function in $v$, we obtain $p(h(\BFV)) \geq p(h(V_1,0,\ldots, 0))$, which, combined with $p(V_1) > 0$ given $V_1 > 0$, implies that $A_\ell \geq A_1$. 

Therefore, taking $\alpha\to 0^+$ gives $\lim_{\alpha\to 0^+}\beta_{\cp,\alpha}  \leq \lim_{\alpha\to 0^+}\beta_{\np,\alpha}$. The inequality is strict if both $p(\cdot)$ and $h(\cdot)$ are strictly monotone increasing.
\Halmos
\endproof

Second, in Theorem~\ref{thm:eff}, we show that in the low prevalence setting, the Dorfman procedure using \CP achieves no lower effective efficiency than using \NP, up to a constant multiplier. This multiplier is determined by the viral load distribution among infected individuals, the test sensitivity function, and the pooling method. 
\begin{theorem}
    $\lim_{\alpha\to 0^+}\dfrac{\gamma_{\cp,\alpha}}{\gamma_{\np,\alpha}}\geq (1 + \delta)^{-1}$, where $\delta = \dfrac{\PP_{\cp,\alpha}(Y=1, \Dt=0\mid S>0)}{\PP_{\cp,\alpha}(Y=1, \Dt > 0\mid S>0)}$ and $\Dt = \sum_{i=1}^{n} W_i$.\label{thm:eff}
\end{theorem}
Here, $\Dt$ represents the number of positives identified if individual tests are performed on the samples in the pool.\footnote{Here, $\Dt$ differs from $D$ in that only individual tests are performed on the samples, i.e., the pooled test is not conducted. As a result, $\Dt\leq D$ a.s.} Thus, the constant $\delta$ can be understood as the odds of follow-up test failures. It is the ratio between the probabilities of a positive-containing correlated pool testing positive at the pooled testing stage but failing to identify any positives individually, versus testing positive and having at least one positive identified individually.

In a special case where a test result reports whether the sample viral load exceeds a threshold value and sample viral loads are diluted by a factor equal to the pool size, \cp consumes no more tests per positive case identified than \np, as formulated in Corollary~\ref{Cor:thm2_det_threshold_case}.

\begin{corollary}\label{Cor:thm2_det_threshold_case}
Suppose the sensitivity function is $p(v)=\mathbbm{1}\{v\geq u_0\}$ for some non-negative constant $u_0$ and the viral load in the pooled sample is $h(\BFv)=\frac{1}{n}\sum_{i=1}^{n}v_i$. Then, $\lim_{\alpha\to 0^+}\dfrac{\gamma_{\cp,\alpha}}{\gamma_{\np,\alpha}}\geq 1$.
\end{corollary}

In the real world, the PCR test sensitivity, albeit not exactly a step function of the viral load $v$, closely resembles the one in Corollary~\ref{Cor:thm2_det_threshold_case} in that it increases rapidly from zero to one within a narrow range of $v$.
(See, e.g., Figure~\ref{fig:indiv_test_sensitivity} in Section~\ref{subsec:pcr}.)
Appendix~\ref{appdx:delta} further shows that, under a realistic test sensitivity function, viral load distribution, and pool size, $\delta$ is at most on the order of $10^{-4}$ and the bound in Theorem~\ref{thm:eff} is almost equal to one.

\subsection{Revisiting Efficiency}\label{subsec:revisit_eff}
We now revisit the conventional efficiency metric studied in the literature, i.e., the number of people screened per test. We make two key arguments. First, \CP can have lower efficiency in reality, contrary to the findings in existing studies that model test errors as independent of viral loads. Second, our effective efficiency metric is a better metric than efficiency for evaluating pooling designs for epidemic control.  

To support our statements, we first observe that efficiency can be expressed for any prevalence $\alpha$ in terms of $\beta_{\pool,\alpha}$ and $\gamma_{\pool,\alpha}$ as follows:
\begin{align}
    \text{efficiency}_{\pool,\alpha} = \frac{n}{1 + n\EE_{\pool,\alpha}[Y]}
    = \frac{\gamma_{\pool,\alpha}}{(1-\beta_{\pool,\alpha})\alpha}.
    \label{eq:eff_as_eta_beta}
\end{align}

Existing literature assuming perfect or fixed sensitivity finds that within-pool correlation leads to better efficiency \citep{comess2021statistical, augenblick2020group, lendle2012group, deckert2020simulation, lin2020positively, basso2021effect}. 
However, this does not hold in general under concentration-dependent test errors.
In fact, \cp's improved sensitivity can detract from its efficiency. 
Correlated pooling can identify positive-containing pools that would have tested negative under naive pooling due to the dilution effect.
This results in more follow-up tests (i.e., a higher $n\EE_{\pool,\alpha}[Y]$ in Equation~\ref{eq:eff_as_eta_beta}). 
This effect can outweigh the reduction in the number of positive-containing pools caused by correlation, leading to a \textit{lower} efficiency than naive pooling. 
Indeed, Appendix~\ref{appdx:example} shows a stylized example where correlated pooling has lower efficiency in pools of size two. 
The same phenomenon can occur whenever the dilution effect prevents a test from identifying a single positive in a pool, but allows it to detect two or more positives.
Under low prevalence, positive pools created by \NP typically have just one positive sample, testing negative. Correlated pooling will create more pools with multiple positives, leading to positive pooled test results and requiring more follow-up tests.

Although \CP can decrease efficiency, we argue that efficiency should not be the sole criterion for evaluating a pooling procedure for epidemic control. 
A pooling procedure with low sensitivity would incur few follow-up tests, resulting in high efficiency, but it would miss a large number of positives, failing to control the epidemic. 
This defies the purpose of epidemic mitigation.
However, maximizing sensitivity alone brings us to the opposite extreme of using individual tests, incurring high test consumption.  Appendix~\ref{sec:results} dives deeper into this tradeoff between sensitivity and efficiency.

Our effective efficiency metric precisely balances this tradeoff.
In fact, Equation~\ref{eq:eff_as_eta_beta} shows that it is proportional to the product of sensitivity and efficiency.
As discussed in Section~\ref{subsec:metric_of_interest}, effective efficiency quantifies the rate of identifying positives under constraints on test budget. 
As such, it characterizes the true utility of consuming one test.
Therefore, under limited test budget, one should choose the pooling procedure that maximizes the effective efficiency to optimize the epidemic control performance.
We contextualize this argument using an SIR model \citep{kermack1927contribution} in Appendix~\ref{subsec:SIR}.

\section{Case Study}\label{sec:exp}
We conduct a case study using an agent-based simulation to elucidate the implications of correlated pooling under a realistic concentration-dependent test-error model for decision-making.
0We mimic the decision-making process of a policy-maker facing an emerging epidemic, who uses simulation to evaluate and select policies for population-wide screening.
We first show that modeling concentration-dependent test errors, compared to assuming a fixed test sensitivity, is essential for accurately quantifying the benefit offered by correlation. 
Then, we argue that a policy-maker who does not account for the naturally occurring within-pool correlation would underestimate the power of population-wide screening and make suboptimal policy choices.
Moreover, we demonstrate that taking measures to enhance within-pool correlation can further improve epidemiological outcomes.
Separately, Appendix~\ref{appdix:static_sim} uses simulation to study how correlation influences the fundamental performance characteristics of pooled testing in a simplified setting without epidemic dynamics. 

\subsection{Motivation and Summary of Findings}\label{subsec:motivation}
Consider a policy-maker facing an emerging epidemic. To curb virus spread, they consider using pooled testing followed by isolation of individuals testing positive.\footnote{
In practice, large-scale screening can complement other mitigation measures, such as contact tracing. Positives missed in contact tracing can be found in screening.} 
They utilize an agent-based simulation to make decisions such as choosing between lockdown and pooled testing or designing the pooled testing policy.
They vary the \textit{policy} (pool size and testing frequency)\footnote{
Within the scope of this paper, we assume that the same screening frequency and pool size apply throughout the period.
Several practical constraints call for a screening policy to remain unchanged over time: labs would face difficulties in altering their established pooling protocols, and public health workers would have to adjust their operations to varying testing frequencies. 
An interesting future direction is to study adaptive schemes that adjust the screening frequency and pool size based on evolving prevalence and network structure while accounting for correlation.
} in simulation and focus on the cumulative number of infections and test consumption,\footnote{In Section~\ref{subsec:revisit_eff}, we argued for maximizing the effective efficiency when designing a pooling procedure. 
However, cumulative infections and test consumption are of more direct interest in policy-making.} considering a policy \textit{attainable} if its test consumption is below a threshold. 
Among the set of attainable policies, the policy-maker chooses the one that minimizes the cumulative number of infections according to their favorite modeling assumptions. 
If the minimal number of cumulative infections is below a threshold, the policy-maker implements it, keeping the economy open. 
Otherwise, if no policy is attainable or the optimal attainable cumulative number of infections exceeds the tolerance, they issue a lockdown.

We consider two broad types of policy implications of our work, one related to modeling correlation, and the second related to actively enhancing correlation.

\paragraph{Modeling correlation.}
Correlation in pools occurs naturally due to interactions within communities at neighborhoods, schools, workplaces, and households. A policy-maker might choose to actively model this correlation when making a decision ({\it correlation-aware}), or choose to ignore this correlation and treat infection status as independent ({\it correlation-oblivious}). 

We also consider a policy-maker's decision on whether to model concentration-dependent test errors, due to its interaction with correlation. 
We consider a policy-maker as choosing between a model that accurately represents how an assay's sensitivity depends on the concentration of the analyte (\textit{assay-aware}) or a model that assumes an idealized assay whose sensitivity is fixed at 80\% (\textit{assay-oblivious}). We choose 80\% because our PCR model is calibrated to have an average sensitivity of 80\% for a representative population (Appendix~\ref{appdx:PCR}).

\begin{figure}
     \FIGURE
     {\includegraphics[width=0.5\textwidth]{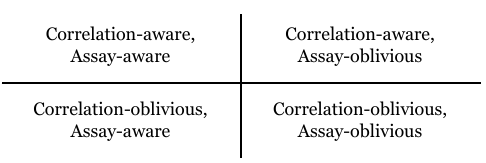}}
     {Modeling choices of correlation and test errors. The top left is the closest to reality. \label{fig:modeling_choice_quadrant}}
     {}
 \end{figure}

We then present evidence that a policy-maker should be \textit{both} correlation-aware \textit{and} assay-aware (i.e., in the top left quadrant of the table in Figure~\ref{fig:modeling_choice_quadrant}). Ignoring either aspect leads to predictions for sensitivity and test efficiency that are significantly different from reality (Section~\ref{subsec:corr_assay_modeling}) and significantly suboptimal decisions (Section~\ref{subsec:corr_assay_decisions}). 

Moreover, the impact of modeling correlation on outcomes (being correlation-aware versus correlation-oblivious) depends strongly on whether we are assay-aware (Section~\ref{subsec:corr_assay_modeling}). In the more realistic assay-aware model the impact is strong, while in the unrealistic assay-oblivious model the impact is much weaker. 

\paragraph{Enhancing correlation.}
In addition to highlighting the importance of modeling correlation, our theoretical findings also suggest benefits in intentionally enhancing correlation. Members within the same household exhibit an even stronger correlation in infections, compared to those in the same community, due to their close and extended daily interactions. Thus, increasing the chance that samples from the same household are pooled together could enhance correlation and increase the performance of pooled testing. Practical measures to achieve this include encouraging household members to get tested together, or mailing test kits to each household for household members to self-collect and pool samples together. We call such policy-makers \textit{correlation-enhancing}. 

We present evidence in this case study that such efforts would deliver benefits in terms of infection control and test consumption (Section~\ref{subsec:corr_as_intervention}).
While we acknowledge that preserving household-induced correlation in pools is more challenging than allowing community-induced correlation to occur naturally, we view one of the benefits of our work as helping to quantify the benefits of such an approach.

\subsection{Simulation Setup}\label{subsec:sim_setup}
We conduct realistic agent-based simulation to understand the policy implications of accurately modeling correlation (Section~\ref{subsec:sim_corr}) using concentration-dependent test errors (Section~\ref{subsec:pcr}). We show that failure to model either part leads to suboptimal policy decisions.

\subsubsection{Screening and pooling in a social network}
We use the \texttt{SEIRSplus} library \citep{seirsplus} to simulate epidemic progression on a realistic social network comprising households of different sizes and community structures.
We simulate 10,000 individuals in households, whose sizes follow the United States' distribution of household sizes.\footnote{We obtained qualitatively similar results using a network of 5,000 individuals.}
Each household forms a complete subgraph, complemented by inter-household edges. 
We divide the population into equally-sized fixed screening groups, screen one group every day, and rotate through all groups repeatedly. 
On each day, we allocate the individuals in the screening group into pools using one of the pooling methods and conduct two-stage Dorfman testing. 
Positive cases are isolated, with isolated individuals excluded from screening and contact with others.
We simulate candidate screening policies that screen every one to seven days with pool sizes of 5, 10, 15, and 20, resulting in 28 policies in total.\footnote{More infrequent testing would consume fewer tests but lead to even more infections. Since testing every seven days results in at least 40\% of the population getting infected for all pooling methods, we assume policy-makers do not consider lower frequencies.
Practically, pools of size 5 to 25 have been used in large-scale screening \citep{wuhan2020, cornell2020, caschools2021, han2022sars, mendoza2021implementation}, so we choose 5, 10, 15, and 20 as representative pool sizes.}

\subsubsection{Correlation in infections}\label{subsec:sim_corr}
To support our case study, we simulate within-pool correlation in three different ways: \textit{naive pooling} (\np), \textit{community-correlated pooling} (\texttt{CCP}), and \textit{household-correlated pooling} (\texttt{HCP}).
Among them, we assume that \texttt{CCP} accurately captures the natural within-pool correlation arising in realistic large-scale screening and that \texttt{HCP} accurately models outcomes when pooling is enhanced by encouraging household members to be pooled together. \np is inaccurate and models the beliefs of a correlation-oblivious decision-maker.

Both the assignment of screening groups and the formation of community and household-correlated pools are implemented using a node embedding and clustering procedure (Appendix~\ref{appdix:node_clustering_desc}).
This approach tends to assign individuals with close network proximity to the same screening group.
For \texttt{CCP} and \texttt{HCP}, we use the same procedure to allocate individuals within a screening group to pools. 
We design our algorithm such that household-correlated pools mostly contain members of the same households. 
This, combined with rapid virus spread within households, results in high within-pool correlation for household-correlated pools. In contrast, community-correlated pools exhibit weaker within-pool correlation. (See Appendix~\ref{appdx:validate_node_clustering} for numerical evidence.)
To focus on the effect of correlation rather than a change in the marginal distribution of infection status, we implement \texttt{NP} by randomly permuting the individuals and placing them sequentially into pools regardless of the social network structure. 

To study the impact of being correlation-aware in modeling (but not actively enhancing it), we will compare decisions made by the correlation-oblivious decision-maker who bases their decisions on the \texttt{NP} simulation with those made by the correlation-aware decision-maker who bases their decisions on the \texttt{CP} simulation. The quality of their decisions will be evaluated by the (more accurate) \texttt{CP} simulation.

To study the impact of intentionally enhancing correlation, we will compare the predictions of the \texttt{HCP} simulation, which corresponds to a simulation where correlation has been intentionally enhanced by pooling households together, with those of the \texttt{CCP} simulation, where the only correlation is that occurring naturally due to community-level correlation.

\begin{figure}
 \FIGURE
 {\includegraphics[width=0.95\textwidth]{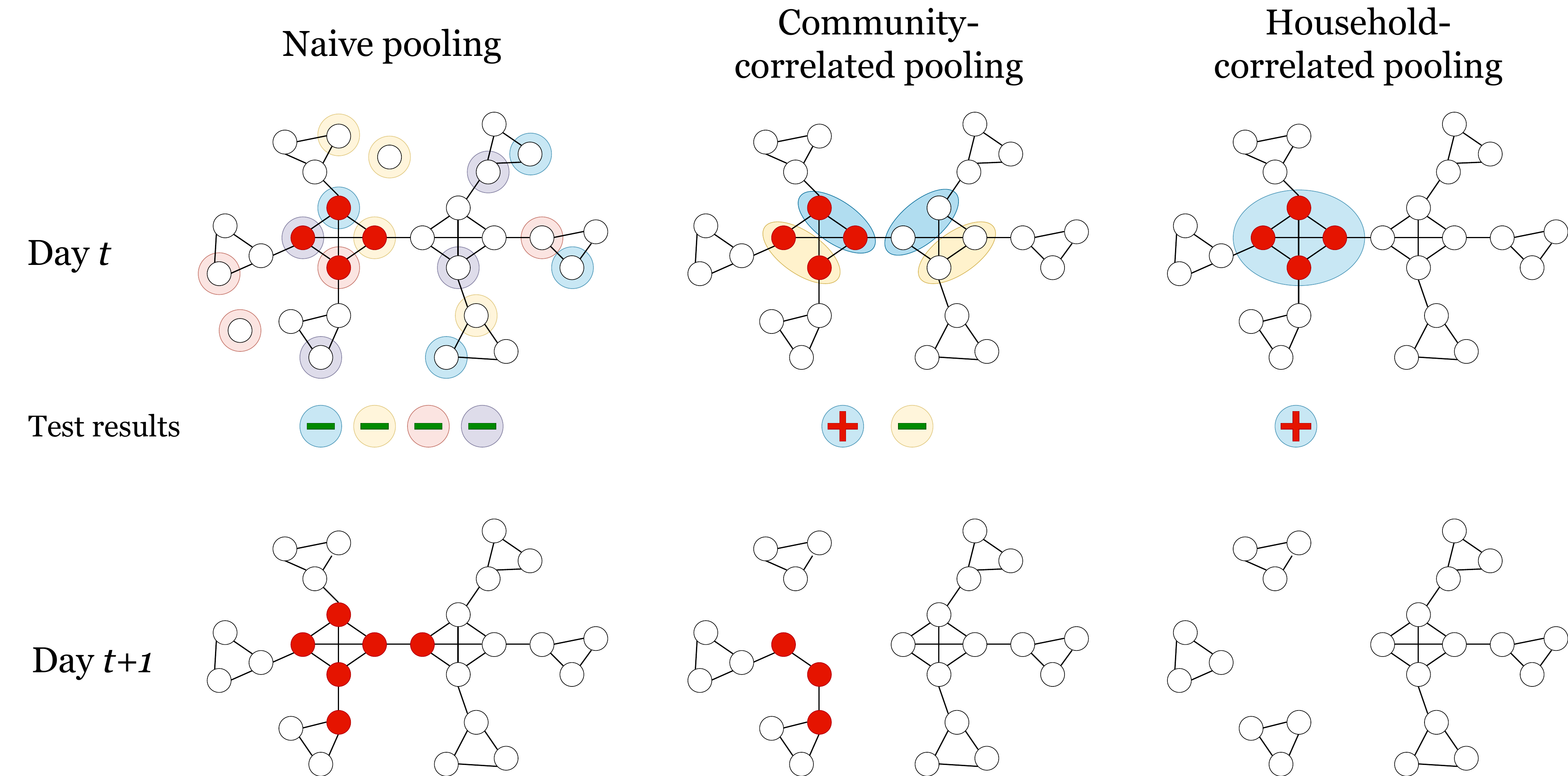}}
 {Schematic illustration of \texttt{NP}, \texttt{CCP}, and \texttt{HCP} on a simple social network with one infected four-person household, all using pools of size four. \vspace{0.1in} \label{fig:pooling_network_schematic}}
 {Each node represents one individual and each colored ellipse represents one pool. 
 Under \texttt{NP}, the four infected individuals are placed into four different pools. With none of the pools testing positive due to dilution, they are not isolated and generate two new infections the next day. 
Under \texttt{CCP}, the four infected individuals are placed into two pools (blue and yellow). Only the blue pool tests positive. The two infected individuals in the yellow pool are not identified and generate one new infection the next day.
Under \texttt{HCP}, all four infected individuals are placed into the same pool (blue), identified, and isolated.  
}
\end{figure}

Figure~\ref{fig:pooling_network_schematic} illustrates the qualitative differences in pooling and testing between the three pooling methods and their implications for epidemic control. 
Under naive pooling, infected members in the same household are dispersed across different pools, which lowers their detection probability due to the dilution effect.
The missed positive cases then spread the disease further. 
On the other hand, if some or all of the members in the same household are placed into the same pool under correlated pooling, the viral load in the pooled sample is higher, raising the detection probability. 
Promptly identifying and isolating the positive cases prevents them from further spreading the disease. 

\subsubsection{Concentration-dependent test errors}\label{subsec:pcr}
It is commonly observed that PCR test sensitivity depends on the sample's viral load and that samples in pooled tests are diluted by a factor of the pool size \citep{zenios1998pooled, laverack2023cornell}.
A policy-maker may either correctly model this dependency (\textit{assay-aware}) or assume a constant test sensitivity (\textit{assay-oblivious}). 

\begin{figure}
 \FIGURE
 {\begin{subfigure}[b]{0.45\textwidth}
         \centering
         \includegraphics[width=\textwidth]{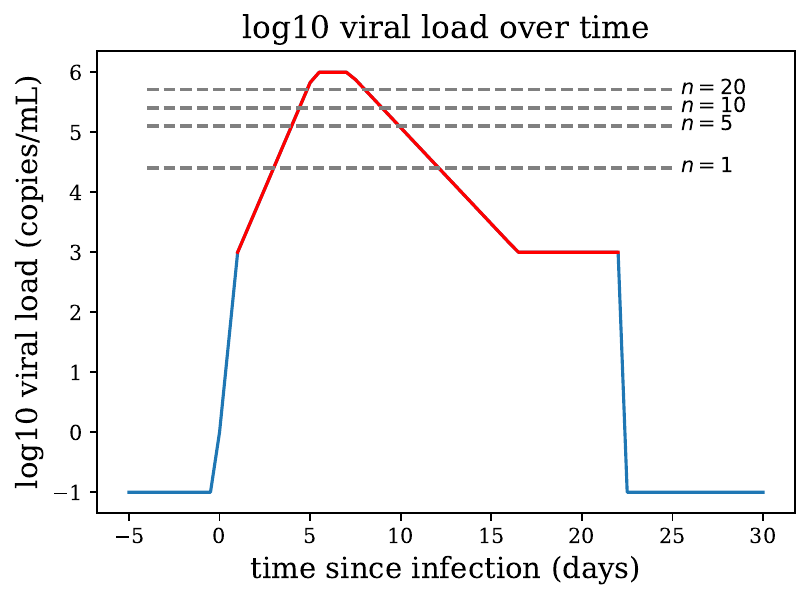}
         \caption{}
         \label{fig:temporal_VL}
     \end{subfigure}
     \begin{subfigure}[b]{0.45\textwidth}
         \centering
         \includegraphics[width=\textwidth]{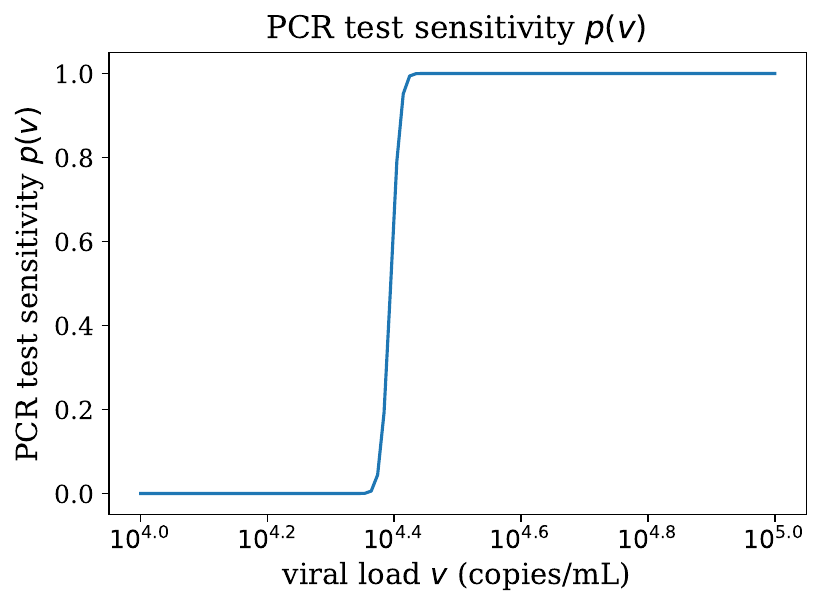}
         \caption{}
         \label{fig:indiv_test_sensitivity}
     \end{subfigure}}
 {(a) Example log10 viral load over time for an infected individual and their 80\% detection threshold when diluted in a size-$n$ pool for different $n$. 
 (b) PCR test sensitivity function $p(v)$.  \label{fig:PCR_VL} \vspace{0.1in}}
 {(a) The period during which the individual is infectious is marked in red. 
 The horizontal dashed lines indicate the threshold values of log10 viral load in a positive sample such that a size-$n$ pool containing this positive sample and $n-1$ negative samples is detectable with probability 80\%, for $n=1,5,10,20$.
 }
 \end{figure}

Accurately modeling concentration-dependent test errors requires modeling the viral load and the PCR tests. 
Viral load of an infected individual typically rises then falls during the course of infection \citep{xu2020clinical,liu2020viral}.
Following \citet{brault2021group}, we model the log10 viral load of an infected individual as a piecewise linear function over several stages: linear increase, constant peak level, linear decrease, constant tail level, and linear decrease to zero. We assume an individual is infectious when their viral load is above a certain threshold. 
To account for heterogeneity across infections, we randomly sample the duration of each stage for each infected individual. 
Figure~\ref{fig:temporal_VL} shows an example log10 viral load trajectory. 
At the start of the simulation, we assume that half of the initial infections are at the beginning of infectivity, and the other half at the onset of the peak.

We develop a realistic PCR model that captures the relationship between PCR test results and sample viral loads, accounting for both the dilution effect and the stochasticity of sample handling.
We assume the pooled viral load is diluted by a factor of the pool size. 
Figure~\ref{fig:temporal_VL} also illustrates how the detection threshold for a positive sample increases if that sample is diluted with other negative samples in a pooled test.

We simulate how a sample undergoes multiple steps of processing (e.g., subsampling and extraction) before entering the PCR machine, where each step introduces stochasticity into the amount of viral RNA that remains \citep{wyllie2020saliva, basu2017digital}. More details are given in Appendix~\ref{appdx:PCR}.
Our modeling of the PCR test is one instantiation of the general test sensitivity function $p(v)$ discussed in Section~\ref{sec:theory} (Figure~\ref{fig:indiv_test_sensitivity}).\footnote{In our realistic model, $p(v) = 0$ for very small $v$, while we assume $p(v)>0$ for $v>0$ in Section~\ref{sec:theory}. Nevertheless, our theoretical and simulation results are still consistent despite this small discrepancy.} 
While our case study focuses on PCR tests, our findings are likely applicable to a range of other tests, such as antibody tests \citep{zenios1998pooled} and other amplification-based tests \citep{westreich2008optimizing}.

\subsection{Overview of Simulation Results}\label{sec:results_overview}
We first provide an overview of the simulation results before discussing their policy implications in Section~\ref{subsec:corr_as_modeling_choice} and~\ref{subsec:corr_as_intervention}. The simulation outcomes are aligned with theoretical results in Section~\ref{sec:theory}. 

\begin{figure}
\FIGURE
{\includegraphics[width=\textwidth]{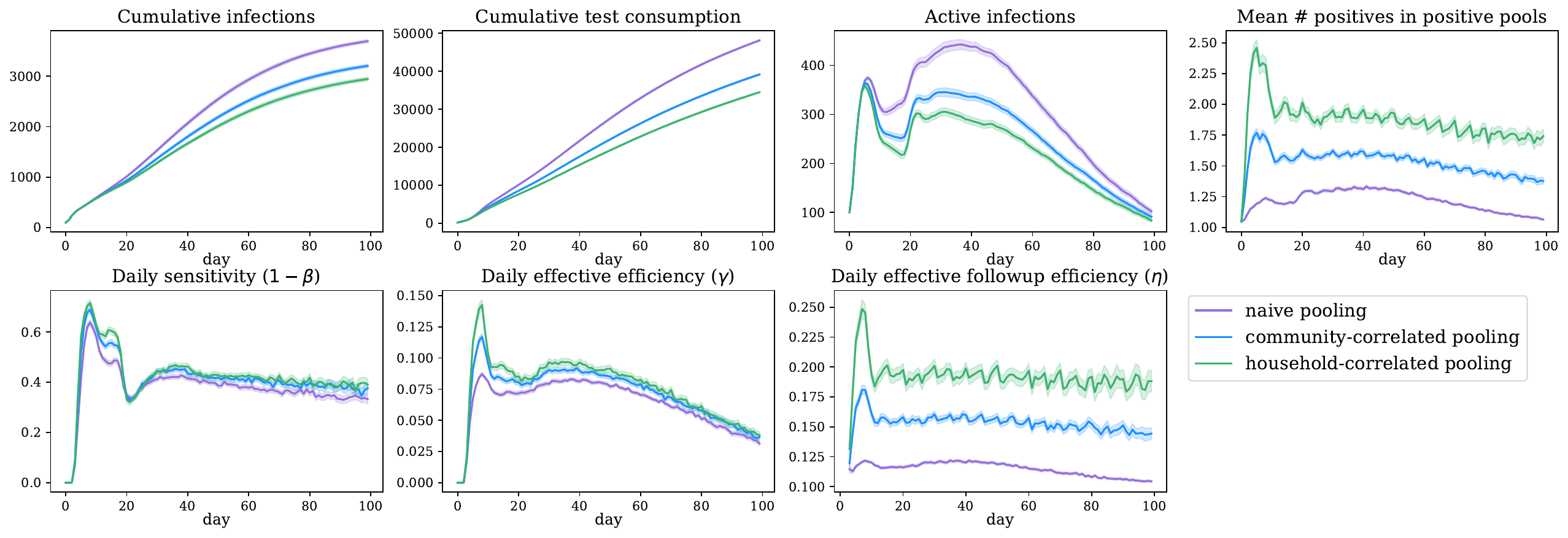}}
 {Simulated epidemic metrics over a 100-day period for \texttt{NP} (purple), \texttt{CCP} (blue), and \texttt{HCP} (green). \label{fig:dynamic_sim_results}}
 {The screening policy is to screen every five days using a pool size of ten. Results are averaged over 200 replications, and error bars are two standard errors of the mean (SEM). }
 \end{figure}

The qualitative differences between \np, \texttt{CCP}, and \texttt{HCP} discussed in Section~\ref{subsec:sim_corr} are evidenced by our simulation.
Figure~\ref{fig:dynamic_sim_results} describes the epidemic progression over a 100-day period, during which we employ different pooling methods under a representative policy of screening every five days using pools of size ten, while accurately modeling concentration-dependent test errors.
We focus on two primary performance metrics, namely the \textit{cumulative number of infections} and \textit{cumulative test consumption}, both of which we aim to minimize. They provide a high-level summary of the epidemic control effort, directly of interest to decision-makers. 
In addition, we report the metrics studied theoretically in Section~\ref{sec:theory}, including the sensitivity $1-\beta$, effective efficiency $\gamma$, and an auxiliary metric, effective follow-up efficiency $\eta$ (defined in Appendix~\ref{appdx:proof_of_thm_2}). We discuss these metrics in detail in Appendix~\ref{appdx:sim_dynamics}.
In all these metrics, \texttt{HCP} outperforms \texttt{CCP}, which, in turn, outperforms \texttt{NP}, validating our theoretical findings. 
Notably, even a modest gap in the daily sensitivity (i.e., the fraction of positive individuals identified among those screened on a given day) leads to significantly wider gaps in cumulative infections over time. This demonstrates that even a small improvement in sensitivity can have a compounding effect on epidemic control.

\begin{figure}
     \FIGURE
     {\includegraphics[width=0.9\textwidth]{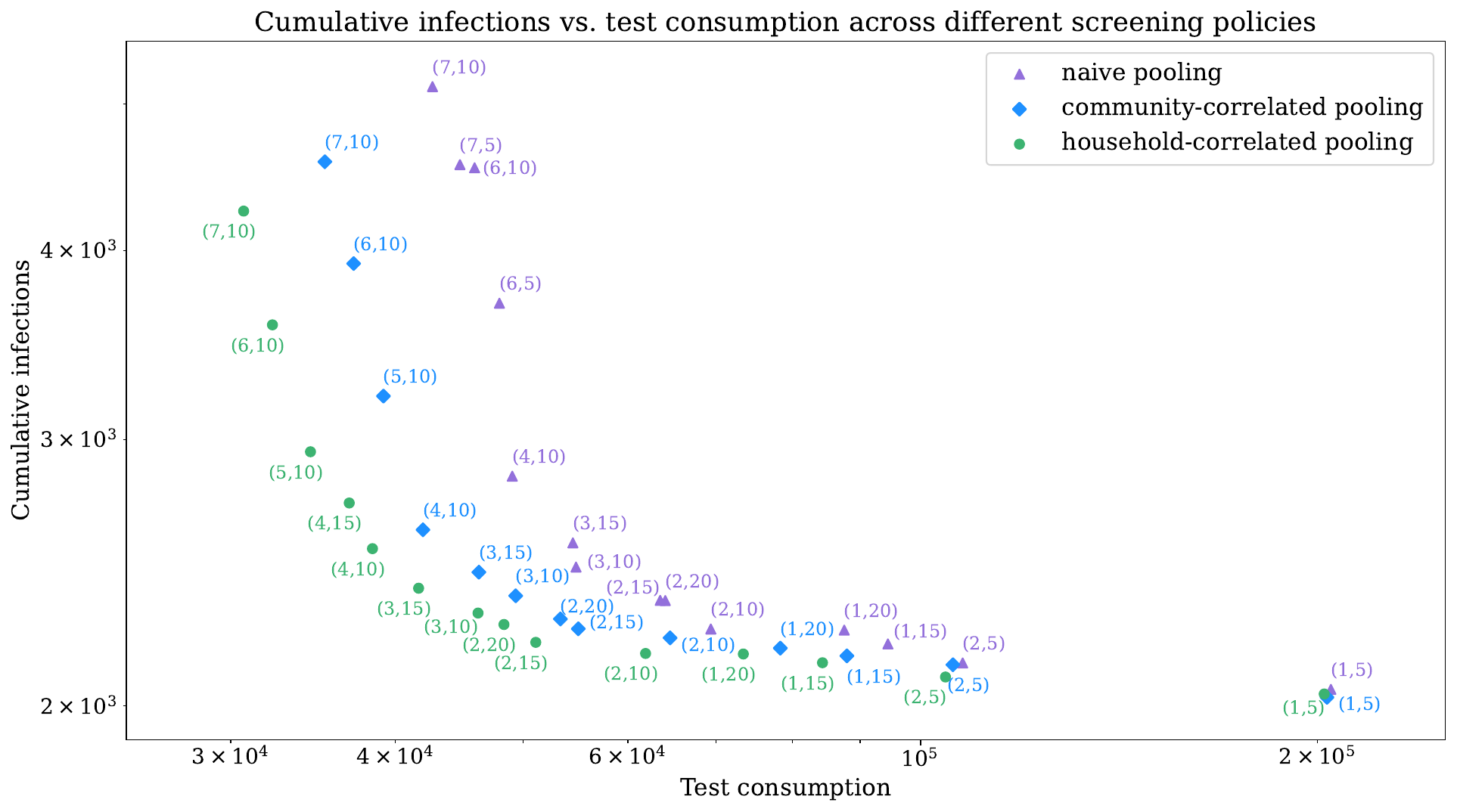}}
     {Cumulative infections and test consumption under Pareto-optimal policy choices for each of \texttt{NP} (purple triangle), \texttt{CCP} (blue diamond), and \texttt{HCP} (green dot). \vspace{0.1in} \label{fig:performance_diff_scatter}}
     {Each point represents a choice of screening frequency and pool size, indicated by a tuple in the annotation. Results are averaged across 200 replications. Error bars are not displayed to avoid clutter. Error bars for test consumption have little overlap across all points. For the same policy, error bars for cumulative infections under \texttt{NP} and \texttt{CCP} do not overlap if the policy screens every three or more days; error bars for cumulative infections under \texttt{CCP} and \texttt{HCP} do not overlap if the policy screens every four or more days. When the population is screened more frequently, most positives are identified promptly for all pooling methods, making the advantage offered by correlation less significant.}
 \end{figure}

Figure~\ref{fig:performance_diff_scatter} presents a landscape of Pareto-optimal screening policies, illustrating the trade-off between cumulative infections and test consumption as modeled by each pooling method.
For a given policy, the ranking of the three pooling methods remains consistent with the results shown in Figure~\ref{fig:dynamic_sim_results}.
By comparing the cumulative number of infections across different pooling methods 
under varying test availability, we argue that modeling correlation is crucial for policy-making and that intentionally enhancing it can offer dramatic benefits.
We discuss such implications in more detail in Section~\ref{subsec:corr_as_modeling_choice} and~\ref{subsec:corr_as_intervention}, zooming in on several representative regions in Figure~\ref{fig:performance_diff_scatter}.

\subsection{Policy Implication I: Correlation as a Modeling Choice}\label{subsec:corr_as_modeling_choice}

First, we demonstrate that it is important to be both correlation-aware and assay-aware, i.e., to model naturally arising within-pool correlation and concentration-dependent test errors. 
Ignoring either aspect leads to predictions for sensitivity and test efficiency that differ significantly from reality (Section~\ref{subsec:corr_assay_modeling}) and significantly suboptimal decisions (Section~\ref{subsec:corr_assay_decisions}).

\subsubsection{Correlation and assay-awareness for accurately modeling reality.}\label{subsec:corr_assay_modeling}

Both correlation-awareness and assay-awareness are essential for accurately predicting the infections and test consumption of a policy decision. Deviating in either dimension leads to inaccurate projections. 

\begin{figure}
     \FIGURE
     {\includegraphics[width=0.8\textwidth]{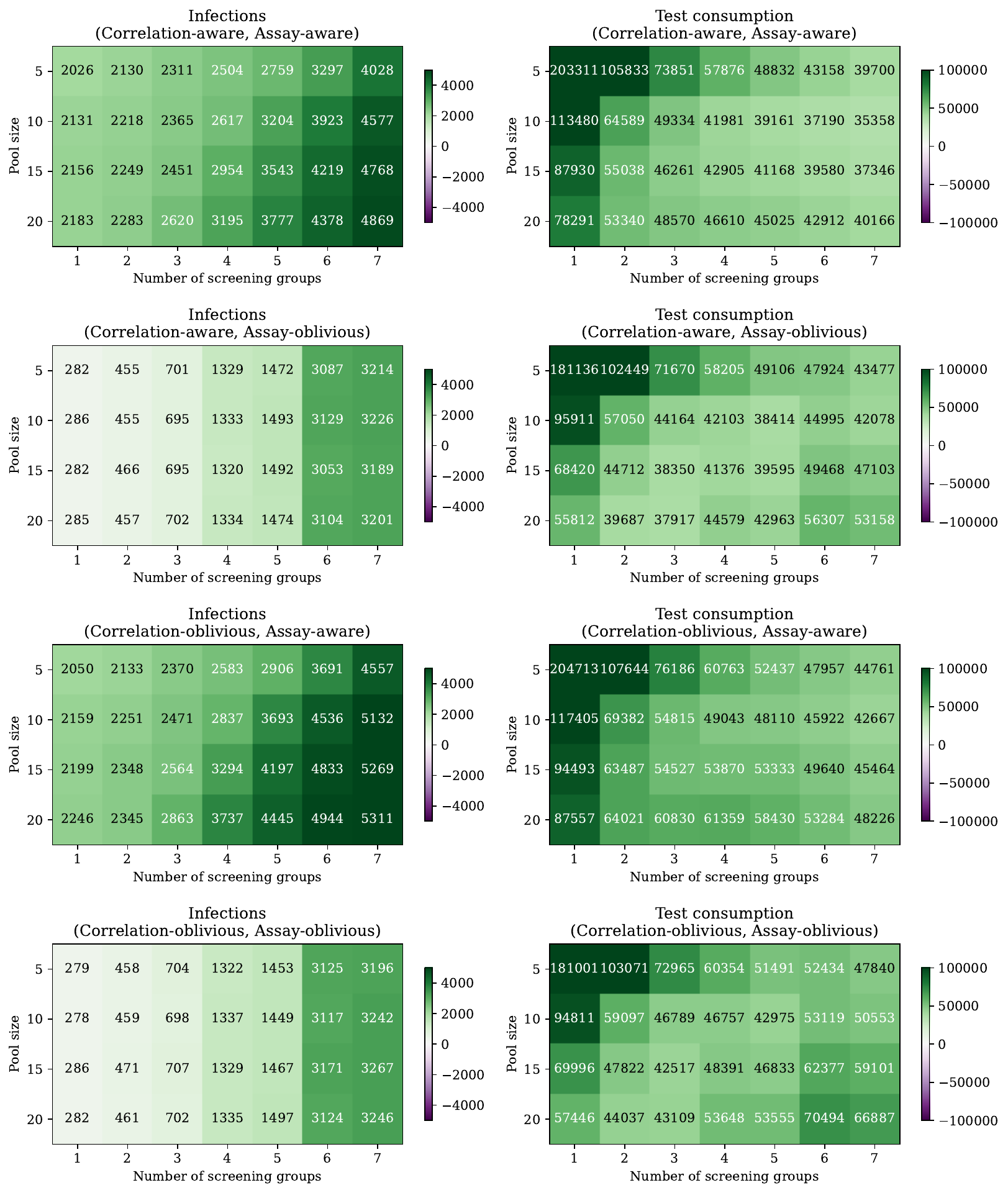}}
     {Cumulative infections and test consumption across all screening policies predicted by policy-makers that model correlation and test errors accurately or inaccurately. \label{fig:heatmap_infections_tests_new}}
     {}
 \end{figure}

Figure~\ref{fig:heatmap_infections_tests_new} shows the average predicted infections and test consumption of all the policies (combinations of pool size and screening frequency) that we simulate, across four policy-makers modeling correlation and test errors accurately or inaccurately (Figure~\ref{fig:modeling_choice_quadrant}). The predictions of the correlation-aware and assay-aware model are significantly different from any of the other three models that fail to capture at least one of correlation and assay-awareness. For example, a correlation-oblivious assay-aware policy-maker consistently overestimates the infections and test consumption, as validated in Figure~\ref{fig:dynamic_sim_results} for an example policy.

Moreover, Figure~\ref{fig:heatmap_infections_tests_new} shows that the difference between correlation-aware and correlation-oblivious modeling depends strongly on whether realistic concentration-dependent test errors are modeled. Under the assay-oblivious model, there is little difference between correlation-aware and correlation-oblivious results, while under the more realistic assay-aware model, there is a strong difference.
Hence, adopting a simplified test error model greatly skews the understanding of the benefit of correlation. 

\subsubsection{Correlation and assay-awareness for optimal decision-making.}\label{subsec:corr_assay_decisions}

Not only is being correlation-aware \textit{and} assay-aware essential for accurately modeling reality, it also underpins optimal decision-making. 
We show that missing either aspect leads to suboptimal decisions.

\paragraph{Assay-aware, correlation-oblivious}

We compare the decisions of a correlation-oblivious and a correlation-aware policy-maker, assuming both of them are assay-aware. 
We study two important decisions: (1) lockdown versus screening, and (2) choice of screening frequency and pool size.
The correlation-oblivious policy-maker, informed by analyses ignoring correlation, tends to make overly conservative decisions compared to the correlation-aware policy-maker. 

The first decision any policy-maker faces during an emerging pandemic is whether to issue a lockdown or to keep the society open while conducting screening. Lockdowns entail huge economic losses and are generally undesirable, but the feasibility of keeping society open depends on resource availability and the policy-maker's risk tolerance.
Suppose $4\times 10^4$ PCR tests are available over 100 days for pooled and individual testing combined.
Based on the \texttt{NP} simulation, the correlation-oblivious policy-maker decides that no screening policy is attainable and thus issues a lockdown. (As in Figure~\ref{fig:performance_diff_scatter}, all policies in the \texttt{NP} simulation use more than $4\times 10^4$ tests.) 
However, the correlation-aware policy-maker, assuming \texttt{CCP}, finds that 
screening every five days with a pool size of ten incurs the fewest infections under the testing budget. They adopt this screening policy while keeping the economy open.

Even with a higher test supply that permits repeated screening under \texttt{NP}, \texttt{NP} can overestimate the cumulative infections, leading to overly cautious decisions. 
Suppose the test supply is $4.5\times 10^4$. The optimal \texttt{NP} policy is screening every seven days with a pool size of five, projected to yield around $4.5\times 10^3$ cumulative infections on average. 
On the other hand, the optimal \texttt{CCP} policy, screening every four days with a pool size of ten, results in $2.6\times 10^3$ cumulative infections, significantly lower than with \texttt{NP}.
Suppose the policy-maker cannot tolerate more than 30\% of the population infected due to resource constraints like intensive care unit (ICU) availability. In this scenario, the correlation-oblivious policy-maker mistakenly issues a lockdown, while the correlation-aware policy-maker conducts screening and keeps the economy open. 

\begin{figure}
    \FIGURE
    {\begin{subfigure}[b]{0.483\textwidth}
         \centering
         \includegraphics[width=\textwidth]{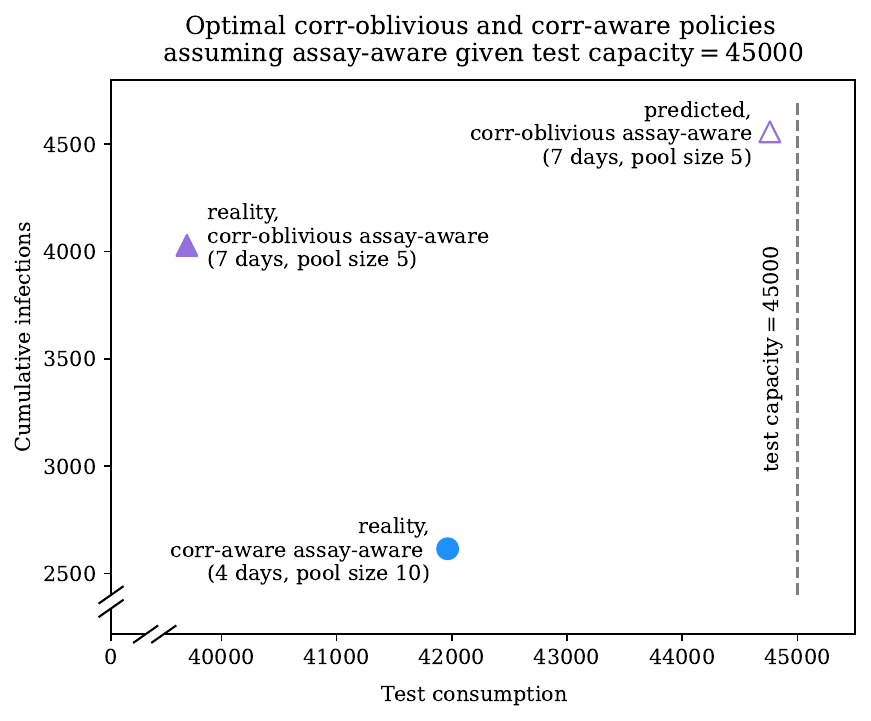}
         \caption{}
         \label{fig:NP_CCP_zoomin_45000_concdep}
     \end{subfigure}
     \begin{subfigure}[b]{0.49\textwidth}
         \centering
         \includegraphics[width=\textwidth]{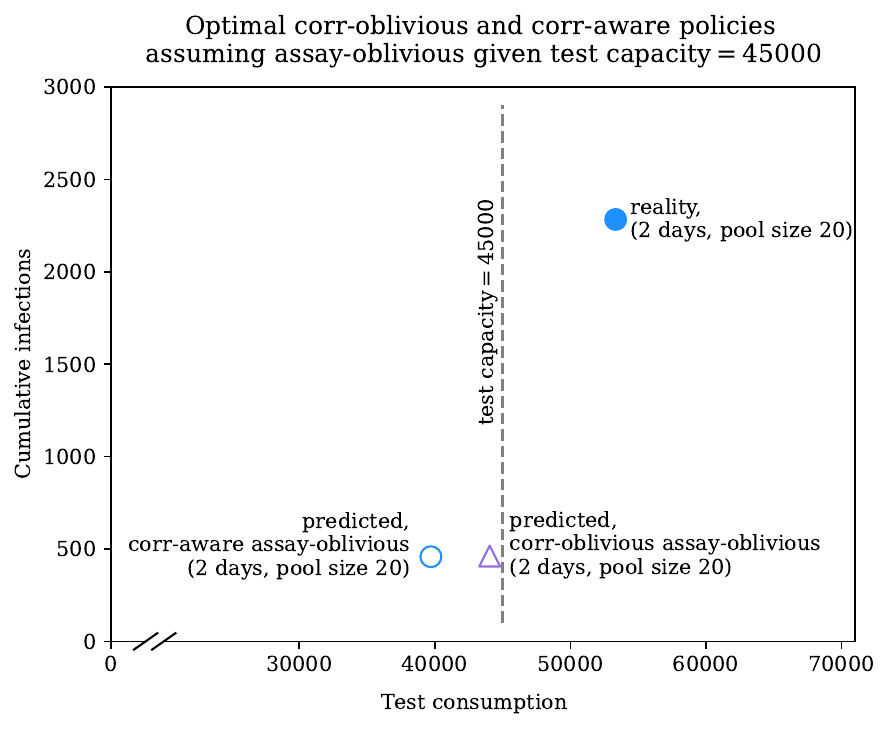}
         \caption{}
         \label{fig:NP_CCP_zoomin_45000_constant80}
     \end{subfigure}}
    {Prediction and reality of optimal correlation-oblivious and correlation-aware policies given $4.5\times 10^4$ test capacity over 100 days, under (a) assay-aware and (b) assay-oblivious test error models. \label{fig:NP_CCP_zoomin_45000}}
    {Results are averaged over 200 replications; error bars are small and are omitted.
    (a) Assume policy-makers are assay-aware. The empty purple triangle indicates the predicted outcome of the optimal attainable correlation-oblivious policy (screening every seven days with a pool size of five) using a concentration-dependent test error model accounting for the dilution effect. The actual outcome of this policy, modeled by community correlated pooling, is indicated by the solid purple triangle. 
    The solid blue dot indicates the outcome of the optimal attainable correlation-aware policy (screening every four days with a pool size of ten) using the same concentration-dependent test error model.
    (b) Assume policy-makers are assay-oblivious and assume PCR test sensitivity is fixed at 80\%. The empty purple triangle (blue circle) indicates the predicted outcome using the optimal attainable correlation-oblivious (correlation-aware) policy. Both correlation-oblivious and correlation-aware policy-makers in this case decide to screen every two days using pools of size 20. However, as test errors are dependent on sample viral loads in reality, this policy incurs much more infections and test consumption than predicted.
    }
\end{figure}

If a policy-maker does opt for screening, they need to decide on a screening frequency and pool size that minimizes the infections to the extent that the test capacity permits. 
We argue that the correlation-oblivious policy-maker tends to choose a suboptimal screening policy compared to the correlation-aware policy-maker. 
Suppose the test supply is $4.5\times 10^4$, as before, but the correlation-oblivious policy-maker can tolerate the predicted $4.5\times 10^3$ infections --- they decide to screen every seven days with a pool size of five. As before, the correlation-aware policy-maker screens every four days with a pool size of ten. Since we assume \texttt{CCP} to reflect the reality, the actual outcome for the correlation-oblivious policy follows \texttt{CCP}'s outcome for the same policy, at about $4\times 10^3$ infections on average, while the correlation-aware policy incurs $2.6\times 10^3$ infections (Figure~\ref{fig:NP_CCP_zoomin_45000_concdep}). 
Since \texttt{NP} underestimates the effective efficiency (Figure~\ref{fig:dynamic_sim_results}, ``daily effective efficiency"), the correlation-oblivious policy-maker underestimates the highest attainable screening frequency.
In reality, their policy consumes 6\% fewer tests but incurs 54\% more infections than the correlation-aware policy.

\paragraph{Correlation-aware, assay-oblivious}
Incorporating correlation in modeling only provides benefits if the policy-maker is assay-aware. 
Modeling test sensitivity as fixed neglects correlation's benefit and generates inaccurate predictions that lead to poor policy decisions.
We consider the same scenario as above with $4.5\times 10^4$ test supply over 100 days, but now we focus on policy-makers assuming 80\% fixed test sensitivity in their simulations (Figure~\ref{fig:NP_CCP_zoomin_45000_constant80}). 
(Recall that the PCR test model is calibrated to have an average sensitivity of 80\% for a representative population.)
As a result of the fixed sensitivity, correlation does not affect sensitivity and thus does not affect infections. Correlation also only mildly impacts test consumption. Being assay-oblivious, both the correlation-oblivious and the correlation-aware policy-makers choose to screen every two days with a pool size of 20. However, they overestimate the sensitivity and underestimate the test consumption of this policy (Figure~\ref{fig:NP_CCP_zoomin_45000_constant80}).\footnote{They would estimate the procedural sensitivity to be $64\%$ assuming both the pooled test and individual test have 80\% sensitivity, but the actual sensitivity of the pooled test is less than 80\% due to dilution, resulting in a procedural sensitivity lower than 64\%.} While the policy is projected to yield around 500 infections and stay under the test capacity limit, it in fact incurs $2.3\times 10^3$ infections and uses more tests than available. 
Thus, being assay-oblivious in modeling leads to poor policy decisions and severe consequences.

\subsection{Policy Implication II: Correlation as an Intervention}\label{subsec:corr_as_intervention}

In addition to modeling the natural correlation, enhancing correlation by pooling households together can further boost epidemic control performance.
We compare a correlation-enhancing policy-maker with one who does not enhance correlation. 
(Both are also assumed to be correlation aware.)
The outcomes from their decisions are given by 
\texttt{HCP} and \texttt{CCP}, respectively. 
Based on arguments in Section~\ref{subsec:corr_as_modeling_choice}, we assume both policy-makers are assay-aware.

In Figure~\ref{fig:dynamic_sim_results}, \texttt{HCP} (green) further reduces both the cumulative number of infections and cumulative test consumption compared to \texttt{CCP} (blue). 
On Day 100, \texttt{HCP} predicts $2.9\times 10^3$ infections on average, which is 9\% fewer than \texttt{CCP} ($3.2\times 10^3$ infections) and 22\% fewer than \texttt{NP} ($3.7\times 10^3$ infections).
The source of the difference between \texttt{HCP} and \texttt{CCP} is the same in nature as that between \texttt{CCP} and \texttt{NP}: The stronger within-pool correlation under \texttt{HCP} improves the overall sensitivity, translating to more effective epidemic mitigation (Figure~\ref{fig:pooling_network_schematic}). 

If the correlation-enhancing policy-maker executes \texttt{HCP} in reality, they achieve better epidemiological outcomes than the correlation-aware policy-maker who does not, and much better ones than the correlation-oblivious policy-maker. 
The same arguments regarding decision-making in Section~\ref{subsec:corr_assay_decisions} apply, and the advantage provided by within-pool correlation is even more pronounced for \texttt{HCP}.

First, the correlation-enhancing policy-maker may keep the economy open at a lower test supply, as shown by the gap between the \texttt{HCP} and \texttt{CCP} outcomes in Figure~\ref{fig:performance_diff_scatter}. For example, if the test supply is $3.2\times 10^4$, the policy-maker who does not enhance correlation must issue a lockdown while the correlation-enhancing policy-maker chooses to conduct screening.

\begin{figure}
     \FIGURE
     {\includegraphics[width=0.5\textwidth]{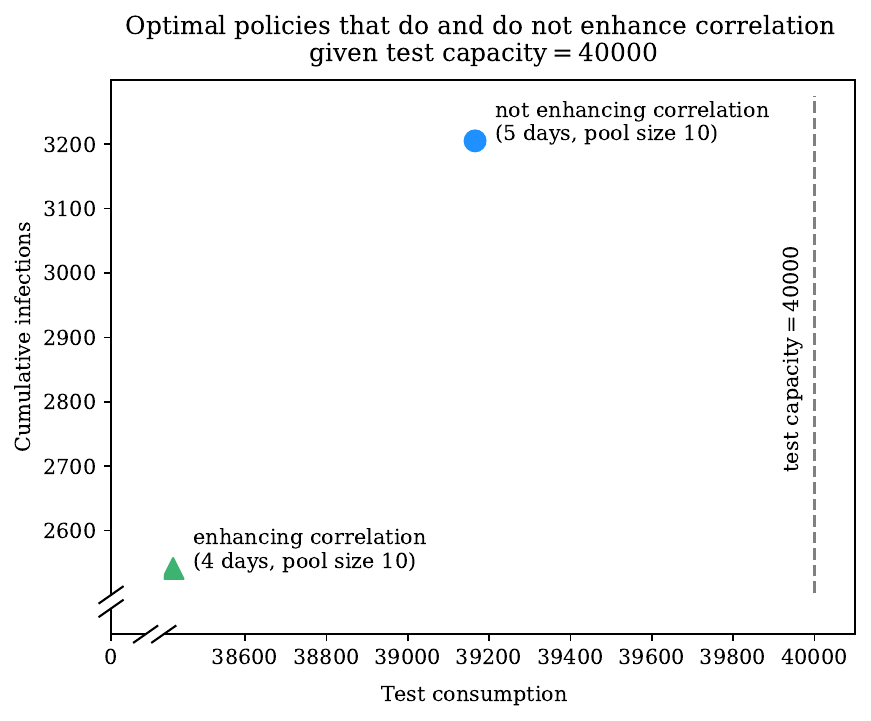}}
     {Optimal policies that do and do not enhance correlation, given $4\times 10^4$ test capacity over 100 days. \label{fig:CCP_HCP_zoomin_40000}}
     {The solid blue dot indicates the outcome of the optimal policy that does not enhance correlation (screening every five days with a pool size of ten). 
    The solid green triangle indicates the outcome of the optimal attainable correlation-enhancing policy (screening every four days with a pool size of ten). By construction, we assume the predicted outcomes from both \texttt{CCP} and \texttt{HCP} align with their actual outcomes. Results are averaged over 200 replications; error bars are small and are omitted. }
 \end{figure}

Furthermore, the correlation-enhancing policy-maker may achieve better epidemiological outcomes than their counterpart who does not enhance correlation if both conduct screening. For example, given a test supply of $4\times 10^4$, the policy-maker not enhancing correlation screens every five days with a pool size of ten, incurring $3.2\times 10^3$ infections on average. The correlation-enhancing policy-maker, taking measures to strengthen the correlation in pools, screens every four days with a pool size of ten, incurring $2.6\times 10^3$ infections on average, a 20\% reduction compared to the one who does not enhance correlation (Figure~\ref{fig:CCP_HCP_zoomin_40000}).

These results suggest that, when possible, it is worth taking explicit measures to strengthen the correlation within pools. 
For example, one can encourage individuals from the same household to get tested at the same location and time slot. One can also mail sample collection kits to each household and ask them to self-collect and combine their samples. 
While we recognize the logistical challenges of implementing these measures, our model provides a general framework for predicting their benefits for epidemic control, allowing policy-makers to make informed cost-benefit assessments.

\subsection{Discussion}\label{sec:discussion}
Our insight in this section is not limited to COVID-19 and applies to epidemic control in general. 
The practical impact of modeling concentration-dependent test errors depends on two key factors: assay characteristics and disease transmissibility.

First, the presence (and sometimes quantity) of molecules associated with an infectious disease, such as a particular nucleic acid sequence, antibodies, or antigen, is often tested using molecular assays. 
For example, PCR assays are used to detect nucleic acids, such as SARS-CoV-2 RNA \citep{van2020comparison}, malaria DNA \cite{hsiang2010pcr}, and hepatitis B DNA \citep{chatterjee2014sensitivity}; chemiluminescent immunoassays (CLIA) and enzyme-linked immunosorbent assays (ELISA) are used to detect antibodies for SARS-CoV-2 \citep{ghaffari2020covid} and HIV \citep{chang2020comparative}.
The sensitivity of such assays typically depends on the concentration of the molecule being detected.\footnote{For such concentration-dependent assays, the relative magnitude of the molecule concentration and assay detection threshold governs how much the dilution effect harms sensitivity.
Theoretically, if an assay can remain highly sensitive even if the sample is substantially diluted (e.g., the droplet digital PCR \citep{suo2020ddpcr}), correlation may not improve sensitivity significantly and accurately modeling the dilution becomes less important.} 
Indeed, the dilution effect has been observed in pooled testing for various diseases, including HIV \citep{kemper1998effects}, malaria \citep{bharti2009malaria, hsiang2010pcr}, and hepatitis B \citep{chatterjee2014sensitivity}. 
For these diseases, correlation in infection status arises among household or community members due to the nature of transmission, e.g., through body fluids or the presence of a common vector in a geographical area. 
Therefore, correlated pooling would likely help improve the sensitivity of screening for these diseases.

Second, the transmissibility of the disease determines the extent to which the improved sensitivity benefits epidemic control.  
For viruses transmitted through intimate contacts, such as HIV and hepatitis B, the benefit may be limited as a missed positive generates a limited number of secondary infections. 
However, for highly transmissible viruses such as SARS-CoV-2, a small improvement in sensitivity translates to a huge reduction in cumulative infections, as shown in Figure~\ref{fig:dynamic_sim_results}. 

Therefore, our overall insight is broadly valuable: 
when designing group testing strategies for COVID-19 and other infectious diseases, accounting for correlation while modeling concentration-dependent test errors enables policy-makers to identify the positives more accurately and contain the epidemic more effectively.

\section{Conclusion}\label{sec:discussion}
In this paper, we proved that under a general correlation structure and a concentration-dependent test error model, \CP achieves asymptotically higher sensitivity but can degrade test efficiency compared to \NP using the same pool size.
We identified an alternative measure of test resource usage, the number of positives found per test consumed, which we argued is better aligned with infection control, and showed that \CP outperforms \NP on this measure.
We validated and contextualized our theoretical results in a realistic agent-based epidemic simulation. 
We argued that policy-makers evaluating group testing protocols for large-scale screening should model test errors realistically, account for the naturally arising within-pool correlation, and intentionally maximize it when possible.

Our work can be extended in several directions in future research.
First, while we focus on the Dorfman procedure when understanding the impact of correlation on pooled testing in the presence of the dilution effect, similar phenomena likely arise in other testing algorithms. In particular, correlation likely improves the sensitivity of tests within these procedures as well. 
We anticipate that follow-on work can show that correlation improves the performance of these other test procedures in the presence of the dilution effect.
Second, the index case and the secondary cases within the same household could become infected at different times. It would be interesting to explore asynchronous testing protocols that both utilize the correlation and optimize the timing to maximize the overall probability of detecting the infected members.
Third, it would be meaningful to incorporate sampling noise, where the sample viral load could be zero for an infected individual. The additional transmission due to undetected individuals may counteract the benefits offered by correlated pooling, and such consideration is of practical interest for large-scale epidemic control. This could be addressed using latent variable models.
Last but not least, we could model correlation's benefit for reducing the test turnaround time, demonstrated to be important for epidemic control \citep{larremore2021test}. Since positives are clustered in fewer pools in correlated pooling, fewer follow-up tests are required, which reduces the time required to obtain test results and notify the positive cases. This effect, combined with the improved test sensitivity and efficiency, would further strengthen correlated pooling's advantage in epidemic control.

\ACKNOWLEDGMENT{The authors are grateful to Diego Diel and Jeff Pleiss for conversations on the implementation details of PCR tests. The authors also thank Stephen Chick for providing valuable feedback. 
This work was conducted under the support of Cornell University when the authors served in the Cornell COVID mathematical modeling team.
Additional support was provided by Air Force Office of Scientific Research FA9550-19-1-0283 and National Science Foundation grant DMS2230023. 
J.W. and Y.Z. contributed equally to this paper.
}

\clearpage
\bibliographystyle{informs2014} 
\bibliography{references} 

\ECSwitch


\ECHead{\centering Online Appendices for \\Correlation Improves Group Testing:\\Modeling Concentration-Dependent Test Errors}

\begin{APPENDICES}
\vspace{2em}
\begin{center}
    \begin{tabular}{c}
        Jiayue Wan \\
        \footnotesize{School of Operations Research \& Information Engineering, Cornell University, NY 14850, \EMAIL{jw2529@cornell.edu}} \\
        \\
        Yujia Zhang \\
        \footnotesize{Center for Applied Mathematics, Cornell University, NY 14850, \EMAIL{yz685@cornell.edu}} \\
        \\
        Peter I. Frazier \\
        \footnotesize{School of Operations Research \& Information Engineering, Cornell University, NY 14850, \EMAIL{pf98@cornell.edu} }\\
    \end{tabular}
\end{center}

\vspace{2em}

\section{Convergence of Population-Level Metrics}
We begin by presenting a lemma which states that under Assumption~\ref{assu:corr_decay}, the population-level quantities converge to constants as $N\to\infty$.
\begin{lemma}\label{lemma:convergence}
 Under Assumption~\ref{assu:corr_decay}, for $\alpha >0$ and $\pool\in\{\np,\cp\}$, random variables $\overline{D}$, $\overline{S}$, and $\overline{Y}$ under $\PP^{(N)}_{\pool,\alpha}$ converge in probability to $\EE_{\pool,\alpha}[D]$, $\EE_{\pool,\alpha}[S]$, and $\EE_{\pool,\alpha}[Y]$, respectively, as $N\to\infty$.
\end{lemma}

\proof{Proof of Lemma~\ref{lemma:convergence}.}
For succinctness, we abbreviate $\PP_{\pool,\alpha}^{(N)}(\cdot)$, $\EE_{\pool,\alpha}^{(N)}[\cdot]$, $\Var_{\pool,\alpha}^{(N)}(\cdot)$, $\Cov_{\pool,\alpha}^{(N)}(\cdot,\cdot)$  as $\PP^{(N)}(\cdot)$, $\EE^{(N)}[\cdot]$, $\Var^{(N)}(\cdot)$ and $\Cov^{(N)}(\cdot, \cdot)$, respectively, in this proof. We break down the proof of Lemma~\ref{lemma:convergence} into two parts, where we first show that $\Var^{(N)}(\overline{Z})\to 0$ for a population-level quantity $\overline{Z}$, and then show that $\overline{Z}$ converges in probability.

1. \underline{$\Var^{(N)}(\overline{Z})\to 0$.}
Consider a population-level quantity $\overline{Z}$, where $\overline{Z} = \frac{1}{|\A|}\sum_{j=1}^{|\A|}Z_j$, $Z_j$ is one of the pool-level quantities, $S_j$, $Y_j$ or $D_j$. We note that $Z_j$ is upper bounded by some positive constant $C_g>0$ that does not involve $N$. We have that 
\begin{align*}
    \Var^{(N)}(\overline{Z}) &= \text{Var}^{(N)}\left(\frac{1}{|\A|}\sum_{j=1}^{|\A|}Z_j\right)\\
    &=\frac{1}{|\A|^2}\left(\sum_{j=1}^{|\A|}\Var^{(N)}(Z_j) + \sum_{j=1}^{|\A|}\sum_{k\neq j}\Cov^{(N)}(Z_j, Z_k)\right).
\end{align*}

In order to bound $\Var(\overline{Z})$, we first provide an upper bound on $|\Cov(Z_j, Z_k)|$ where $j\neq k$. By definition, 
\begin{align*}
    \Cov^{(N)}(Z_j, Z_k) &= \EE^{(N)}[Z_jZ_k] - \EE^{(N)}[Z_j]\EE^{(N)}[Z_k]\nonumber\\
    &= \EE^{(N)}[(\EE^{(N)}[Z_j\mid \BFU_{A_k}] - \EE^{(N)}[Z_j])Z_k].\nonumber
\end{align*}

Now, applying the definition of $\Delta_{\pool,\alpha}^{(N)}$, we find that 
\begin{align*}
|\Cov^{(N)}(Z_j, Z_k)| &\leq \EE^{(N)}\left[Z_k\cdot \Delta_{\pool,\alpha}^{(N)}(Z_j, k)\right]\nonumber\\
&\leq C_g \cdot \Delta_{\pool,\alpha}^{(N)}(Z_j, k).
\end{align*}

This allows us to bound the variance of $\overline{Z}$:
\begin{align*}
    \Var^{(N)}(\overline{Z}) &\leq \frac{1}{|\A|^2}\left(\sum_{j=1}^{|\A|}\Var(Z_j) + \sum_{j=1}^{|\A|}\sum_{k\neq j}|\Cov(Z_j, Z_k)|\right)\\
    &\leq \frac{1}{|\A|^2}\left(|\A|\cdot \frac14 C_g^2 + \sum_{j=1}^{|\A|}\sum_{k\neq j}  C_g\cdot \Delta_{\pool,\alpha}^{(N)}(Z_j, k)\right).
\end{align*}

For any $\epsilon >0$, we have that 
\begin{align*}
    \Var^{(N)}(\overline{Z}) &\leq \frac{1}{|\A|^2}\left(|\A|\cdot \frac14 C_g^2 + \sum_{j=1}^{|\A|} C_g\cdot \left((N - 1 - m_{\pool, \alpha}^{(N)}(\epsilon, Z_{1:|\A|}))\cdot \epsilon  + m_{\pool, \alpha}^{(N)}(\epsilon, Z_{1:|\A|}) \cdot \frac{1}{4}C_g^2\right)\right)\\
    &= \frac{1}{|\A|^2}\left(|\A|\cdot \frac14 C_g^2 +|\A| \cdot C_g\cdot \left((N - 1 - m_{\pool, \alpha}^{(N)}(\epsilon, Z_{1:|\A|}))\cdot \epsilon  + m_{\pool, \alpha}^{(N)}(\epsilon, Z_{1:|\A|}) \cdot \frac{1}{4}C_g^2\right)\right)\\
    &\leq \frac{C_g^2}{4|\A|} + \frac{C_g N \epsilon}{|\A|} + \frac{C_g^3}{4|\A|}\cdot m_{\pool, \alpha}^{(N)}(\epsilon, Z_{1:|\A|})\\
    &= \frac{nC_g^2}{4N} + nC_g \cdot \epsilon + \frac{nC_g^3}{4}\cdot \frac{m_{\pool, \alpha}^{(N)}(\epsilon, Z_{1:|\A|})}{N}.
\end{align*}

Let $\epsilon_N$ be a sequence satisfying Assumption~\ref{assu:corr_decay}, i.e., $\epsilon_N\downarrow 0$ and $\lim_{N\rightarrow \infty}\frac{1}{N}m_{\pool, \alpha}^{(N)}(\epsilon, Z_{1:|\A|})=0$. Taking the limit $N\rightarrow \infty$ of the expression above, we have
\begin{align*}
    \lim_{N\rightarrow\infty}\Var(\overline{Z})\leq \lim_{N\to\infty} \frac{nC_g^2}{4N} + nC_g \cdot \epsilon_N + \frac{nC_g^3}{4}\cdot \frac{m_{\pool, \alpha}^{(N)}(\epsilon, Z_{1:|\A|})}{N} = 0.
\end{align*}

2. \underline{Proof of convergence}. 
First, it is straightforward that for any $N$
\begin{align*}
    \EE^{(N)}[\overline{Z}] = \EE^{(N)}\left[\frac{1}{|\A|}Z_j\right] = \EE^{(N)}[Z_J] = \EE^{(N)}[Z].
\end{align*}
Because $\EE^{(N)}[Z]$ converges to $\EE[Z]$ as $N$ goes to infinity, it follows that $\EE^{(N)}[\overline{Z}]\to \EE[Z]$. 

Fix $\epsilon > 0$. By definition of limit, there exists some $N_1\in\mathbb{N}$ such that for all $N\geq N_1$, 
$$
\left|\EE^{(N)}[Z] - \EE[Z]\right|<\frac{1}{2}\epsilon.
$$

Observing that for all $N\geq N_1$,
\begin{align*}
    |\overline{Z} - \EE[Z]| &= \left|\overline{Z} - \EE^{(N)}[Z] + \EE^{(N)}[Z] - \EE[Z]\right|\\
    &\leq \left|\overline{Z} - \EE^{(N)}[Z]\right| + \left|\EE^{(N)}[Z] - \EE[Z]\right|\\
    & < \left|\overline{Z} - \EE^{(N)}[Z]\right|+ \frac{1}{2}\epsilon,
\end{align*}
we have that 
\begin{align*}
    \PP^{(N)}\left(|\overline{Z} - \EE[Z]| > \epsilon\right) &\leq \PP^{(N)}\left(\left|\overline{Z} - \EE^{(N)}[Z]\right| + \frac{1}{2}\epsilon > \epsilon\right)\\
    &=\PP^{(N)}\left(\left|\overline{Z} - \EE^{(N)}[Z]\right| > \frac{1}{2}\epsilon \right)\\
    &\leq \frac{\Var^{(N)}(\overline{Z})}{(\frac{1}{2}\epsilon)^2}
\end{align*}
by Chebyshev's inequality. Now, in part 1 of the proof, we have shown that $\Var^{(N)}(\overline{Z})\to 0$ as $N\to\infty$. Therefore, for any $\delta > 0$, there exists some $N_2\in \NN$, such that for all $N\geq N_2$,
$$
\Var^{(N)}(\overline{Z})<\delta\left(\frac{1}{2}\epsilon\right)^2. 
$$
It follows that for all $N\geq \max\{N_1,N_2\}$,
$$
\PP^{(N)}(|\overline{Z} - \EE[Z]|>\epsilon)\leq \delta.
$$

By definition of limit, we have $\PP^{(N)}(|\overline{Z} - \EE[Z]|>\epsilon)\to 0$ as $N\to\infty$. Because this holds true for any $\epsilon >0$, we conclude that $\overline{Z}$ converges to $\EE[Z]$ in probability. \Halmos 
\endproof

By applying the continuous mapping theorem to Lemma~\ref{lemma:convergence}, we establish that  Proposition~\ref{prop:convergence_of_metrics} holds true.

\section{Proofs of Propositions \ref{proposition:asymp_indep} and \ref{proposition:corr_pool}}

For proofs in the section, the subscript $\pool$ (or $\alpha$) is dropped when a quantity/operator does not depend on the pooling method (or prevalence level).

We first show that sample viral loads within pool $J$ are identically distributed, under both $\PP_{\np,\alpha}$ and $\PP_{\cp,\alpha}$.

\begin{lemma}
\label{lemma:id}
The viral loads $V_i: i=1,\cdots, n$ in pool $J$ are identically distributed under $\PP_{\np, \alpha}$. They also follow the same distribution under $\PP_{\cp, \alpha}$.
\end{lemma}

It is worth noting that our model does accommodate heterogeneity in viral load across individuals. This property of identical distribution described in Lemma~\ref{lemma:id} arises from applying an independent random permutation to shuffle the samples within each pool after their formation, facilitating the proofs thereafter.
\proof{Proof of Lemma~\ref{lemma:id}.}
Let $I$ denote the population index of an arbitrary individual from the naive pool. Because naive pools are formed by picking individuals uniformly at random from the population, $I\sim U(\{1,\cdots,N\})$. That is, $\PP_{\np,\alpha}^{(N)}(I=i)=1/N$ for all $i=1,\cdots,N$. The cdf of the viral load of this sample is
\begin{align*}
    \PP_{\np,\alpha}^{(N)}(V_I \leq v) &= \sum_{i=1}^N \PP_{\np}^{(N)}(I=i) \PP_{\alpha}^{(N)}(V_i\leq v)\\
    &= \sum_{i=1}^N \frac{1}{N} \PP_\alpha^{(N)}(V_i\leq v).
\end{align*}

Suppose the correlated pool being studied is the $J$th of the $|\mathcal{A}|$ correlated pools. Because it is chosen randomly from the $|\mathcal{A}|$ pools, $\PP^{(N)}(J=j')=1/|\mathcal{A}|$ for all $j'=1,\cdots,|\mathcal{A}|$. Now consider an arbitrary individual from this pool, and suppose this individual is the $i$th of this pool. Recall that we reordered samples in each pool by performing an independent random permutation of 1 through $n$, denoted by $\pi$. 
Then, the index of $i$ before the permutation is uniformly distributed over $\{1,\cdots,N\}$, that is, $P^{(N)}(\pi(i')=i)=\frac{1}{n}$ for all $i'=1,\cdots,n$. Hence, the cdf of the sample viral load of this arbitrary individual from the correlated pool is given by
\begin{align*}
    \PP_{\cp,\alpha}^{(N)}(V_{J,i}\leq v) &= \sum_{j'=1}^{|\mathcal{A}|} \sum_{i'=1}^n 
    \PP^{(N)}(J=j')\PP^{(N)}(\pi(i')=i)\PP_{\alpha}^{(N)}(V_{j',i'}\leq v)\\
    &= \frac{1}{|\mathcal{A}|}\frac{1}{n} \sum_{j'=1}^{|\mathcal{A}|} \sum_{i'=1}^n \PP_{\alpha}^{(N)}(V_{j',i'}\leq v)\\
    &= \frac{1}{N} \sum_{j'=1}^{|\mathcal{A}|} \sum_{i'=1}^n \PP_{\alpha}^{(N)}(V_{j',i'}\leq v)\\
    &= \sum_{k=1}^N \frac{1}{N} \PP_{\alpha}^{(N)}(V_k\leq v),
\end{align*}
where the last equality follows from the observation that this double sum is equivalent to summing over all individuals in $\{1,\cdots,N\}$. This is identical to the cdf of the viral load of an individual chosen uniformly at random from the naive pool. 

Because $\PP_{\np,\alpha}^{(N)}(V_{J,i}\leq v) = \PP_{\cp,\alpha}^{(N)}(V_{J,i'}\leq v)$ for all $N$ and $i,i'\in \{1,\cdots, n\}$, taking $N$ to the limit of infinity keeps the equality, i.e., 
$$
\PP_{\np,\alpha}^{(N)}(V_{J,i}\leq v) = \PP_{\cp,\alpha}^{(N)}(V_{J,i'}\leq v),\quad \forall i,i'\in\{1,\cdots, n\}.
$$
We are done. \Halmos
\endproof

\subsection{Proof of Proposition~\ref{proposition:asymp_indep}}
\proof{Proof of Proposition~\ref{proposition:asymp_indep}.}
For succinctness, let random variables $[1],[2],\cdots,[n]$ be the population indices of the individuals placed into this randomly chosen naive pool $J$.

To prove the proposition, we want to show that the joint cdf of viral loads in a naive pool factors into a product of cdf's of individual viral loads as $N\rightarrow \infty$. Let $\BFu\in\mathbb{R}_{\geq 0}^n$. We first use the law of conditional probability to expand the joint cdf:
\begin{align}
    &\PP_{\np,\alpha}^{(N)}(U_{[1]}\leq u_1, \cdots, U_{[n-1]}\leq u_{n-1}, U_{[n]}\leq u_{n})\nonumber\\
    =~& \PP_{\np,\alpha}^{(N)}(U_{[1]}\leq u_1, \cdots, U_{[n-1]}\leq u_{n-1})\cdot \PP_{\np,\alpha}^{(N)}(U_{[n]}\leq u_{n}\mid U_{[1]}\leq u_1, \cdots, U_{[n-1]}\leq u_{n-1}).\label{eq:prop2_main}
\end{align}

To analyze the conditional probability in the second term of Equation~\ref{eq:prop2_main}, we first make the following claim: For all $\BFj\subset\{1,\cdots,N\}$ with $|\BFj|=n-1$ and $i\notin\BFj$,
\begin{align}
    \left|\PP_{\alpha}^{(N)}(U_{i}\leq u_{n}\mid U_{j_1}\leq u_1, \cdots, U_{j_{n-1}}\leq u_{n-1}) - \PP_{\alpha}^{(N)}(U_i \leq u_n)
    \right|\leq \Lambda_{\alpha}^{(N)}(i,\BFj).\label{eq:prop2_claim}
\end{align}

To prove Claim~\ref{eq:prop2_claim}, we first expand and bound its conditional probability:
\begin{align*}
    &\PP_{\alpha}^{(N)}(U_{i}\leq u_{n}\mid U_{j_1}\leq u_1, \cdots, U_{j_{n-1}}\leq u_{n-1})\\
    =~ & \frac{\PP_{\alpha}^{(N)}(U_{i}\leq u_{n}, U_{j_1}\leq u_1, \cdots, U_{j_{n-1}}\leq u_{n-1})}{\PP_{\alpha}^{(N)}(U_{j_1}\leq u_1, \cdots, U_{j_{n-1}}\leq u_{n-1})}\\
    =~& \frac{\int_{\BFz\in[0,u_1]\times\cdots \times[0,u_{n-1}]} \PP_{\alpha}^{(N)}(U_i\leq u_n\mid U_{j_1}=z_1,\cdots, U_{j_{n-1}}=z_{n-1}) f(U_{j_1}=z_1,\cdots, U_{j_{n-1}}=z_{n-1})\diff\BFz'}{\int_{\BFz\in[0,u_1]\times\cdots \times[0,u_{n-1}]}f(U_{j_1}=z_1',\cdots, U_{j_{n-1}}=z'_{n-1})\diff\BFz'}\\
    \in~&\left[\inf_{\BFz\in\mathbb{R}_{\geq 0}^{n-1}} \PP_{\alpha}^{(N)}(U_i\leq u_n\mid U_{j_1}=z_1,\cdots, U_{j_{n-1}}=z_{n-1}) , \  \sup_{\BFz\in\mathbb{R}_{\geq 0}^{n-1}} \PP_{\alpha}^{(N)}(U_i\leq u_n\mid U_{j_1}=z_1,\cdots, U_{j_{n-1}}=z_{n-1}) \right].
\end{align*}

Then, the proof of Claim~\ref{eq:prop2_claim} becomes straightforward:
\begin{align*}
    &\left|\PP_{\alpha}^{(N)}(U_{i}\leq u_{n}\mid U_{j_1}\leq u_1, \cdots, U_{j_{n-1}}\leq u_{n-1}) - \PP_{\alpha}^{(N)}(U_i \leq u_n)
    \right|\\
    \leq ~& \max\left\{\left|\inf_{\BFz\in\mathbb{R}_{\geq 0}^{n-1}} \PP_{\alpha}^{(N)}(U_i\leq u_n\mid U_{j_1}=z_1,\cdots, U_{j_{n-1}}=z_{n-1}) - \PP_{\alpha}^{(N)}(U_i \leq u_n)\right|,\right.\\
    &\hspace{5em}   \left.\left|\sup_{\BFz\in\mathbb{R}_{\geq 0}^{n-1}} \PP_{\alpha}^{(N)}(U_i\leq u_n\mid U_{j_1}=z_1,\cdots, U_{j_{n-1}}=z_{n-1})- \PP_{\alpha}^{(N)}(U_i \leq u_n)\right|\right\}\\
    \leq~& \sup_{\BFz\in\mathbb{R}_{\geq 0}^{n-1}}\left| \PP_{\alpha}^{(N)}(U_i\leq u_n\mid U_{j_1}=z_1,\cdots, U_{j_{n-1}}=z_{n-1})- \PP_{\alpha}^{(N)}(U_i \leq u_n)\right|\\
    \leq~& \sup_{\substack{u_n\in\mathbb{R}_{\geq 0}\\\BFz\in\mathbb{R}_{\geq 0}^{n-1}}}\left| \PP_{\alpha}^{(N)}(U_i\leq u_n\mid U_{j_1}=z_1,\cdots, U_{j_{n-1}}=z_{n-1})- \PP_{\alpha}^{(N)}(U_i \leq u_n)\right|\\
    =~& \Lambda_{\alpha}^{(N)}(i,\BFj).
\end{align*}

Claim~\ref{eq:prop2_claim} enables a closer analysis of the conditional probability in Equation~\ref{eq:prop2_main}. Using the law of iterated expectations where we condition on $[1],[2],\cdots,[n]$ (hereafter abbreviated as $[1:n]$), we have that
\begin{align}
    &\PP_{\np,\alpha}^{(N)}(U_{[n]}\leq u_{n}\mid U_{[1]}\leq u_1, \cdots, U_{[n-1]}\leq v_{n-1})\nonumber \\
    =~& \EE_{\np,\alpha}^{(N)}\left[\PP_{\alpha}^{(N)}(U_{[n]}\leq u_{n}\mid U_{[1]}\leq u_1, \cdots, U_{[n-1]}\leq u_{n-1}, [1:n])\right]\nonumber\\
    \leq~& \EE_{\np,\alpha}^{(N)}\left[\PP_{\alpha}^{(N)}(U_{[n]}\leq u_{n}\mid [n]) + \Lambda_{\alpha}^{(N)}([n], [1:n-1]) \right]\nonumber\\
    =~& \PP_{\np,\alpha}^{(N)}(U_{[n]}\leq u_{n}) + \EE_{\np,\alpha}^{(N)}\left[\Lambda_{\alpha}^{(N)}([n], [1:n-1]) \right].\label{eq:prop2_LIE_ub}
\end{align}
We consider two cases for the expectation in the second term. For any $\epsilon>0$, $\Lambda_{\alpha}^{(N)}([n], [1:n-1])$ could either be less than $\epsilon$, or greater than $\epsilon$ but upper bounded by 1. That is, 
\begin{align}
    \EE_{\np,\alpha}^{(N)}\left[\Lambda_{\alpha}^{(N)}([n], [1:n-1]) \right]&\leq 1\cdot \PP_{\np,\alpha}^{(N)}(\Lambda_{\alpha}^{(N)}([n], [1:n-1])>\epsilon) + \epsilon\cdot \PP_{\np,\alpha}^{(N)}(\Lambda_{\alpha}^{(N)}([n], [1:n-1])\leq\epsilon)\nonumber\\
    &\leq \PP_{\np,\alpha}^{(N)}(\Lambda_{\alpha}^{(N)}([n], [1:n-1])>\epsilon) + \epsilon.\label{eq:prop2_LIE_ub_2nd_term}
\end{align}
Now, we unpack the first term in this expression.
\begin{align*}
    \PP_{\np,\alpha}^{(N)}(\Lambda_{\alpha}^{(N)}([n], [1:n-1])>\epsilon) &= \EE_{\np,\alpha}^{(N)} \left[\PP_{\np,\alpha}^{(N)} (\Lambda_{\alpha}^{(N)}([n],[1:n-1])>\epsilon\mid [1:n-1])\right]\\
    &= \EE_{\np,\alpha}^{(N)} \left[\frac{|\{i:\Lambda_{\alpha}^{(N)}(i,[1:n-1])>\epsilon\}|}{N-(n-1)}\mid [1:n-1]\right]\\
    &\hspace{0.25em}\text{because under $\PP_{\np,\alpha}^{(N)}$, $[n]$ takes values other than $[1:n-1]$ with equal probability}\\
    &\leq \EE_{\np,\alpha}^{(N)}\left[\frac{d_{\alpha}^{(N)}(\epsilon)}{N-(n-1)}\mid [1:n-1]\right]\quad\text{by Equation~\ref{eq:pop_model_d_def}}\\
    &= \frac{d_{\alpha}^{(N)}(\epsilon)}{N-(n-1)}.
\end{align*}
Plugging this result back to Equations~\ref{eq:prop2_LIE_ub_2nd_term} and~\ref{eq:prop2_LIE_ub}, we have the following for each $\epsilon>0$:
\begin{align}
    &\PP_{\np,\alpha}^{(N)}(U_{[n]}\leq u_{n}\mid U_{[1]}\leq u_1, \cdots, U_{[n-1]}\leq u_{n-1})\nonumber \\
    =~& \PP_{\np,\alpha}^{(N)}(U_{[n]}\leq u_{n}) + \EE_{\np,\alpha}^{(N)}\left[\Lambda_{\alpha}^{(N)}([n], [1:n-1]) \right]\nonumber\\
    \leq~& \PP_{\np,\alpha}^{(N)}(U_{[n]}\leq u_{n}) + \frac{d_{\alpha}^{(N)}(\epsilon)}{N-(n-1)} + \epsilon.\label{eq:bound_for_iterative_decomposition}
\end{align}

We can apply Bound~\ref{eq:bound_for_iterative_decomposition} to iteratively decompose and bound the full joint cdf in Equation~\ref{eq:prop2_main}. 
\begin{align*}
    &\PP_{\np,\alpha}^{(N)}(U_{[1]}\leq u_1, \cdots, U_{[n-1]}\leq u_{n-1}, U_{[n]}\leq u_{n})\nonumber\\
    =~& \PP_{\np,\alpha}^{(N)}(U_{[1]}\leq u_1, \cdots, U_{[n-1]}\leq u_{n-1})\cdot \PP_{\np,\alpha}^{(N)}(U_{[n]}\leq u_{n}\mid U_{[1]}\leq u_1, \cdots, U_{[n-1]}\leq u_{n-1})\\
    \leq~& \PP_{\np,\alpha}^{(N)}(U_{[1]}\leq u_1, \cdots, U_{[n-1]}\leq u_{n-1}) \cdot\left(\PP_{\np,\alpha}^{(N)}(U_{[n]}\leq u_{n}) + \frac{d_{\alpha}^{(N)}(\epsilon)}{N-(n-1)} + \epsilon\right)\\
    \leq~& \PP_{\np,\alpha}^{(N)}(U_{[1]}\leq u_1, \cdots, U_{[n-2]}\leq u_{n-2}) \cdot\left(\PP_{\np,\alpha}^{(N)}(U_{[n]}\leq u_{n}) + \frac{d_{\alpha}^{(N)}(\epsilon)}{N-(n-1)} + \epsilon\right)\\
    &\hspace{50mm}\cdot\left(\PP_{\np,\alpha}^{(N)}(U_{[n-1]}\leq u_{n-1}) + \frac{d_{\alpha}^{(N)}(\epsilon)}{N-(n-2)} + \epsilon\right)\\
    \leq~& \cdots\\
    \leq~& \prod_{k=1}^n \left(\PP_{\np,\alpha}^{(N)}(U_{[k]}\leq u_{k}) + \frac{d_{\alpha}^{(N)}(\epsilon)}{N-(n-k)} + \epsilon\right)\\
    \leq~& \prod_{k=1}^n \left(\PP_{\np,\alpha}^{(N)}(U_{[k]}\leq u_{k}) + \frac{d_{\alpha}^{(N)}(\epsilon)}{N-n} + \epsilon\right).
\end{align*}

Let $\epsilon_N$ be a sequence satisfying Assumption~\ref{assu:corr_decay_indep}, i.e., $\epsilon_N\downarrow 0$ and $\lim_{N\rightarrow \infty}\frac{1}{N}d_{\alpha}^{(N)}(\epsilon_N)=0$. Taking the limit $N\rightarrow \infty$ of the expression above, we have
\begin{align*}
    \lim_{N\rightarrow\infty}\PP_{\np,\alpha}^{(N)}(U_{[1]}\leq u_1, \cdots, U_{[n-1]}\leq u_{n-1}, U_{[n]}\leq u_{n}) &\leq \lim_{N\rightarrow\infty }\prod_{k=1}^n \left(\PP_{\np,\alpha}^{(N)}(U_{[k]}\leq u_{k}) + \frac{d_{\alpha}^{(N)}(\epsilon_N)}{N-n} + \epsilon_N\right)\\
    &= \lim_{N\rightarrow\infty }\prod_{k=1}^n \PP_{\np,\alpha}^{(N)}(U_{[k]}\leq u_{k}).
\end{align*}

Similarly, we can use the other direction of Inequality~\ref{eq:prop2_claim} to derive a lower bound counterpart to Inequality~\ref{eq:prop2_LIE_ub}: 
\begin{align}
    &\PP_{\np,\alpha}^{(N)}(U_{[n]}\leq u_{n}\mid U_{[1]}\leq u_1, \cdots, U_{[n-1]}\leq u_{n-1})\nonumber \\
    \geq~& \EE_{\np,\alpha}^{(N)}\left[\PP_{\np,\alpha}^{(N)}(U_{[n]}\leq u_{n}\mid [n]) - \Lambda_{\alpha}^{(N)}([n], [1:n-1]) \right]\nonumber\\
    =~& \PP_{\np,\alpha}^{(N)}(U_{[n]}\leq u_{n}) - \EE_{\np,\alpha}^{(N)}\left[\Lambda_{\alpha}^{(N)}([n], [1:n-1]) \right].\label{eq:prop2_LIE_lb}
\end{align}

Applying Inequality~\ref{eq:prop2_LIE_lb} to Equation~\ref{eq:prop2_main}, we derive a lower bound for the joint cumulative distribution function:
\begin{align*}
    &\PP_{\np,\alpha}^{(N)}(U_{[1]}\leq u_1, \cdots, U_{[n-1]}\leq u_{n-1}, U_{[n]}\leq u_{n})\geq \prod_{k=1}^n \left(\PP_{\np,\alpha}^{(N)}(U_{[k]}\leq u_{k}2) - \frac{d_{\alpha}^{(N)}(\epsilon)}{N-n} - \epsilon\right).
\end{align*}

For the same sequence of $\epsilon_N$ satisfying Assumption~\ref{assu:corr_decay_indep}, we have 
\begin{align*}
    \lim_{N\rightarrow\infty}\PP_{\np,\alpha}^{(N)}(U_{[1]}\leq u_1, \cdots, U_{[n-1]}\leq u_{n-1}, U_{[n]}\leq u_{n}) &\geq \lim_{N\rightarrow\infty }\prod_{k=1}^n \left(\PP_{\np,\alpha}^{(N)}(U_{[k]}\leq u_{k}) - \frac{d_{\alpha}^{(N)}(\epsilon_N)}{N-n} - \epsilon_N\right)\\
    &= \lim_{N\rightarrow\infty }\prod_{k=1}^n \PP_{\np,\alpha}^{(N)}(U_{[k]}\leq u_{k}).
\end{align*}

Since the lower and upper bounds coincide, we have that
\begin{align*}
    \lim_{N\rightarrow\infty}\PP_{\np,\alpha}^{(N)}(U_{[1]}\leq u_1, \cdots, U_{[n-1]}\leq u_{n-1}, U_{[n]}\leq u_{n}) = \lim_{N\rightarrow\infty }\prod_{k=1}^n \PP_{\np,\alpha}^{(N)}(U_{[k]}\leq u_{k}),
\end{align*}
i.e., 
$$
    \PP_{\np,\alpha}(U_{[1]}\leq u_1, \cdots, U_{[n-1]}\leq u_{n-1}, U_{[n]}\leq u_{n}) = \prod_{k=1}^n \PP_{\np,\alpha}(U_{[k]}\leq u_{k}),
$$
which concludes the proof.
\Halmos 
\endproof

\subsection{Proof of Proposition~\ref{proposition:corr_pool}}
\label{appdx:corr_pool}

\proof{Proof of Proposition~\ref{proposition:corr_pool}.}
For succinctness, we abbreviate the probability operator $\PP_{\cp,\alpha}^{(N)}(\cdot)$ and the expectation operator $\EE_{\cp,\alpha}^{(N)}[\cdot]$ as $\PP(\cdot)$ and $\EE[\cdot]$ in Appendix~\ref{appdx:corr_pool}.

For a generic pool $j\in\{1,\cdots,|\mathcal{A}|\}$, let $I(j)$ be the sample in pool $A_j$ with nonzero infection probability and the smallest population index, $I(j) = \min\{i:\PP(U_i >0) >0, i\in A_j\}$. If such a sample does not exist in $A_j$, then $I(j) = \infty$. 
Let $C_{I(j)}$ denote the set of $I(j)$'s close contacts and $K(j)$ denote an individual selected uniformly at random from $C_{I(j)}$. 
Let $S_j=\sum_{i\in A_j} \mathbbm{1}\{U_{i}>0\}$. Since the pooling assignment $\mathcal{A}$ is a random variable, $A_j$, $I(j)$ and $C_{I(j)}$ are all random.
We make the following observation: if sample $I(j)$ is positive, sample $K(j)$ is positive, and $K(j)$ is also in pool $j$, then pool $j$ must contain more than one positive. Therefore,
\begin{align*}
    \PP(S_j > 1) &= \PP(S_j > 1\mid I(j) <\infty)\cdot\PP(I(j) < \infty)\\
    &\geq \PP(U_{I(j)}>0, U_{K(j)}>0, K(j)\in A_j\mid I(j) <\infty)\cdot \PP\left(\sum_{i\in A_j}\PP(U_i>0) > 0\right)\\
    &= \PP(U_{I(j)}>0\mid I(j) <\infty) \cdot \PP(U_{K(j)}>0\mid U_{I(j)}>0, I(j) <\infty) \cdot\\
    &\hspace{10em}\PP(K(j)\in A_j\mid U_{I(j)}>0, U_{K(j)}>0, I(j) <\infty)\cdot
    \PP\left(\sum_{i\in A_j}\PP(U_i>0) > 0\right)\\
    &= \PP(U_{I(j)}>0\mid I(j)<\infty) \cdot \PP(U_{K(j)}>0\mid U_{I(j)}>0, I(j)<\infty) \cdot \PP(K(j)\in A_j)\cdot \PP\left(\sum_{i\in A_j}\PP(U_i>0) > 0\right)\\
    &\hspace{10mm} \text{since pooling assignment is assumed to be independent of viral loads}\\
    &\geq  \epsilon_0\alpha \cdot c_1 \cdot c_2\cdot \PP\left(\sum_{i\in A_j}\PP(U_i>0) > 0\right) \quad\text{by Assumption~\ref{assu:justify_CP}}.
\end{align*}

We generalize this result to a pool $J$ selected uniformly at random from all pools.
\begin{align}
\PP(S > 1) &= \sum_{j=1}^{|\mathcal{A}|}
\PP(S_j > 1) \PP(J = j)\nonumber\\
&=\frac{1}{|\mathcal{A}|}\sum_{j=1}^{|\mathcal{A}|}\PP(S_j > 1)\nonumber\\
&\geq \frac{1}{|\mathcal{A}|}\sum_{j=1}^{|\mathcal{A}|}\epsilon_0\alpha\cdot c_1\cdot c_2\cdot \PP\left(\sum_{i\in A_j}\PP(U_i>0) > 0\right)\nonumber\\
&= \epsilon_0\alpha \cdot c_1\cdot c_2\cdot \frac{1}{|\mathcal{A}|}\sum_{j=1}^{|\mathcal{A}|}\PP\left(\sum_{i\in A_j}\PP(U_i>0) > 0\right).\label{eq:proof_cond1_S1}
\end{align}

On the other hand, for a fixed pooling assignment $\mathcal{A}$, the probability that a generic pool $j$ contains one or more positives can be bounded above:
\begin{align}
    \PP(S_j>0\mid \mathcal{A}) &= \PP\left(\bigcup_{i\in A_j}\mathbbm{1}\{U_{i}>0\} \mid \mathcal{A} \right)\nonumber\\
    &\leq \sum_{i\in A_j} \PP(U_i>0\mid \mathcal{A}) \quad\quad\text{by the union bound}\nonumber\\
    &= \sum_{i\in A_j} \PP(U_i>0\mid \mathcal{A}) \cdot \mathbbm{1}\left\{\sum_{i\in A_j} \PP(U_i>0)>0\mid\mathcal{A}\right\}\nonumber\\
    &= \sum_{i\in A_j} \PP(U_i>0) \cdot \mathbbm{1}\left\{\sum_{i\in A_j} \PP(U_i>0)>0\mid\mathcal{A}\right\}\nonumber\\
    &\hspace{20mm}\quad\text{since viral load does not depend on pooling assignment}\nonumber\\
    &\leq \Pi_0\alpha\cdot n \cdot \mathbbm{1}\left\{\sum_{i\in A_j} \PP(U_i>0)>0\mid\mathcal{A}\right\} \quad\text{by Assumption~\ref{assu:justify_CP}}.\label{eq:proof_cond1_S0_pool}
\end{align}

We now generalize the result in Equation~\ref{eq:proof_cond1_S0_pool} to a pool $J$ selected uniformly at random from all pools and all pooling assignments:
\begin{align}
    \PP(S > 0) &= \sum_{j=1}^{|\mathcal{A}|}\EE_{\mathcal{A}}\left[\PP(S_j > 0\mid \mathcal{A})\right] \cdot \PP(J = j)\nonumber\\
    &= \frac{1}{|\mathcal{A}|} \sum_{j=1}^{|\mathcal{A}|} \EE_{\mathcal{A}}\left[\PP(S_j > 0\mid \mathcal{A})\right]\nonumber\\
    &\leq \frac{1}{|\mathcal{A}|} \sum_{j=1}^{|\mathcal{A}|} \Pi_0\alpha\cdot n \cdot \EE_{\mathcal{A}}\left[\mathbbm{1}\left\{\sum_{i\in A_j} \PP(U_i>0)>0\mid\mathcal{A}\right\}\right]\nonumber\\
    &= \Pi_0\alpha\cdot n \cdot \frac{1}{|\mathcal{A}|} \sum_{j=1}^{|\mathcal{A}|} \PP\left(\sum_{i\in A_j} \PP(U_i>0)>0\right) .\label{eq:proof_cond1_S0}
\end{align}

Combining Equations~\ref{eq:proof_cond1_S1} and~\ref{eq:proof_cond1_S0}, we find
\begin{align*}
    \PP(S > 1\mid S > 0) &= \frac{\PP(S > 1)}{\PP(S > 0)}\\
    &\geq \frac{\epsilon_0 \alpha \cdot c_1\cdot c_2}{\Pi_0\alpha \cdot n} \\
    &= \frac{\epsilon_0\cdot c_1 \cdot c_2}{\Pi_0\cdot n},
\end{align*}
which is a positive constant that does not depend on $\alpha$ or $N$. Taking $N$ to the limit of infinity proves the proposition.
\Halmos
\endproof

\section{Proofs of Theorems~\ref{thm:fnr}, \ref{thm:eff} and Corollary~\ref{Cor:thm2_det_threshold_case}}

\subsection{Proof of Theorem~\ref{thm:fnr}}\label{appdx:thm_fnr_proof}
\proof{Proof of Theorem~\ref{thm:fnr}.}
For $\pool\in\{\np,\cp\}$, we have that the overall false negative rate is given by
\begin{align*}
    \beta_{\pool,\alpha}  &= 1 - \frac{\EE_{\pool,\alpha}[\text{\# positives identified in a pool}]}{\EE_{\pool,\alpha}[\text{\# positives in a pool}]}\\
    &= 1 - \frac{\EE_{\pool,\alpha}[D]}{n\alpha} \\
    &= 1 - \frac{\EE_{\pool,\alpha}[\sum_{i=1}^{n}YW_i]}{n\alpha}\\
    &= 1 - \frac{1}{n\alpha}\cdot \sum_{i=1}^{n}\EE_{\pool,\alpha}[YW_i]\\
    &= 1 - \frac{1}{n\alpha}\cdot \sum_{i=1}^{n}\EE_{\pool,\alpha}[\EE_{\pool,\alpha}[YW_i\mid V_{1:n}]]\\
    &= 1 - \frac{1}{n\alpha}\cdot \sum_{i=1}^{n}\EE_{\pool,\alpha}[\EE_{\pool,\alpha}[Y\mid V_{1:n}]\EE_{\pool,\alpha}[W_i\mid V_i]].
\end{align*}

In both \CP and \NP, all $V_i$'s are identically distributed by Lemma~\ref{lemma:id}, which follows that $(\EE_{\pool,\alpha}[Y\mid V_{1:n}], \EE_{\pool,\alpha}[W_i | V_{i}])$ are also identically distributed. Hence, we obtain that
\begin{align}
    \beta_{\pool,\alpha} &= 1 - \frac{1}{n\alpha}\cdot n\cdot \EE_{\pool,\alpha}[\EE_{\pool,\alpha}[Y\mid V_{1:n}]\EE_{\pool,\alpha}[W_1\mid V_1]]\nonumber\\
    &= 1 - \frac{1}{\alpha}\cdot \EE_{\pool,\alpha}\left[p\left(h(\BFV)\right) p(V_1)\right]\quad\text{where $\BFV = (V_1,\ldots, V_n)$}\nonumber\\
    &= 1 - \frac{1}{\alpha}\EE_{\pool,\alpha}\left[p\left(h(\BFV)\right) p(V_1)\mid V_1 > 0\right]\PP_\alpha(V_1> 0)\nonumber\\
    &= 1 - \EE_{\pool,\alpha}\left[p\left(h(\BFV)\right) p(V_1)\mid V_1 > 0\right]\label{eq:simplified_beta}.
\end{align}

For \NP, the $V_i$'s are i.i.d. Hence, 
\begin{align*}
    \beta_{\np,\alpha} &= 1 - \sum_{\ell=1}^{n}\EE_{\np,\alpha}\left[p\left(h(\BFV)\right) p(V_1)\mid V_1 > 0, S=\ell\right]\PP_{\np,\alpha}\left(S =\ell\mid V_1>0\right)\quad\text{recall that $S = \sum_{i=1}^{n}\mathbbm{1}\{V_i > 0\}$}\\
    &= 1 - \sum_{\ell=1}^{n}\EE_{\np,\alpha}\left[p\left(h(\BFV)\right) p(V_1)\mid V_1 > 0, S=\ell\right]{n-1\choose \ell - 1}\alpha^{\ell - 1}(1-\alpha)^{n-\ell}.
\end{align*}

Taking $\alpha\to 0^+$ gives
\begin{align*}
    \lim_{\alpha\to 0^+}\beta_{\np,\alpha} &= \lim_{\alpha\to 0^+}\Bigg( 1 - \EE_{\np,\alpha}\left[p\left(h(\BFV)\right) p(V_1)\mid V_1 > 0, S=1\right]{n-1\choose 1 - 1}\alpha^{1 - 1}(1-\alpha)^{n-1}\Bigg)\\
    &=1 - \EE\left[p\left(h(V_1,0,\ldots, 0)\right) p(V_1)\mid V_1 > 0\right].
\end{align*}

Similarly, we derive $\beta_{\cp,\alpha}$ for correlated pooling. Following Equation~\ref{eq:simplified_beta} we have that
\begin{align*}
    \beta_{\cp,\alpha} &= 1 - \sum_{\ell=1}^{n}\EE_{\cp,\alpha}\left[p\left(h(\BFV)\right) p(V_1)\mid V_1 > 0, S=\ell\right]\PP_{\cp,\alpha}\left(S =\ell\mid V_1>0\right)\\
    &\overset{\Delta}{=} 1 - \sum_{\ell = 1}^n A_{\ell}\cdot \PP_{\cp,\alpha}(S=\ell\mid S > 0)
\end{align*}
where $A_\ell\overset{\Delta}{=}\EE_{\cp,\alpha}\left[p\left(h(\BFV)\right) p(V_1)\mid V_1 > 0, S=\ell\right]$

When $\ell = 1$, $A_1 = \EE\left[p\left(h(V_1,0,\ldots, 0)\right) p(V_1)\mid V_1 > 0\right]$. When $\ell \geq 2$, we have $h(\BFV) \geq h(V_1,0,\ldots, 0)$ because there exists at least one $i\neq 1$ such that $V_i > 0$ and $h(\cdot)$ is monotone increasing as described in Section~\ref{subsec:model_setup}.  Assuming $p(v)$ is a monotone increasing function in $v$, we obtain $p(h(\BFV)) \geq p(h(V_1,0,\ldots, 0))$, which, combined with $p(V_1) > 0$ given $V_1 > 0$, implies that $A_\ell \geq A_1$. 

Therefore, taking $\alpha\to 0^+$ gives
\begin{align*}
    \lim_{\alpha\to 0^+}\beta_{\cp,\alpha} &=1 - \lim_{\alpha\to 0^+}\sum_{\ell=1}^{n}A_\ell \cdot  \PP_{\cp,\alpha}(S =\ell\mid S > 0)\\
    &= 1 - \sum_{\ell=1}^{n}A_\ell \cdot \lim_{\alpha\to 0^+} \PP_{\cp,\alpha}(S =\ell\mid S > 0) \quad\text{$A_\ell$'s do not involve $\alpha$ because they condition on $S=\ell$}\\
    & \leq 1 - \sum_{\ell=1}^{n}A_1 \cdot \lim_{\alpha\to 0^+} \PP_{\cp,\alpha}(S =\ell\mid S > 0)\\
    &= 1 - A_1\\
    &=\lim_{\alpha\to 0^+}\beta_{\cp,\alpha}.
\end{align*}

If $p(v)$ is strictly increasing in $v$, $A_\ell>A_1$ for $l>1$. By Proposition~\ref{proposition:corr_pool}, there exists $\ell\geq 2$ such that $\lim_{\alpha\to 0^+} \PP_{\cp,\alpha}(S =\ell\mid S > 0) > 0$. It follows that $\lim_{\alpha\to 0^+}\beta_{\cp,\alpha}<\lim_{\alpha\to 0^+}\beta_{\np,\alpha}$.
\Halmos
\endproof

\subsection{Proof of Theorem~\ref{thm:eff}}\label{appdx:proof_of_thm_2}

To prove Theorem~\ref{thm:eff} we first investigate an auxiliary metric whose structure admits study more easily.

\begin{definition}[Effective Follow-up Efficiency]\label{def:eff_followup_eff}
Let $\eta$ denote the number of positive cases identified per follow-up test consumed. That is, 
$\eta=\dfrac{\overline{D}}{n\overline{Y}}.$
\end{definition}

To better understand the behavior of $\gamma$, we can rewrite the expression of $\gamma$ in Definition~\ref{def:eff_eff} as 
\begin{equation}
\gamma = \left(\frac{1}{\overline{D}} + \frac{n\overline{Y}}{\overline{D}}\right)^{-1} = \left(\frac{1}{n\alpha(1-\beta)} + \frac{1}{\eta}\right)^{-1}. \label{eq:num_test_per_pos_rewrite}
\end{equation}

Analogous to Proposition~\ref{prop:convergence_of_metrics} we have that $\eta$ converges in probability to $\frac{\EE_{\pool,\alpha}[D]}{n\EE_{\pool,\alpha}[Y]}$
(denoted $\eta_{\pool,\alpha}$), as $N\to\infty$, for $\alpha>0$ and $\pool\in \{\np,\cp\}$.

We present Lemma~\ref{lemma:eff}, which provides a bound on the ratio of effective follow-up efficiency under \CP and \NP.
\begin{lemma}
$\lim_{\alpha\to 0^+}\dfrac{\eta_{\cp,\alpha}}{\eta_{\np,\alpha}}\geq (1 + \delta)^{-1}$ where $\delta = \dfrac{\PP_{\cp,\alpha}(Y=1, \Dt=0\mid S>0)}{\PP_{\cp,\alpha}( Y=1, \Dt > 0\mid S>0)}$ and $\Dt = \sum_{i=1}^{n} W_i$.\label{lemma:eff}
\end{lemma}

\proof{Proof of Lemma~\ref{lemma:eff}.}
We first derive $\eta_{\np,\alpha}$ for naive pooling. By similar arguments in the Proof of Theorem~\ref{thm:fnr}, the denominator of $\eta_{\np,\alpha}$ is given by
\begin{align}
    n\EE_{\np,\alpha}[Y] &= n\EE_{\np,\alpha}\left[p\left(h(\BFV)\right)\right]\nonumber\\
    &= n\sum_{\ell=1}^n \EE_{\np,\alpha}\left[p\left(h(\BFV)\right)\mid S=\ell\right]\PP_{\np,\alpha}\left(S=\ell\right)\nonumber\\
    &= n\sum_{\ell=1}^n \EE_{\np,\alpha}\left[p\left(h(\BFV)\right)\mid S=\ell\right] \binom{n}{\ell}\alpha ^\ell (1-\alpha)^{n-\ell}\nonumber\\
    &= n\alpha \cdot \sum_{\ell=1}^n \EE_{\np,\alpha}\left[p\left(h(\BFV)\right)\mid S=\ell\right] \binom{n}{\ell}\alpha ^{\ell-1} (1-\alpha)^{n-\ell}.\label{eq:EY_for_eta_0}
\end{align}

The numerator of $\eta_{\np,\alpha}$ is given by
\begin{align}
    \EE_{\np,\alpha}[D] &=  \EE_{\np,\alpha}\left[\sum_{i=1}^{n}YW_i\right]\nonumber\\
    &=n\alpha\cdot \EE_{\np,\alpha}[p(h(\BFV))p(V_1)\mid V_1 > 0]\nonumber\\
    &=n\alpha \cdot \sum_{\ell=1}^{n}\EE_{\np,\alpha}\left[p\left(h(\BFV)\right) p(V_1)\mid V_1 > 0, S = \ell\right]{n - 1\choose \ell - 1}\alpha^{\ell - 1} (1-\alpha)^{n-\ell}.\label{eq:ED_for_eta_0}
\end{align}

By definition of $\eta_{\np,\alpha}$ and Equations \ref{eq:ED_for_eta_0} and \ref{eq:EY_for_eta_0}, taking $\alpha\to 0^+$ gives 
\begin{align}
    \lim_{\alpha\to 0^+}\eta_{\np,\alpha} 
    &=\lim_{\alpha\to 0^+}\frac{\EE_{\np,\alpha}[D]}{n\EE_{\np,\alpha}[Y]}\nonumber\\
    &=\lim_{\alpha\to 0^+}\frac{\cancel{n\alpha} \cdot \sum_{\ell=1}^{n}\EE_{\np,\alpha}\left[p\left(h(\BFV)\right) p(V_1)\mid V_1 > 0, S = \ell\right]{n - 1\choose \ell - 1}\alpha^{\ell - 1} (1-\alpha)^{n-\ell}}{\cancel{n\alpha} \cdot \sum_{\ell=1}^n \EE_{\np,\alpha}\left[p\left(h(\BFV)\right)\mid S=\ell\right] \binom{n}{\ell}\alpha ^{\ell-1} (1-\alpha)^{n-\ell}}\nonumber\\
    &=\frac{\EE_{\np,\alpha}\left[p\left(h(\BFV)\right) p(V_1)\mid V_1 > 0,S = 1\right]}{\EE_{\np,\alpha}\left[p\left(h(\BFV)\right)\mid S=1\right] \cdot {n\choose 1}}\nonumber\\
    &=\frac{\EE\left[p\left(h(V_1,0,\ldots, 0)\right) p(V_1)\mid V_1 > 0\right]}{n\cdot \EE\left[p\left(h(V_1,0,\ldots, 0)\right)\mid V_1 > 0\right] }\quad\text{because $V_i$'s are iid}\label{eq:lim_eta_0_counterexample}\\
    &\hspace{10mm} \text{(the denominator is nonzero because $p(v)>0\ \forall v>0$)}\nonumber\\
    &= \frac{\EE[p(h(V_1,0,\ldots, 0))W_1\mid V_1 > 0]}{n\cdot \EE[p(h(V_1,0,\ldots, 0))\mid V_1 > 0]}\nonumber\\
    &= \frac{1}{n}\cdot \frac{\EE[p(h(V_1,0,\ldots, 0))\mid V_1 > 0, W_1 = 1]\cdot\PP(W_1=1\mid V_1 > 0)}{\sum_{j=0,1}\EE[p(h(V_1,0,\ldots, 0))\mid V_1 > 0, W_1 = j]\cdot\PP(W_1 = j\mid V_1 > 0) }\nonumber\\
    &= \frac{1}{n}\cdot \left(1 + \frac{\EE[p(h(V_1,0,\ldots, 0))\mid V_1 > 0, W_1 = 0]}{\EE[p(h(V_1,0,\ldots, 0))\mid V_1 > 0, W_1 = 1]}\cdot \frac{\PP(W_1=0\mid V_1 > 0)}{\PP(W_1=1\mid V_1 > 0)}\right)^{-1}\label{eq:limit_of_eta_0}.
\end{align}

Then, we derive $\eta_{\cp,\alpha}$ for correlated pooling. 
\begin{align}
    \eta_{\cp,\alpha} &= \frac{\EE_{\cp,\alpha}[D]}{n\EE_{\cp,\alpha}[Y]}\nonumber\\
    &= \frac{\EE_{\cp,\alpha}[\sum_{i=1}^{n}YW_i]}{n\EE_{\cp,\alpha}[Y]}\nonumber\\
    &= \frac1n\cdot \frac{\EE_{\cp,\alpha}[Y\Dt\mid S>0]\cancel{\PP_{\cp,\alpha}(S > 0)}}{\EE_{\cp,\alpha}[Y\mid S > 0]\cancel{\PP_{\cp,\alpha}(S > 0)}}\label{eq:lim_eta_1_counterexample}\\
    &= \frac1n\cdot \frac{\EE_{\cp,\alpha}[Y\Dt\mid \Dt > 0]\PP_{\cp,\alpha}(\Dt > 0\mid S > 0)}{\PP_{\cp,\alpha}(Y=1\mid \Dt>0)\PP_{\cp,\alpha}(\Dt>0\mid S>0) + \PP_{\cp,\alpha}(Y = 1\mid \Dt = 0, S > 0)\PP_{\cp,\alpha}(\Dt=0\mid S>0)}\nonumber\\
    &\geq  \frac1n\cdot \frac{\PP_{\cp,\alpha}(Y=1\mid \Dt > 0)\PP_{\cp,\alpha}(\Dt > 0\mid S > 0)}{\PP_{\cp,\alpha}(Y=1\mid \Dt>0)\PP_{\cp,\alpha}(\Dt>0\mid S>0) + \PP_{\cp,\alpha}(Y = 1\mid \Dt = 0, S > 0)\PP_{\cp,\alpha}(\Dt=0\mid S>0)}\nonumber\\
    &\hspace{30mm}\text{because $\EE_{\cp,\alpha}[Y\Dt\mid \Dt > 0] \geq \EE_{\cp,\alpha}[Y\mid \Dt>0] = \PP_{\cp,\alpha}(Y=1\mid \Dt>0)$}\label{eq:YS_D_geq_Y}\\
    &\hspace{30mm} \text{(both terms in the denominator are nonzero because $p(v)>0\ \forall v>0$)}\nonumber\\
    &= \frac1n\left(1 + \frac{\PP_{\cp,\alpha}(Y=1\mid \Dt=0, S>0)}{\PP_{\cp,\alpha}(Y=1\mid \Dt > 0)}\cdot \frac{\PP_{\cp,\alpha}(\Dt = 0\mid S>0)}{\PP_{\cp,\alpha}(\Dt > 0\mid S > 0)}\right)^{-1}\nonumber\\
    &= \frac1n\left(1 + \frac{\PP_{\cp,\alpha}(Y=1, \Dt=0\mid S>0)}{\PP_{\cp,\alpha}(Y=1, \Dt > 0\mid S>0)}\right)^{-1} \label{eq:bound_on_eta_1}.
\end{align}

Upper-bounding Equation~\ref{eq:limit_of_eta_0} by $\dfrac1n$ and using Equation~\ref{eq:bound_on_eta_1} gives the desired result.
\Halmos
\endproof

Then, the proof of Theorem~\ref{thm:eff} follows Lemma~\ref{lemma:eff} in a straightforward manner.
\proof{Proof of Theorems~\ref{thm:eff}.}
By Equation~\ref{eq:num_test_per_pos_rewrite}, we have that for $\pool\in\{\cp,\np\}$
\begin{equation}
\gamma_{\pool,\alpha} = \left(\frac{1}{n\alpha(1-\beta_{\pool,\alpha})} + \frac{1}{\eta_{\pool,\alpha}}\right)^{-1}. \label{eq:eff_efficiency_formula}
\end{equation}

Hence, using the results shown in Theorems~\ref{thm:fnr}~and~\ref{thm:eff}, we find that
\begin{align*}
    \gamma_{\cp,\alpha} &= \left(\frac{1}{n\alpha(1-\beta_{\cp,\alpha})} + \frac{1}{\eta_{\cp,\alpha}}\right)^{-1}\\
    &\geq \left(\frac{1}{n\alpha(1-\beta_{\np,\alpha})} + \frac{1}{(1+\delta)^{-1}\eta_{\np,\alpha}}\right)^{-1}\\
    &\geq \left((1+\delta)\left(\frac{1}{n\alpha(1-\beta_{\np,\alpha})} + \frac{1}{\eta_{\np,\alpha}}\right)\right)^{-1}\\
    &= (1+\delta)^{-1}\gamma_{\np,\alpha},
\end{align*}
which concludes the proof.\Halmos
\endproof

\subsection{Proof of Corollary~\ref{Cor:thm2_det_threshold_case}}
\proof{Proof of Corollary~\ref{Cor:thm2_det_threshold_case}.}

We apply the threshold sensitivity function to the calculation of $\lim_{\alpha\to 0^+}\eta_{\np,\alpha}$ and $\lim_{\alpha\to 0^+}\eta_{\cp,\alpha}$. In Equation~\ref{eq:limit_of_eta_0}, the first term on the numerator in the parenthesis is given by
\begin{align*}
    \EE\left[p\left(\frac{1}{n}V_1\right)\mid V_1 > 0, W_1 = 0\right] &= \EE\left[\mathbbm{1}\left\{\frac{1}{n}V_1\geq u_0\right\}\mid V_1 > 0, V_1<u_0\right]=0,
\end{align*}
which implies $\lim_{\alpha\to 0^+}\eta_{\np,\alpha}=1/n$. 

In Equation~\ref{eq:bound_on_eta_1}, the numerator of the last term is given by
\begin{align*}
    \PP_{\cp,\alpha}(Y=1, \Dt=0\mid S>0) &= \PP_{\cp,\alpha}(\bar{V}_n \geq u_0, V_i<u_0\ \forall i\text{ s.t. }V_j>0 \mid S>0)=0,
\end{align*}
which implies $\eta_{\cp,\alpha}\geq \frac{1}{n}$. Hence, $\lim_{\alpha\to 0^+}\dfrac{\eta_{\cp,\alpha}}{\eta_{\np,\alpha}}\geq 1$, which follows that $\lim_{\alpha\to 0^+}\dfrac{\gamma_{\cp,\alpha}}{\gamma_{\np,\alpha}}\geq 1.$
\Halmos
\endproof

\section{Example Where Correlated Pooling Has Lower Efficiency}\label{appdx:example}

We give an example of sensitivity and viral load distribution under which correlated pooling has lower test efficiency, contrary to the claims in the literature. 

Consider a sensitivity function $p$ such that $p(0)=0$, $p(1) = 1$, and $p(1/2)=q$ for some $q\in(0,1)$. Suppose that any pooled test is subject to dilution by a factor equal to the pool size, two. We examine a correlated pool consisting of two samples with the joint viral load distribution given in Table~\ref{tab:counterex_VL_dist}. By Lemma~\ref{lemma:id} and Proposition~\ref{proposition:asymp_indep}, the corresponding naive pool contains two samples whose viral loads are independent with the same marginal distribution as that in Table~\ref{tab:counterex_VL_dist}. 
\begin{table}[h]
\TABLE
{Joint viral load distribution in the correlated pool.\label{tab:counterex_VL_dist}}
{\addtolength{\tabcolsep}{7pt}\begin{tabular}{ccc}
\hline
 & $V_2=0$ & $V_2=1$ \\ \hline
$V_1=0$ & $1-\alpha$ & $0$  \\ 
$V_1=1$ & $0$ & $\alpha$ \\ \hline
\end{tabular}\addtolength{\tabcolsep}{-7pt}}{}
\end{table}

For $\pool\in\{\np,\cp\}$ and prevalence $\alpha$, we have $\text{efficiency}_{\pool,\alpha}$, the number of individuals screened per test consumed, given by the following expression:
    \begin{equation}
    \text{efficiency}_{\pool,\alpha} = \frac{n}{1 + n\EE_{\pool, \alpha}[Y]}\quad\text{for } \pool\in\{\np,\cp\}. \label{eq:appdx_eff_as_eta_beta}
\end{equation}

To derive efficiency, we compute the expected value of $Y$, the pooled test outcome:
\begin{align}
    \EE_{\np,\alpha}[Y] &= \alpha^2\cdot 1 + 2\alpha(1-\alpha)\cdot q + (1-\alpha)^2\cdot 0 = \alpha^2 + 2q\cdot\alpha(1-\alpha)\nonumber\\
    \EE_{\cp,\alpha}[Y] &= \alpha\cdot 1 + (1-\alpha)\cdot 0 = \alpha.\label{eq:EY_counter}
\end{align}

Plugging Equation~\ref{eq:EY_counter} into Equation~\ref{eq:appdx_eff_as_eta_beta} gives the expressions for efficiency under naive and correlated pooling:
\begin{align*}
    \text{efficiency}_{\np,\alpha} &= \left(\frac{1}{2} + \alpha^2 + 2q\cdot\alpha(1-\alpha) \right)^{-1}\nonumber\\
    \text{efficiency}_{\cp,\alpha} &= \left(\frac{1}{2} + \alpha\right)^{-1}.
\end{align*}

We observe that when $q\in(0,1/2)$, for any $\alpha\in (0, 1)$, $\text{efficiency}_{\np,\alpha} > \text{efficiency}_{\cp,\alpha}$. 

\section{Supplemental Information for the Dynamic Simulation}

We provide implementation details for our simulation in Section~\ref{sec:exp}.
In Section~\ref{appdix:network_simulation_description} we describe the setup of the network-based epidemic simulation with large-scale screening using pooled testing. 
In Section~\ref{appdx:VL_progression} we model the viral load progression over time within an infected individual.
In Section~\ref{appdx:PCR} we describe a realistic model for PCR testing.

\subsection{Screening and Pooling in a Social Network}\label{appdix:network_simulation_description}

\subsubsection{Network Generation}
We build on the \texttt{SEIRSplus} \citepAP{seirsplus} library to simulate generalized SEIRS disease dynamics on a contact network. 
We generate population-wide contact networks with realistic household and community structures using the library's built-in implementation of the FARZ algorithm \citepAP{fagnan2018modular}.
Given input distributions of age and household size, the FARZ algorithm creates communities within the same age group and households across age groups that comply to the desired distributions. 
Each household is fully connected.
We set the population size to be 10,000 and the household size and age distributions to be those mimicking the United States in \texttt{SEIRSplus} (Tables~\ref{tab:US_household_size_dist_dynamic} and~\ref{tab:age_dist_dynamic}).
The documentation of \texttt{SEIRSplus} \citepAP{seirsplus} provides further details of the FARZ implementation.

\begin{table}[!htbp]
    \centering
    \TABLE{U.S. household size distribution.  \label{tab:US_household_size_dist_dynamic}}
    {\addtolength{\tabcolsep}{1pt}
    \begin{tabular}{llllllll}
        \toprule
        Household size & 1 & 2 & 3 & 4 & 5 & 6 & 7\\
        Weight & 0.284 & 0.345 & 0.151 & 0.128 & 0.058 & 0.023 & 0.013 \\ \bottomrule
    \end{tabular}
    \addtolength{\tabcolsep}{-1pt}}
{}
\end{table}

\begin{table}[!htbp]
    \centering
    \TABLE{U.S. age distribution.  \label{tab:age_dist_dynamic}}
    {\addtolength{\tabcolsep}{1pt}
    \begin{tabular}{llllllllll}
        \toprule
        Age & 0-9 & 10-19 & 20-29 & 30-39 & 40-49 & 50-59 & 60-69 & 70-79 & 80+\\
        Weight & 0.121 & 0.131 & 0.137 & 0.133 & 0.124 & 0.131 & 0.115 & 0.070 & 0.038 \\ \bottomrule
    \end{tabular}
    \addtolength{\tabcolsep}{-1pt}}{}
\end{table}

To mimic a certain level of social distancing amid a pandemic, we downsample the edges generated by the FARZ algorithm. For each node, we sample a number $n_e$ from Exponential(1/50), select $n_e$ edges uniformly at random, and discard the rest. 
We ensure that each household is still fully connected.

\subsubsection{Epidemic Dynamics and Interventions}

The epidemic follows the classical SEIR dynamics \citepAP{biswas2014seir} with additional compartments for isolation. More description can be found in the documentation of \texttt{SEIRSplus} \citepAP{seirsplus}. 
We simulate repeated population-wide screening as an intervention. 
Given a choice of screening frequency, the population is divided into equally sized screening groups. 
Each individual is assigned to a specific screening group and one group is tested on each day using pooled testing. 
Positive individuals identified in screening are isolated and isolated individuals do not participate in screening.
Isolation lasts for at most 14 days, after which the subject would be released. 

We set the simulation parameter \texttt{alpha}, governing susceptibility to infection, to be 2, and we increase the intra-household edge weight to 10 to mimic faster transmission within households than between households. 

\subsubsection{Pooling Based on Node Clustering}\label{appdix:node_clustering_desc}

To create screening groups from the population and correlated pools from each screening group, we generate vector representations for each node and cluster similar nodes using $k$-means clustering.

We use the Python implementation \citepAP{node2vec_github} of \texttt{node2vec} \citepAP{grover2016node2vec} to generate a vector representation (i.e., an \textit{embedding}) for each node that captures the node's structural position in the network and the communities it belongs to. 
We use the following parameters in running \texttt{node2vec}:
\begin{itemize}
    \item embedding dimensions: 32
    \item number of nodes in each walk: 20
    \item number of walkers per node: 10
    \item number of workers per node: 1
    \item window: 10
    \item min\_count: 1
\end{itemize}
Furthermore, to emphasize the household structure in learning the embedding, we set a weight $10^{10}$ of for intra-household edges while keeping the weight to be 1 for other edges.
(This modification only affects the learning of node embedding. It does not affect the transmission dynamics on the network.)

Given the learned embeddings, we partition the nodes into smaller, equally-sized clusters using $k$-means clustering and minimum weight matching, using L2 as the distance metric.  
In particular, to partition $n_{N}$ nodes into $n_{C}$ clusters of size $s$ within embedding space $\mathbb{R}^{d}$ (without loss of generality, assume $n_N=n_C\cdot s$), we perform the following:
\begin{itemize}
    \item First, we run $k$-means clustering to obtain $n_{C}$ cluster centroids. They can be represented in a matrix $C\in\mathbb{R}^{n_C\times d}$ where each row is a centroid. The clusters formed from $k$-means are not necessarily equally-sized.
    \item Let $\tilde{C}\in \mathbb{R}^{n_N\times d}$ be the matrix obtained from repeating $C$ for $s$ times along the row dimension. Mathematically, $\tilde{C} = (\mathbf{1}_{s} \bigotimes I_{n_C})C$, where $\mathbf{1}_{s}$ is the all-1 column vector in $\mathbb{R}^{s}$, $I_{n_C}$ is the identity matrix in $\mathbb{R}^{n_C}$, and $\bigotimes$ denotes the Kronecker product.
    \item Compute $L\in\mathbb{R}^{n_N\times n_N}$ such that $L_{ij}$ is the L2 distance between the $i$th node embedding and the $j$th row in $\tilde{C}$.
    \item Solve the minimum weight matching problem using $L$ as the cost matrix, such that each node is matched to one row in $\tilde{C}$ and the total cost is minimized: $$\min \sum_{i=1}^{n_N}\sum_{j=1}^{n_N} L_{ij} X_{ij},\quad X_{ij}\in\{0,1\}.$$
    Denote the solution as $X^*$. By construction, only one entry is 1 and the rest are 0 in each row and each column of $X^*$.
    \item For each node $i$, let $J(i)$ denote the location of 1 in the $i$th row of $X^*$. Assign node $i$ to the cluster $(J(i) \ mod \ n_C)$. It can be shown that the clusters assigned this way are all equally sized.
\end{itemize}

In our simulation, suppose we screen the size-$N$ population every $k$ days using pools of size $n$. The above procedure has three use cases:
\begin{itemize}
    \item Generating screening groups from the population: partition $N$ nodes into size-$N/k$ clusters.
    \item Generating community-correlated pools: partition the screening group into size-$2n$ clusters; within each cluster, divide them randomly into two size-$n$ pools. We use this to simulate community-induced correlation.
    \item Generating household-correlated pools: partition the screening group directly into size-$n$ clusters. This simulates household-induced correlation.
\end{itemize}

Finally, \NP is implemented by reordering the entire screening group and forming pools sequentially from the permuted group members.

\subsubsection{Validation of Pooling Implementation}\label{appdx:validate_node_clustering}
We present numerical evidence that validates the implementation of community and household-correlated pooling.
We consider the setting of screening every five days on a population of size 10,000 using pools of size 10, consistent with Figure~\ref{fig:dynamic_sim_results}.

First, we validate that the household-correlated pools are more closely connected than the community-correlated pools, and community-correlated pools are more closely connected than naive pools. We quantify the closely-connectedness using a simple metric, namely the number of edges on the subgraph induced by members of a pool. 
In Figure~\ref{fig:hist_num_edges_per_pool}, we plot the distribution of the number of edges within a pool over all realized pools under each pooling method. The median is 6 for community-correlated pools and 12 for household-correlated pools. On the other hand, naive pools mostly have 1-2 edges. 
This stark difference among pooling methods implies that the possibility of a pool containing multiple positives would be significantly higher in correlated pools than in naive pools, especially under low prevalence.

Moreover, Figure~\ref{fig:hist_num_pools_per_household} presents the distribution of the number of pools each household is allocated to. The majority of households are allocated to only one pool.

Therefore, the evidence presented in Figure~\ref{fig:pooling_validation} validates our pooling implementation using node embedding and clustering.

 \begin{figure}[!htbp]
     \FIGURE
     {\begin{subfigure}[b]{0.45\textwidth}
         \centering
         \includegraphics[width=\textwidth]{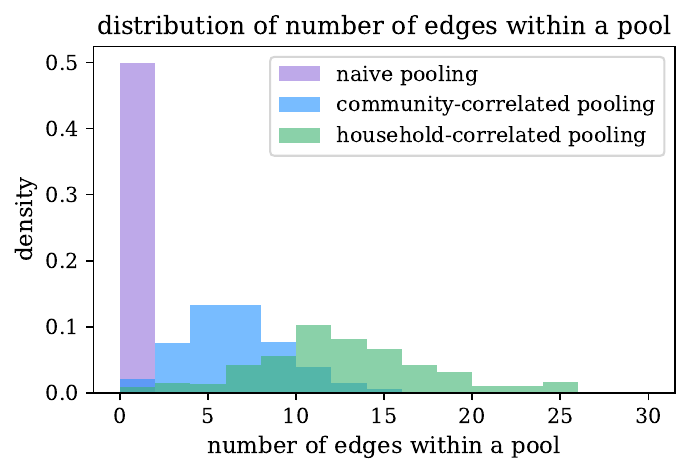}
         \caption{}
         \label{fig:hist_num_edges_per_pool}
     \end{subfigure}
     \begin{subfigure}[b]{0.45\textwidth}
         \centering
         \includegraphics[width=\textwidth]{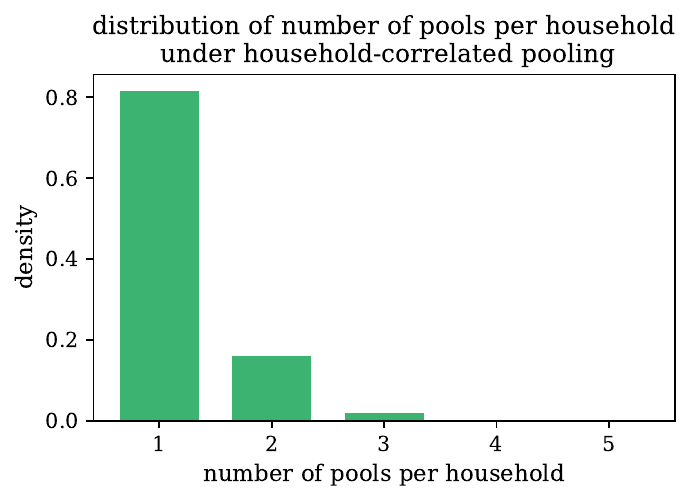}
         \caption{}
         \label{fig:hist_num_pools_per_household}
     \end{subfigure}}
     {Validation of the pooling implementation. \label{fig:pooling_validation}}
     {(a) Distribution of the number of edges within a pool under each pooling method. A larger number implies that the members of the pool are more closely connected. (b) Distribution of the number of pools that each household is allocated to under household-correlated pooling. The majority of households are placed into the same pool.}
 \end{figure}

\subsection{Realistic Viral Load Progression}\label{appdx:VL_progression}

We follow \citeAP{brault2021group} and model the viral load of an infected individual as a piecewise log-linear function.
A similar pattern has been discussed in other studies, such as \citeAP{cleary2021using}. 

In particular, we assume the log10 viral load rises, reaches a plateau value of 6, drops, remains at 3 for a while, then drops to -1. We further assume that the individual is infectious whenever their log10 viral load is at least 3. 

For each infected individual, assuming their infection starts at time 0, the viral load progression is parameterized by the following critical time points:
\begin{itemize}
    \item $t_1$, the time at which the log10 viral load reaches 3;
    \item $t_2$, the time at which the log10 viral load reaches 6;
    \item $t_3$, the time at which the log10 viral load starts declining from 6;
    \item $t_4$, the time at which the log10 viral load reaches 3;
    \item $t_5$, the time at which the log10 viral load drops to $-1$.
\end{itemize}

Figure~\ref{fig:sample_VL_curve_annotated} shows an example progression of log10 viral load. 
We set $t_1=1$ for all infected individuals. 
To create heterogeneity, we sample the duration of each subsequent piece uniformly from an interval, specified in Table~\ref{tab:VL_progression_params}.

\begin{table}[!htbp]
\centering
\TABLE{Parameter values for viral load progression. Unif[$\cdot,\cdot$] denotes a continuous uniform distribution.\label{tab:VL_progression_params}}
{\addtolength{\tabcolsep}{30pt}
\begin{tabular}{cc}
\toprule
 & Sample range \\ \midrule
$t_1$ & 1 \\ 
$t_2-t_1$ & $\text{Unif}[3,5]$ \\ 
$t_3-t_2$ & $\text{Unif}[1, 3]$ \\ 
$t_4-t_3$ & $\text{Unif}[7, 10]$ \\ 
$t_5-t_4$ & $\text{Unif}[5, 6]$ \\ \bottomrule
\end{tabular}
\addtolength{\tabcolsep}{-30pt}}
{}
\end{table}

Among the initial infections at the start of the simulation, we let half of them be at the start of infectivity (i.e., at $t_1$) and the other half to be at the start of the peak (i.e., at $t_2$). 

 \begin{figure}
     \FIGURE
     {\includegraphics[width=0.6\textwidth]{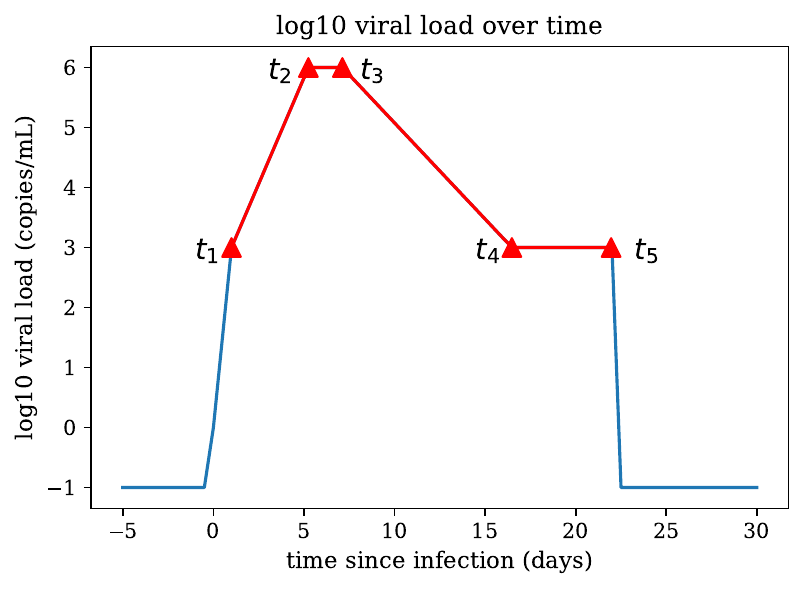}}
     {Example log10 viral load progression for an infected individual. \label{fig:sample_VL_curve_annotated}}
     {The critical time points are marked and annotated. The individual is assumed infectious when their log10 viral load is at least 3 (red).}
 \end{figure}

\subsection{PCR Modeling}
\label{appdx:PCR}

We describe a realistic sensitivity function for the PCR test that captures the randomness in the subsampling and pooling processes, an aspect overlooked by most existing literature studying group testing protocols. 

The first step in a pooled PCR test is the collection of samples from each subject. For SARS-CoV-2 testing, the most common sample types include nasopharyngeal swabs, anterior nares swabs, and saliva. We assume the raw volume of the samples is the same across all subjects, denoted by $V_{sample}$. (Nasopharyngeal and anterior nares swabs can be transported in a fixed amount of viral transport media; saliva samples, whether self-collected or not, can require a prescribed volume.) 

Once the $n$ samples are collected, they are transported to the lab to be prepared for pooling.
Let $V_i$ denote the viral load (i.e., the number of viral RNA copies per unit volume) of the $i$th sample in the pool. If the $i$th sample is negative, then $V_i= 0$. A pipetting robot fetches a volume of ${V_{subsample}}$ from each sample for pooling, so the number of RNA copies selected for pooling is $N_i\sim \text{Binom}\left(V_{sample} \cdot V_i, \frac{V_{subsample}}{V_{sample}}\right)$ for the $i$th sample. 
Compared to an individual test, pooling reduces the subsampling volume by a multiplicative factor of $n$. (That is, the $n$ subsamples, when pooled together, have the same volume as an individual test in the same step.) 
Then, all $n$ subsamples are pooled together and go through an RNA extraction step using glass fiber plates. 
Assuming that each RNA copy attaches to the glass fiber plates independently with probability $\xi$, the number of eluted RNA copies used as templates that enter the PCR machine follows a binomial distribution $M\sim \text{Binom}\left(\sum_{i=1}^{n} N_i, \xi \right)$. Aggregating the binomial subsampling in these steps, we find that $M$ follows a binomial distribution: $M\sim \text{Binom}\left(V_{sample}\cdot\sum_{i=1}^{n}V_i, \frac{V_{subsample}}{V_{sample}}\cdot \xi\right)$.\footnote{The proof of this relation is straightforward, based on two identities: (i) If $X_i\sim \text{Binom}(n_i, p)$ are independent, then $\sum_{i}X_i\sim \text{Binom}(\sum_{i}n_i, p)$; (ii) If $X\sim\text{Binom}(n,p)$ and $Y\mid X\sim\text{Binom}(X,q)$, then $Y\sim\text{Binom}(n, pq)$.}
Finally, we assume the PCR test has a detection threshold $\tau$, a positive integer, such that if $M\geq \tau$, the test returns a positive result; otherwise, negative.\footnote{The detection threshold $\tau$ is not to be confused with the limit of detection (LoD), i.e., the lowest concentration of the target (in copies per volume) that a PCR assay can detect at least 95\% of the time \citepAP{burns2008modelling}. In our model, a higher $\tau$ corresponds to a higher LoD. The way we model the subsampling steps using binomial random variables captures the randomness associated with the definition of LoD.} (As a result, a negative sample is always classified as negative.)  

\begin{table}[h]
\TABLE
{Parameter values used in the realistic PCR model.\label{tab:param}}
{\begin{tabular}{ccc}
\toprule
 Parameter name& Symbol & Parameter value\\ \midrule
 Sample volume & $V_{sample}$&  1 mL\\ 
 Subsample volume & $V_{subsample}$ &  100/pool size (pooled);  100 (individual) \muL \\ 
 Glass fiber binding efficiency & $\xi$ & 0.5  \\ 
Detection threshold & $\tau$ & calibrated to population-average individual test FNR \\ \bottomrule
\end{tabular}}{}
\end{table}

This PCR model enables us to simulate the test outcome given the sample viral loads in a pooled test. Table~\ref{tab:param} gives the parameter values we use in simulation. Among them, the detection threshold $\tau$ is a key quantity that affects the test outcome. Since it varies for different approved assays \citepAP{FDA_LoD}, we choose to not set a single value for it. Instead, we utilize its correspondence with the false negative rate (FNR) of a PCR test: a higher detection threshold leads to a higher false negative rate when testing the same sample, and vice versa. 
In particular, while keeping the other parameters in Table~\ref{tab:param} fixed, we use simulation to calibrate $\tau$ to different values of \textit{population-average individual test FNR} $\bar\beta$, i.e., the expected false negative probability of a PCR test on an individual positive sample whose viral load follows the viral load distribution in the population.
We use a viral load distribution calibrated from a large real-world dataset of infected individuals from \citetAP{brault2021group} (see Table~\ref{tab: GMM_VL} in Section~\ref{sec:VL_distribution}). 
Table~\ref{tab:LoD} describes the calibrated values of $\tau$ corresponding to $\bar{\beta}$ values of 2.5\%, 5\%, 10\% and 20\%. 
We use $\tau=1240$ in our simulation.

\begin{table}[!htbp]
\TABLE
{Population-average individual test FNRs $\bar\beta$ and their corresponding calibrated values of $\tau$ in the PCR model.\label{tab:LoD}}
{\addtolength{\tabcolsep}{25pt}
\begin{tabular}{cc}\toprule
        \\[-1em]
        $\bar{\beta}$ & Calibrated value of $\tau$ \\\midrule
        2.5\% & 108\\
        5\% & 174\\
        10\% & 342\\
        20\% & 1240\\ \bottomrule
    \end{tabular}
    \addtolength{\tabcolsep}{-25pt}}
{} 
\end{table}

\subsection{Analysis of Simulation Dynamics}\label{appdx:sim_dynamics}
Figure~\ref{fig:dynamic_sim_results} shows the projected epidemic progression under a representative policy of screening every five days with a pool size of ten.
We focus on two primary performance metrics, namely the cumulative number of infections and cumulative test consumption, as well as additional metrics studied in Section~\ref{sec:theory}, including the sensitivity $1-\beta$, effective efficiency $\gamma$, and the effective follow-up efficiency $\eta$ (defined in Appendix~\ref{appdx:proof_of_thm_2}).

The mean number of positives in positive pools reflects the distribution of positive cases in positive pools, with higher values indicating better pooled testing performance. 
Daily sensitivity positively correlates with the mean number of positives in positive pools. We observe an initial peak in daily sensitivity because the initial conditions of the simulation assume that early infectious cases have medium to high viral loads. In contrast, later in the time period simulated, sensitivity becomes lower because the prevalence is lower and because many of the positive cases in screening are early in their infection and so have low viral loads. 

Daily effective efficiency also drops over time due to the decreasing prevalence which reduces the proportion of positive pools. In contrast, daily effective follow-up efficiency remains relatively flat because it measures positive cases identified per follow-up test, less impacted by the overall prevalence.

Nonetheless, even lower-sensitivity tests do have a role in screening. As Figure~\ref{fig:dynamic_sim_results} shows, even with a sensitivity of roughly 40\%, the screening strategy is able to dramatically reduce the number of active infections from a peak of 400 to 100 at the end of the time horizon. 

\subsection{Necessity of an Accurate Test Error Model}\label{appdx:necessity_of_dilution}
In Section~\ref{subsec:corr_as_modeling_choice}, we demonstrate that modeling concentration-dependent test errors is important for accurately understanding the benefit offered by within-pool correlation. 
Here we further argue that modeling the dilution effect is also crucial. 
We consider an alternative test error model that depends on the viral loads but does not model the dilution effect, i.e., $p(h(\BFv)) = p(\sum_{i=1}^{n}v_i)$.
We show that this test error model, similar to the ones that assume a fixed sensitivity, also understates the benefits of correlation, compared to realistically modeling the viral loads and the dilution effect.

Figure~\ref{fig:heatmap_diffs_dil_nodil} shows the difference in cumulative infections and test consumption given by naive and community-correlated pooling under the two viral-load-dependent test error models that do and do not model dilution.
The model not accounting for the dilution effect drastically underestimates the difference in cumulative infections between naive and correlated pooling. It also obtains biased estimates for the difference in test consumption. 

Therefore, the results in both Figure~\ref{fig:heatmap_infections_tests_new} and Figure~\ref{fig:heatmap_diffs_dil_nodil} demonstrate that modeling viral loads and modeling the dilution effect are both very important for accurately quantifying the benefit of correlated pooling and making informed decisions for SARS-CoV-2 screening. This insight applies to epidemic control in general. We provide more discussion in Section~\ref{sec:discussion}.

 \begin{figure}
     \FIGURE
     {\includegraphics[width=0.99\textwidth]{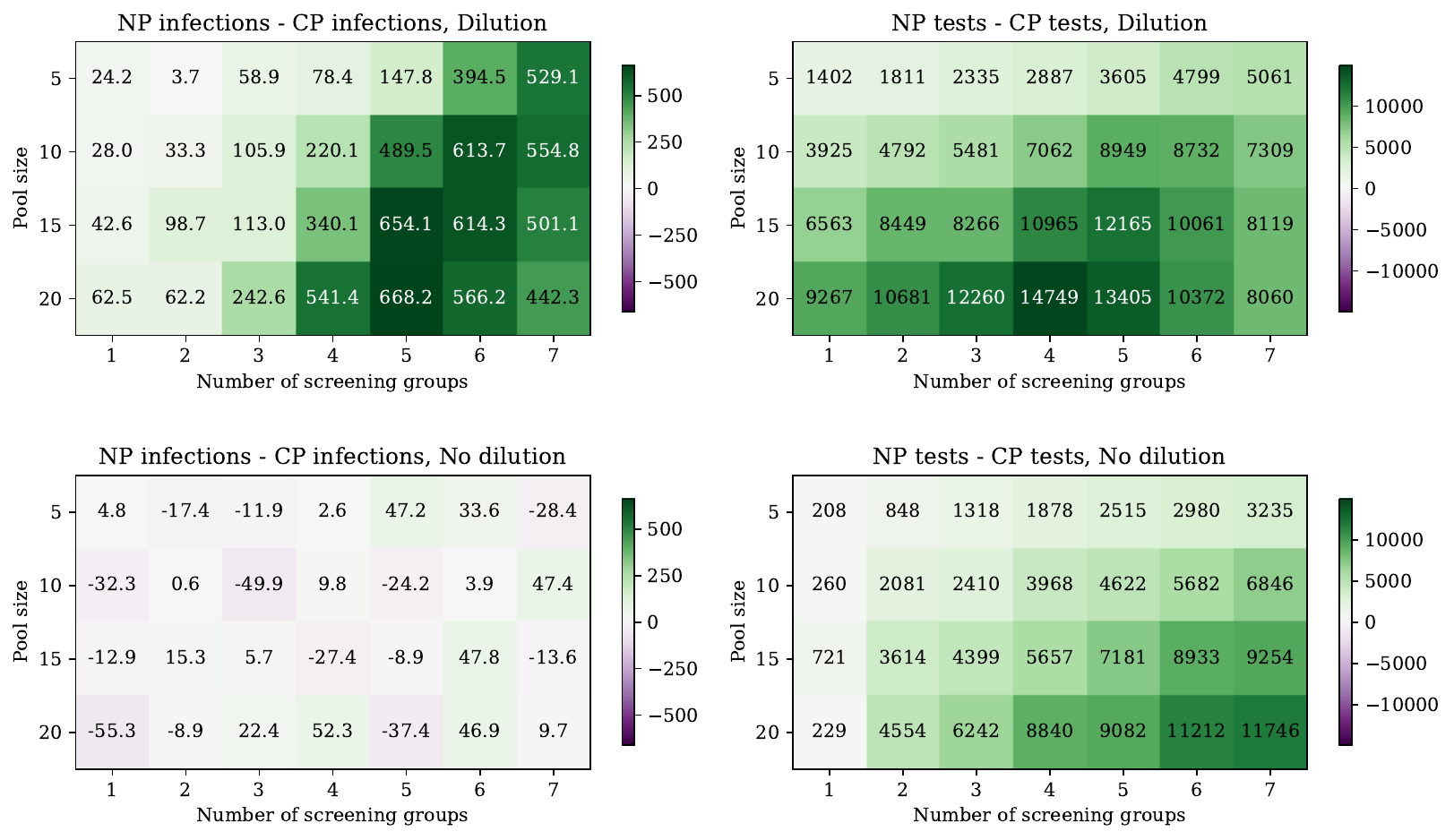}}
     {Difference in cumulative infections and test consumption between naive and community-correlated pooling for concentration-dependent test error models that do (top) and do not (bottom) account for dilution. \label{fig:heatmap_diffs_dil_nodil}}
     {The text annotation of each cell reports the average difference across 200 replications.
    ``Dilution" refers to using $p(h(\BFv)) = p(\sum_{i=1}^{n}v_i)$. 
     ``No dilution" refers to using $p(h(\BFv)) = p(\sum_{i=1}^{n}v_i)$.
    }
 \end{figure}

\section{Static Simulation}\label{appdix:static_sim}

In addition to our dynamic simulation, we thoroughly study how different factors (prevalence, pool size, household size distribution, PCR test sensitivity, and strength of correlation) affect the test performance of \NP and \CP in more controlled settings.
We call these the ``static simulation". 
The results from the static simulation offer important insights into decision-making similar to those from the dynamic simulation. 

We do not explicitly model community-correlated pooling here.
Instead, we assume the within-pool correlation arises only due to transmission within households and we tune a parameter governing the strength of household transmission to vary the within-pool correlation. 
We model dilution by a factor of pool size $n$ in the pooled tests, i.e., $h(\BFv) = \frac{1}{n}\sum_{i=1}^{n}v_i = \bar{v}_n$.

We demonstrate that \CP consistently outperforms \NP in terms of both sensitivity and efficiency. 
Based on an SIR model \citepAP{kermack1927contribution} that incorporates repeated large-scale screening, we show that \CP can stabilize or decrease the number of active infections using fewer tests than \NP.

\subsection{Viral Load Distribution}\label{sec:VL_distribution}

We use the viral load distribution calibrated on a large collection of infected individuals in \citetAP{brault2021group}.
We acknowledge that this distribution is different from the one induced by viral load progression and epidemic dynamics in our dynamic simulation. We opt to use it here because it is well-specified. 

We first specify a probability distribution governing viral loads across infected individuals.
One way to quantify the viral load in a sample is with the so-called Ct value. A PCR test amplifies the viral RNA copies in a sample by approximately doubling them in each cycle of the reaction. The minimum number of cycles required for the RNA copies to reach a detectable threshold is called the \textit{cycle threshold}, denoted Ct \citepAP{heid1996real}. The lower the initial viral load in the sample, the more duplicating cycles it requires to become detectable, and the larger its Ct value is. 

\citeAP{jones2020analysis} obtains empirically measured Ct values from asymptomatic screening conducted in Germany. \citeAP{brault2021group} fits a censored Gaussian mixture model (GMM) to the distribution of Ct values in \citeAP{jones2020analysis}:
\begin{equation}f(x)=\sum_{k=1}^3 \pi_k \frac{f_{\mu_k,\sigma_k}(x)}{F_{\mu_k,\sigma_k}(d_{cens})}\cdot \mathbbm{1}\{x\leq d_{cens}\}.\label{eq:cgmm}
\end{equation}
In Equation~\ref{eq:cgmm}, $f_{\mu_k,\sigma_k}$ and $F_{\mu_k,\sigma_k}$ denote the probability density function and cumulative density function of the $k^{th}$ component with mean $\mu_k$ and standard deviation $\sigma_k$, respectively. The censoring threshold $d_{cens}$ represents the limit of detection of the PCR assay, such that a sample with Ct value exceeding it is not observed. \citeAP{brault2021group} obtains  $d_{cens}=35.6$ and GMM parameter values in Table~\ref{tab: GMM_Ct}.
\begin{table}[h]
\TABLE{Gaussian mixture model parameters for the distribution of Ct values.\label{tab: GMM_Ct}}
{\addtolength{\tabcolsep}{15pt}  
\centering
\begin{tabular}{cccc}
\toprule
 & $\pi_k$ & $\mu_k$ & $\sigma_k$ \\ \midrule
$k=1$ & 0.33 & 20.13 & 3.60 \\ 
$k=2$ & 0.54 & 29.41 & 3.02 \\ 
$k=3$ & 0.13 & 34.81 & 1.31 \\ \bottomrule
\end{tabular}
\addtolength{\tabcolsep}{-15pt}  }
{\textit{Note}. Here, $\pi_k$, $\mu_k$, $\sigma_k$ are the weight, mean and standard deviation of the $k^{th}$ component, respectively.}
\end{table}

The associated \textit{uncensored} GMM model represents the true Ct distribution of the entire population, including those that may not be detected through individual PCR tests.

Moreover, since Ct value is a measurement of the viral load, and viral load is the quantity directly of interest to our simulation, we use a formula given in \citeAP{jones2020analysis} to convert this distribution to that of the $\log_{10}$ of viral load (copies/mL):\footnote{The data reported in \citeAP{jones2020analysis} are based on two PCR assays, the cobas system and the LC480 system, each of which has a conversion formula between Ct and viral load. Since over 60\% of the positives in their screened population were identified with the cobas system and the two conversion formulae are approximately the same, we use the formula for the cobas system here.}
\begin{align*}
    \log_{10}VL&=\log_{10}(1.105\cdot 10^{14}\cdot e^{-0.681C_t})\\
    &=(14+\log_{10}1.105) - \frac{0.681}{\ln{10}}Ct.
\end{align*}
This results in a GMM on the $\log_{10}$ of the viral load with parameters shown in Table~\ref{tab: GMM_VL}. A normally distributed mixture component on the Ct value is equivalent to a normally distributed mixture component with a different mean and variance on the $\log_{10}$ viral load. 

\begin{table}[h]
\TABLE
{Gaussian mixture model parameters for the distribution of $\log_{10}$ viral load (copies/mL) among infected individuals. \label{tab: GMM_VL}}
{\addtolength{\tabcolsep}{20pt}  
\begin{tabular}{cccc}
\toprule
 & $\pi_k$ & $\mu_k$ & $\sigma_k$ \\ \midrule
$k=1$ & 0.33 & 8.09 & 1.06 \\ 
$k=2$ & 0.54 & 5.35 & 0.89 \\
$k=3$ & 0.13 & 3.75 & 0.39 \\ \bottomrule
\end{tabular}
\addtolength{\tabcolsep}{-20pt}  }
{\textit{Note}: Here, $\pi_k$, $\mu_k$, $\sigma_k$ are the weight, mean and standard deviation of the $k^{th}$ component, respectively.}
\end{table}

In our simulation, we assume the viral load of any individual is independent of the viral loads of all other individuals given their infection status, stemming from heterogeneity in the individual biological response to the virus. Hence, for each infected individual, we can sample their viral load from the distribution specified in Table~\ref{tab: GMM_VL}. 

\subsection{Household Size Distribution}

Tables~\ref{tab:house} and \ref{tab:house_var} describe the household size distribution of four different countries from census data and variants of the U.S. household size distribution.
\begin{table}[h]
\TABLE
{Household size distribution of the U.S., China, Australia, and France.\label{tab:house}}
{\addtolength{\tabcolsep}{4pt}
\begin{tabular}{lllllll}
\toprule
 & 1 & 2 & 3 & 4 & 5 & 6+ \\ \midrule
United States (\texttt{US}) &  0.284 & 0.345 & 0.151 & 0.127 & 0.058 & 0.035 \\ 
China (\texttt{CN}) & 0.156 & 0.272 & 0.247 & 0.171 & 0.089 & 0.065 \\ 
Australia (\texttt{AUS}) & 0.244 & 0.334 & 0.162 & 0.159 & 0.067 & 0.034 \\ 
France (\texttt{FR}) & 0.364 & 0.327 & 0.136 & 0.115 & 0.042 & 0.016 \\\bottomrule
\end{tabular}
\addtolength{\tabcolsep}{-4pt}
}
{\textit{Source}: U.S. \protect\citepAP{US_household_sizes}, China \protect\citepAP{china_household_sizes}, Australia \protect\citepAP{australia_household_sizes}, and France \protect\citepAP{france_household_sizes}.}
\end{table}
    
\begin{table}[h]
\TABLE
{Household size distribution variants based on U.S. data. \label{tab:house_var}}
{\addtolength{\tabcolsep}{4pt}
\begin{tabular}{lllllll}
\toprule
 Household size & 1 & 2 & 3 & 4 & 5 & 6+ \\ \midrule
\texttt{US+1} &  0.209 & 0.36 & 0.166 & 0.142& 0.073& 0.05 \\ 
\texttt{US+2} & 0.134 & 0.375 & 0.181 & 0.157 & 0.088 & 0.065\\ 
\texttt{US-1} &  0.359 & 0.33 & 0.136 & 0.112& 0.043 & 0.020\\ 
\texttt{US-2} & 0.434 & 0.315 & 0.121 & 0.097 & 0.028 & 0.005\\ \bottomrule
\end{tabular}
\addtolength{\tabcolsep}{-4pt}  
}
{\textit{Note}. \texttt{US$\pm$1}, \texttt{US$\pm$2} are household distributions with weights $\pm 0.075$, $\pm 0.15$ respectively uniformly allocated to household sizes $>1$ from the weight of household size 1. For example, \texttt{US+1} has weight $0.284-0.075$ on households of size 1, weight $0.345+0.075/5$ on households of size 2, weight $0.151+0.075/5$ on households of size 3, etc. }
\end{table}

\subsection{Experiment Setup}
\subsubsection{Correlated Infections in Households}
We model the population as consisting of households with size $H$ ranging from one to six (since households of size larger than six are rare). We gather the household size distributions of four countries from census data and assume that all probability mass on $H>6$ is allocated to $H=6$ (Table~\ref{tab:house}). We also explore variants of the U.S. census data, in which we either add to or subtract from the weight on household size of one and adjust the weights on other household sizes accordingly (Table~\ref{tab:house_var}). 

A household is said to be infected if one person is infected as the index case in the household. We assume different households are infected independently with probability $p_h$, i.e., correlation through other social groups is considered negligible. Within each infected household, we assume transmissions occur independently with an SAR of $q$. That is, given a positive index case in a size-$h$ household, the remaining $h-1$ members become infected independently with probability $q$. We consider the following possible values for $q$: [0.166, 0.140, 0.193, 0.005, 0.446]. These are the estimated mean, 95\% CI lower and upper bounds, minimum and maximum values of household SAR from 40 studies, respectively, reported by a meta-analysis \citepAP{madewell2020household}.

The distribution of household size $H$ and the choices of $p_h$ and $q$ together yield an expected prevalence in the population, which matches the overall population-level prevalence $\alpha$:
\begin{equation}
  p_h\cdot \EE_H[(1+(H-1)q)] = \alpha. \label{eq:experiments_prevalence_equation}
\end{equation}

We now describe the steps for simulating correlated infections within households, given a fixed population-level prevalence, SAR, and household size distribution:
\begin{enumerate}
    \item Compute the household infection probability $p_h$ using Equation~\ref{eq:experiments_prevalence_equation}.
    \item Generate households with sizes drawn from the household size distribution.
    \item Let each household be infected independently with probability $p_h$, with one member selected uniformly at random as the index case.
    \item In each infected household, generate secondary infections.
    \item Assign to each infection a viral load sampled from the distribution described in Table~\ref{tab: GMM_VL}.
\end{enumerate}

\subsubsection{Pooling Assignment}

Having developed a model for correlated infections in households, we now describe how we allocate samples into pools when using \NP and \CP, under the Dorfman procedure:
\begin{itemize}
\item Naive pooling: We perform an independent random permutation on all the individual samples from the population and place them sequentially into pools regardless of household membership.
\item Correlated pooling: We aim to place samples of individuals from the same household in the same pool. 
A collection of partially full pools is maintained and households are added sequentially.
To add a household, we look for the first unfinished, capacity-permitting pool and place all samples of the household into this pool. If this is infeasible, we split the household across two or more pools.
\end{itemize}

Per the Dorfman procedure, samples in the same pool undergo one pooled test. All individuals in the pools testing positive take follow-up tests. 
We assume the amount of sample collected from each individual is enough so that no re-sampling is required if the follow-up test is necessary. This implies that the viral loads in the subsamples used for the pooled test and follow-up test are equal. 
The subsample for the pooled test is smaller than that for an individual test by a factor of the pool size, which results in dilution in the pooled sample.

\subsection{Simulation Results}\label{sec:results}
We demonstrate the \textit{advantage} of \cp over \np through numerical results under different sets of parameters.
First, we pick a set of parameters as the baseline setting, shown in Table~\ref{tab:nominal}. We consider this as a representative setting for a medium-sized town in the early stage of an epidemic. The choice of pool size is informed by empirical implementations of group testing for COVID-19 \citepAP{wuhan2020, cornell2020, barak2021lessons}.
We set the detection threshold $\tau=174$, corresponding to a population-average individual test FNR of 5\% (Table~\ref{tab:LoD}). The test sensitivity function $p(v)$ is shown in Figure~\ref{fig:pcr_sensitivity_174}.\footnote{Here we use a different detection threshold $\tau$ than in the dynamic simulation. We would like our static simulation to approximate the stylized setting with high sensitivity while the dynamic simulation would allow more test errors. However, the same insights hold if $\tau$ is varied.}
In Section~\ref{appdx:sensitivity_analysis}, we vary these parameters to show that the advantage is robust. 
\begin{table}[!htbp]
    \TABLE
    {Baseline parameter values in the static simulation.\label{tab:nominal}}
    {\addtolength{\tabcolsep}{20pt}\begin{tabular}{cc}\toprule
        Parameter & Value \\\midrule
        Population-level prevalence & 1\%\\
        Pool size & 6\\
        SAR & 16.6\%\\
        Household distribution & \texttt{US}\\
        Population-average individual test FNR & 5\%\\
        Population size & 12000\\\bottomrule
    \end{tabular}\addtolength{\tabcolsep}{-20pt}}{}
\end{table}

 \begin{figure}
     \FIGURE
     {\includegraphics[width=0.5\textwidth]{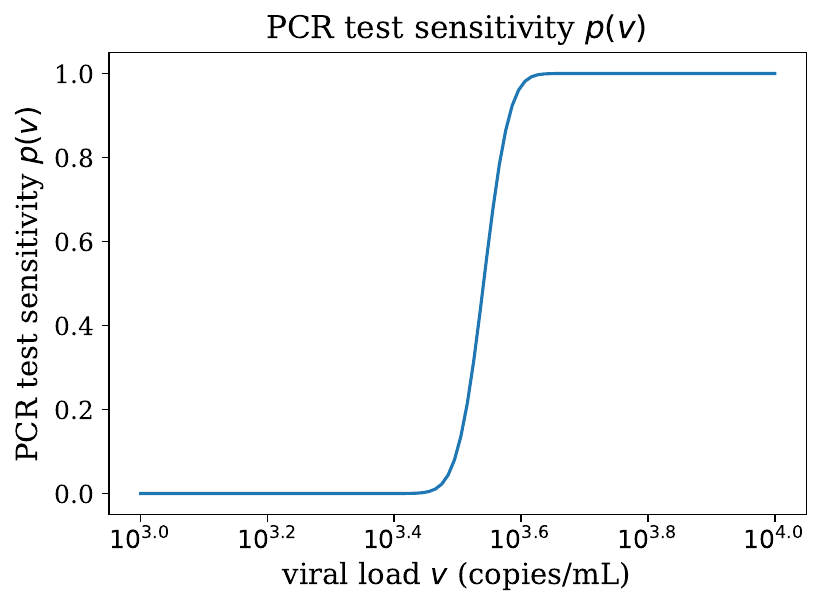}}
     {PCR test sensitivity $p(v)$ used in the static simulation. \label{fig:pcr_sensitivity_174}}
     {}
 \end{figure}

We focus on two metrics to evaluate the performance of a group testing protocol, namely sensitivity (i.e., $1-\mathrm{FNR}$) and efficiency, the number of individuals screened per test. Both are important for epidemic mitigation, as high sensitivity helps identify the positives accurately, while high efficiency permits more frequent screening under limited resources. Here we present efficiency as the metric for test consumption because it is most widely used. The performance in the metric $\gamma_{\texttt{POOL},\alpha}$ proposed in Section~\ref{subsec:metric_of_interest}, the number of positive cases identified per PCR test, can be inferred by taking the product of sensitivity and efficiency. 

The performance of \NP and \CP under the baseline setting over 2000 iterations is shown in Table~\ref{tab:nominal_performance}. As a reference, only using individual testing has a sensitivity of 95\% and an efficiency of 1. 
Correlated pooling has better performance in terms of both sensitivity and efficiency than \NP. This is because \CP in general has more positive cases in a positive-containing pool (due to correlation among samples from the same household). As a result, a sample with low viral load, which might otherwise be missed in \NP, is more likely to be ``rescued'' by other positive samples in the same pool in \CP, leading to higher sensitivity. (This is referred to as the ``hitchhiker effect" in \citetAP{barak2021lessons}.) Meanwhile, the clustering of more positive cases in the same pool also implies a smaller number of pools that contain positive samples and require follow-up tests, resulting in a higher efficiency of \CP. 
\begin{table}[!htbp]
\centering
\TABLE{Performance of naive and correlated pooling in the Dorfman procedure under the baseline parameter setting, averaging over 2000 iterations.  \label{tab:nominal_performance}}
{\addtolength{\tabcolsep}{25pt}
\begin{tabular}{ccc}
\toprule
Pooling method         & Sensitivity & Efficiency \\ \midrule
Naive pooling (\np)            & 81.9\%      & 4.67               \\
Correlated pooling (\cp)      & 86.0\%      & 4.83             \\
Percent advantage of \cp over \np  & 5.02\%        & 3.51\%               \\ \bottomrule
\end{tabular}
\addtolength{\tabcolsep}{-25pt}}{\textit{Note}. The standard errors for the sensitivity and efficiency are within 0.1\% and 0.01, respectively.}
\end{table}

Such improvement has a significant impact on real-world policymaking. We will show in Section~\ref{subsec:SIR} that, when pool size is optimized for both pooling methods separately, \CP enables more effective epidemic control than \NP. 

\subsubsection{Sensitivity Versus Efficiency Across Pool Sizes}
Under the same population-level prevalence, we anticipate test accuracy and efficiency will vary when we choose different pool sizes. Figure~\ref{fig:pareto} reveals the \textit{tradeoff} between sensitivity and efficiency using the two pooling methods under different prevalence levels. 
All parameters other than the prevalence level and the pool size take the values given in Table~\ref{tab:nominal}. In most scenarios (except when under high prevalence \textit{and} large pool size), \CP outperforms \NP in both sensitivity and efficiency. 

\begin{figure}[!htbp]
    \centering
    \caption{Tradeoff between sensitivity and efficiency of \cp and \np for different prevalence levels.}
    \label{fig:pareto}
    \begin{minipage}{0.3\textwidth}
        \centering
        \includegraphics[width=\textwidth]{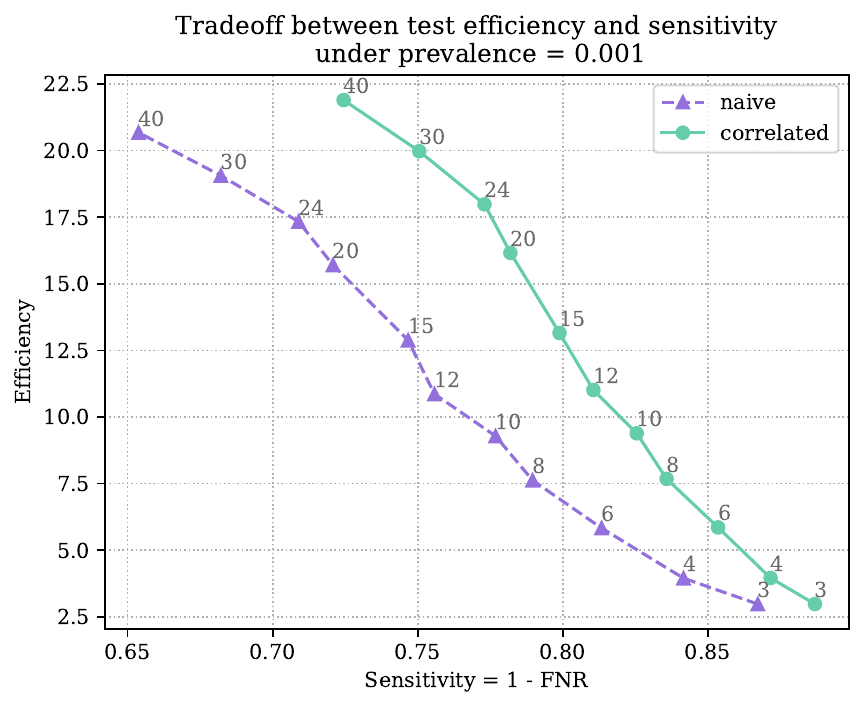}
        \subcaption{$\alpha = 0.1\%$}
        \label{fig:pareto_0.001}
    \end{minipage}
    \hspace{0.03\textwidth} 
    \begin{minipage}{0.3\textwidth}
        \centering
        \includegraphics[width=\textwidth]{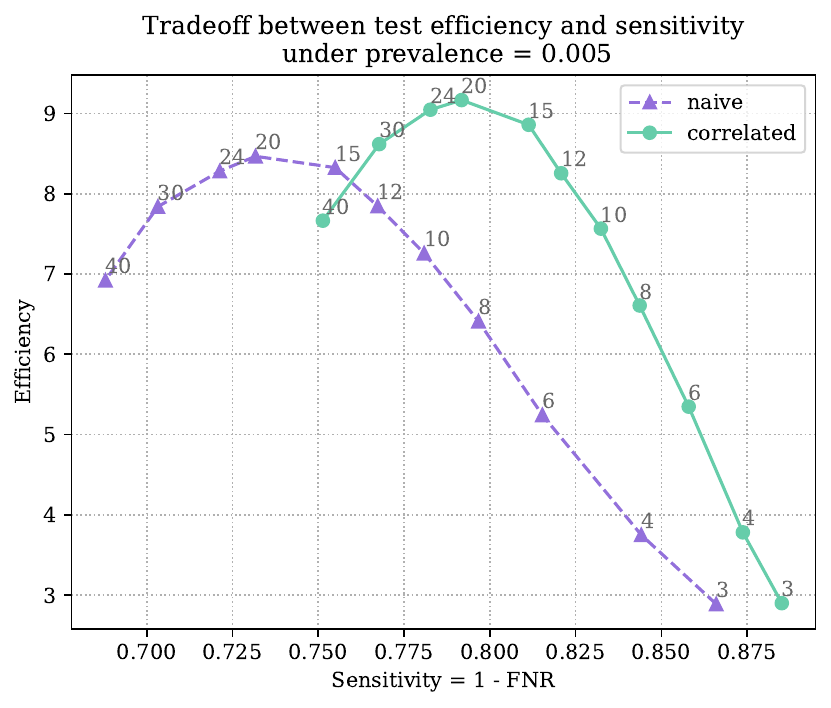}
        \subcaption{$\alpha = 0.5\%$}
        \label{fig:pareto_0.005}
    \end{minipage}
    \hspace{0.03\textwidth}
    \begin{minipage}{0.3\textwidth}
        \centering
        \includegraphics[width=\textwidth]{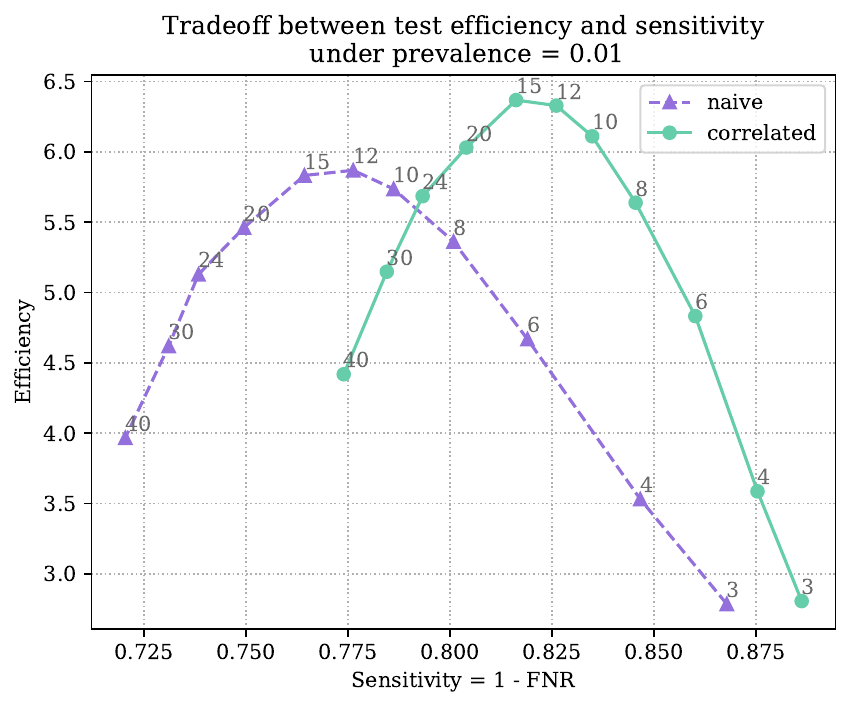}
        \subcaption{$\alpha = 1\%$}
        \label{fig:pareto_0.01}
    \end{minipage}
    \vspace{1em} 
    \begin{minipage}{0.3\textwidth}
        \centering
        \includegraphics[width=\textwidth]{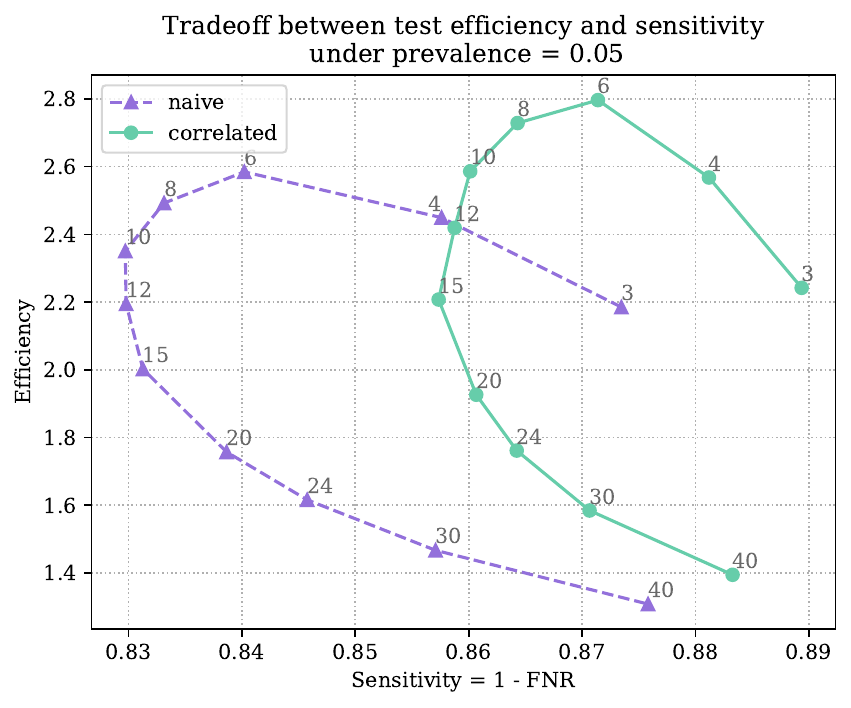}
        \subcaption{$\alpha = 5\%$}
        \label{fig:pareto_0.05}
    \end{minipage}
    \hspace{0.03\textwidth} 
    \begin{minipage}{0.3\textwidth}
        \centering
        \includegraphics[width=\textwidth]{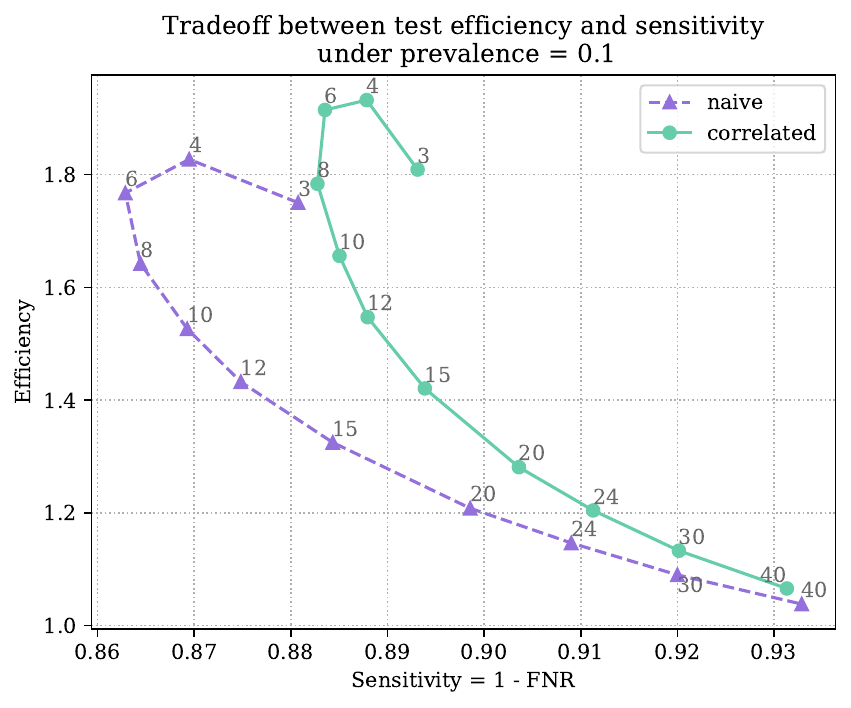}
        \subcaption{$\alpha = 10\%$}
        \label{fig:pareto_0.1}
    \end{minipage}
    \begin{flushleft}\footnotesize{\textit{Note.} As we prefer both higher sensitivity and higher efficiency, a point in the upper right corner of the plot is more preferable. Each point is obtained by taking the average outcome over 2000 replications using a pool size annotated next to the point.}\end{flushleft}
\end{figure}

When prevalence is low (e.g., 0.1\%, Figure~\ref{fig:pareto_0.001}), as pool size increases, sensitivity decreases and efficiency increases. Under low prevalence, most pools have either zero or one positive sample even when the pool size is large. A larger pool size causes a stronger dilution effect, lowering the pooled test sensitivity. Meanwhile, efficiency increases with pool size because fewer pools are needed, and under low prevalence, not many pools require follow-up tests even if they are large.

When prevalence is intermediate (e.g., 0.5\% or 1\%, Figures~\ref{fig:pareto_0.005} or \ref{fig:pareto_0.01}), as pool size increases, sensitivity decreases because of the dilution effect. Efficiency, however, reaches a peak first before declining. This is because a large pool size under intermediate prevalence results in many positive pools. The heightened demand for follow-up tests offsets the savings in the number of pooled tests.

When prevalence is high (e.g., 5\% or 10\%, Figures~\ref{fig:pareto_0.05} or \ref{fig:pareto_0.1}), as pool size increases, sensitivity first decreases and then increases. This is because a larger pool size under high prevalence leads to multiple positive samples in the same pool, offsetting the dilution effect. Efficiency drops dramatically as pool size increases since a majority of pools test positive and most samples require follow-up tests. The efficiency of large pools under 10\% prevalence, for example, is close to 1, indicating little reduction in test consumption compared to individual tests. In this scenario, one should consider using individual testing instead of group testing, as is also suggested in \citeAP{eberhardt2020multi}.

Figure~\ref{fig:exp2_heatmap} visualizes the advantage of \CP over \NP under different prevalence levels and pool sizes. Except when prevalence $\alpha= 10\%$, pool size $n=40$, \CP is more advantageous. The advantages in sensitivity and efficiency are both more significant under low prevalence and when the pool size is large. 

 \begin{figure}[!htbp]
     \FIGURE
     {\includegraphics[width=13cm]{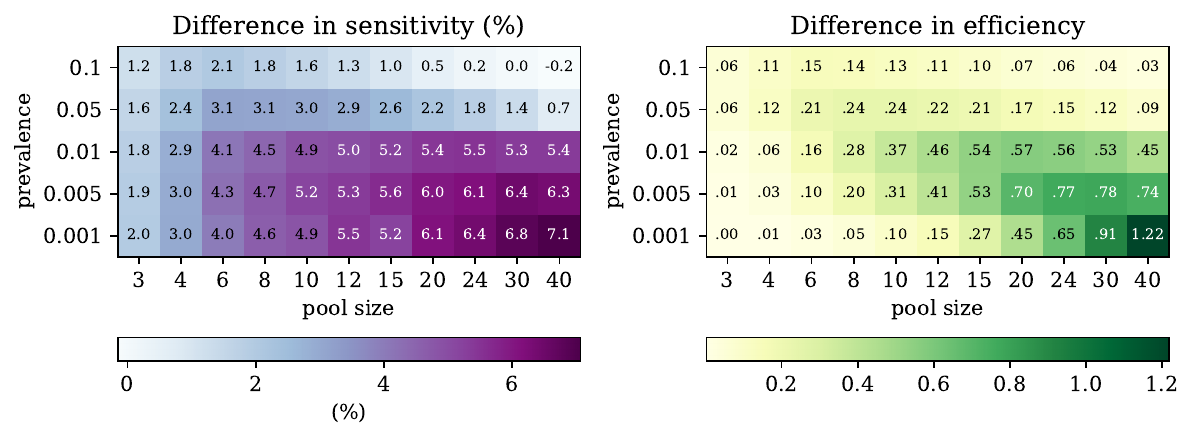}}
     {The advantage of \CP in (left) sensitivity and (right) efficiency, over \NP.\vspace{0.1in} \label{fig:exp2_heatmap}}
     {In both heatmaps, the value in the cell is the metric value of \CP minus that of \NP; a positive value implies that \CP is more advantageous.}
 \end{figure}

\subsubsection{Test Specificity}\label{subsec:specificity}

As discussed in Section~\ref{sec:intro}, false positives pose challenges to large-scale screening, including waste of public health and economic resources, disruption of personal lives, and increased exposure risk during unnecessary treatment. Though false positives are not explicitly included in our modeling, here we argue that they are not a significant concern if pooling is used. In particular, we demonstrate that group testing has substantially lower FPR than individual testing, and, moreover, \CP achieves a lower FPR than \NP. 

For our discussion, we start by assuming that false positives originate mainly from lab contamination that occurs independently across tests. We assume any PCR test on a negative sample has a small constant FPR (e.g., 0.01\% as reported in \citeAP{ontario2020}), much smaller than the probability that a typical positive-containing pool tests positive. Under these assumptions, the probability that a negative sample in an all-negative pool is declared positive is negligible (e.g., $10^{-8}$) compared to when it is in a positive-containing pool. Hence, we estimate the FPR of a testing protocol by the fraction of negative samples that receive individual tests, assuming they are all in positive-containing pools. This can be directly inferred from our simulation results.

First, we compute the fraction of samples in the population receiving individual tests using $\texttt{frac}_{\text{indiv}} = \text{efficiency}^{-1} - 1/n$. Second, we estimate $\texttt{frac}_{\text{pos, indiv}}$, the fraction of samples that are positive \textit{and} receive individual tests, using  ${\alpha\cdot \text{sensitivity}}.\footnote{Note that not all positives receiving individual tests test positive. Hence, this estimate is an underestimate, which eventually leads to an upper bound on FPR.}$ We take the difference of the above two quantities to estimate $\texttt{frac}_{\text{neg, indiv}}$, the fraction of samples that are negative \textit{and} receive individual tests. Multiplying this difference by 0.01\% then gives $\texttt{frac}_{\text{neg, indiv pos}}$, the fraction of samples that are negative \textit{and} test positive in individual tests. Finally, we divide the $\texttt{frac}_{\text{neg, indiv pos}}$ by $1-\alpha$, the fraction of samples that are negative, to obtain the estimate for FPR. 

We summarize the above calculations for \CP and \NP in Table~\ref{tab:FPR_calculation} based on the simulation results for the baseline setting in Table~\ref{tab:nominal_performance}. We see that both pooling methods achieve an FPR on the order of $10^{-6}$, with \CP slightly outperforming \NP. In our regime of discussion, the FPR roughly scales linearly with pool size and prevalence. Hence, for a prevalence of up to 1\% and a pool size of up to 20, we expect the FPR of either pooling method to be at least as good as $10^{-5}$. This is a ten-fold reduction from the FPR of individual testing. Such specificity is sufficiently high in many uses of repeated screening for infection control. 

\begin{table}[!htbp]
\centering
\TABLE{FPR estimates for naive and correlated pooling under the baseline setting.  \label{tab:FPR_calculation}}
{\addtolength{\tabcolsep}{16pt}
\begin{tabular}{ccc}
\toprule
Quantity                & Correlated pooling & Naive pooling \\ \midrule
$\texttt{frac}_{\text{indiv}}$          & 4.03\%                  & 4.75\%             \\
$\texttt{frac}_{\text{pos, indiv}}$     & 0.86\%                  & 0.82\%             \\
$\texttt{frac}_{\text{neg, indiv}}$     & 3.17\%                  & 3.93\%             \\
$\texttt{frac}_{\text{neg, indiv pos}}$ & 3.17E-6                 & 3.93E-6            \\
FPR estimate            & 3.20E-6                 & 3.97E-6            \\ \bottomrule
\end{tabular}
\addtolength{\tabcolsep}{-16pt}}
{}
\end{table}

We also argue that false positives from PCR tests have little impact on efficiency, i.e., they incur only a small number of extra tests. In the pooled stage, 0.01\% of the all-negative pools are expected to test positive and require follow-up tests for their samples. As the number of samples in all-negative pools is upper bounded by $N$, the extra tests due to PCR false positives translate to a less than $10^{-4}$ increment in the number of tests per person. Besides, sensitivity is not affected by false positives of PCR tests.

\subsection{Robustness Analysis}\label{appdx:sensitivity_analysis}

We demonstrate that the advantage of \CP over \NP is robust across different parameter values.
In Figure~\ref{fig:sens}, we show the performance of naive and \CP when varying the population-level prevalence, pool size, population-average individual test FNR, SAR, and household size distribution respectively, while keeping others at the baseline setting. 
Each bar/point in the plots is obtained by taking the average outcome over 2000 replications. In all plots, \CP consistently performs better than \NP in terms of both sensitivity and efficiency.
\begin{figure}[p]
    \centering
    \caption{Sensitivity and efficiency for varying (a) prevalence, (b) pool size, (c) population-average individual test FNR, (d) SAR, and (e) household size distribution, under \CP and \NP.}
    \label{fig:sens}
    \begin{minipage}{0.42\textwidth}
        \centering
        \includegraphics[width=\textwidth]{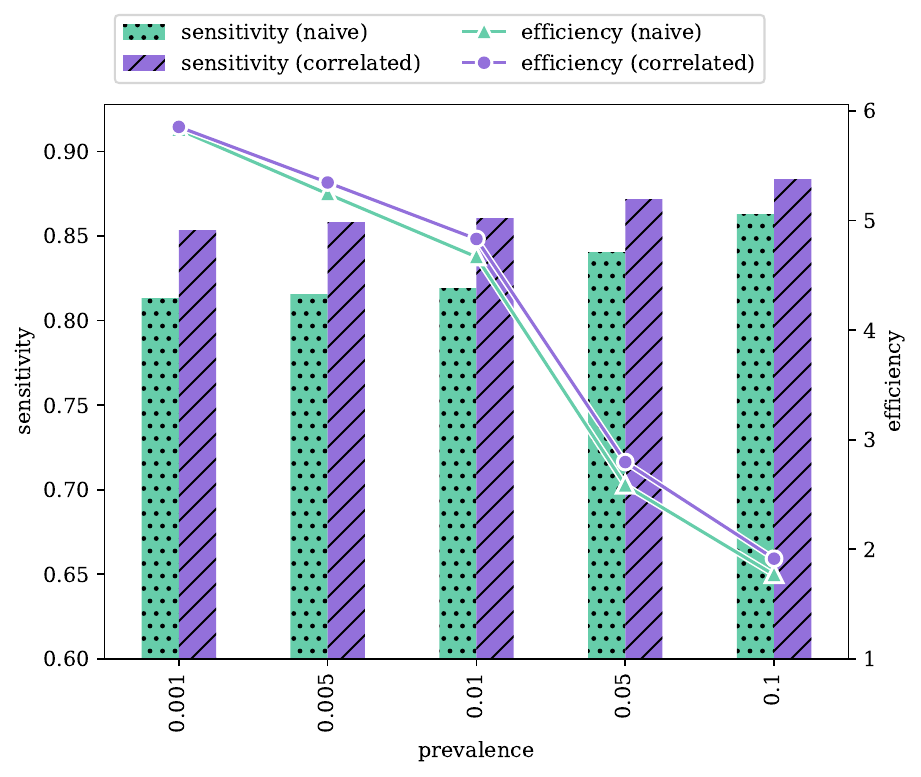}
        \subcaption{Prevalence} 
        \label{fig:sens_prev}
    \end{minipage}
    \hspace{0.03\textwidth} 
    \begin{minipage}{0.42\textwidth}
        \centering
        \includegraphics[width=\textwidth]{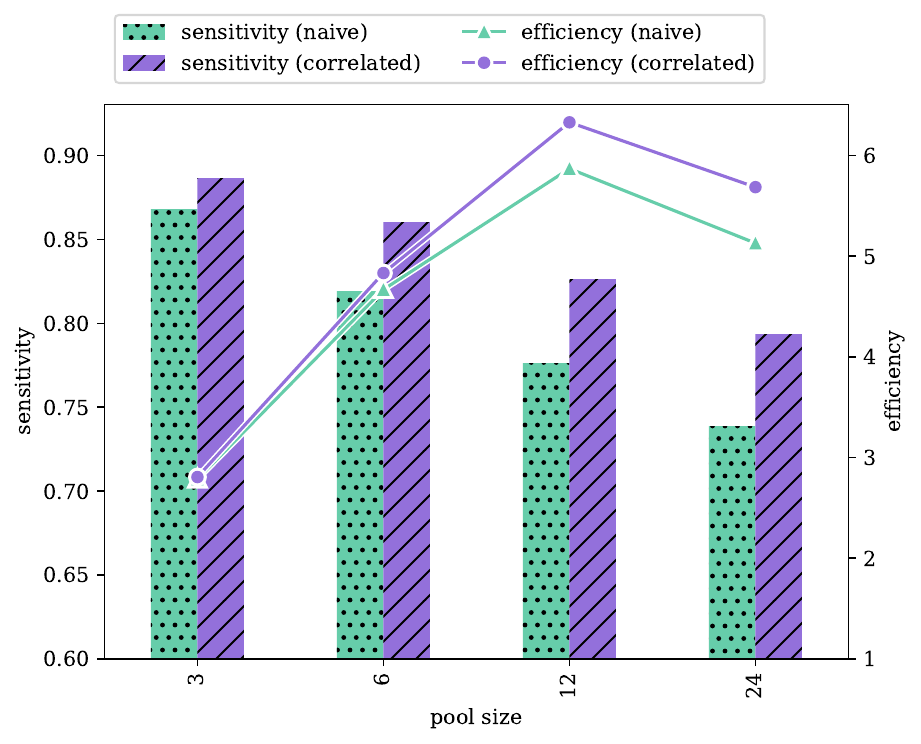}
        \subcaption{Pool size} 
        \label{fig:sens_pool_size}
    \end{minipage}
    
    \vspace{1em} 
    \begin{minipage}{0.42\textwidth}
        \centering
        \includegraphics[width=\textwidth]{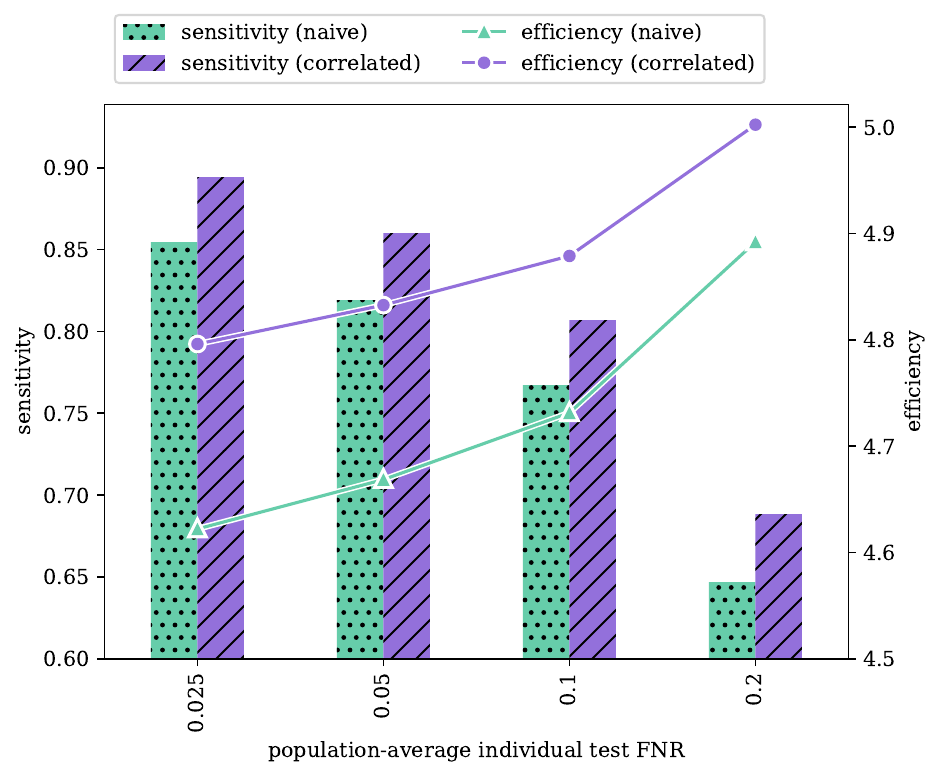}
        \subcaption{\small{Population-average individual test FNR}} 
        \label{fig:sens_fnr}
    \end{minipage}
    \hspace{0.03\textwidth} 
    \begin{minipage}{0.42\textwidth}
        \centering
        \includegraphics[width=\textwidth]{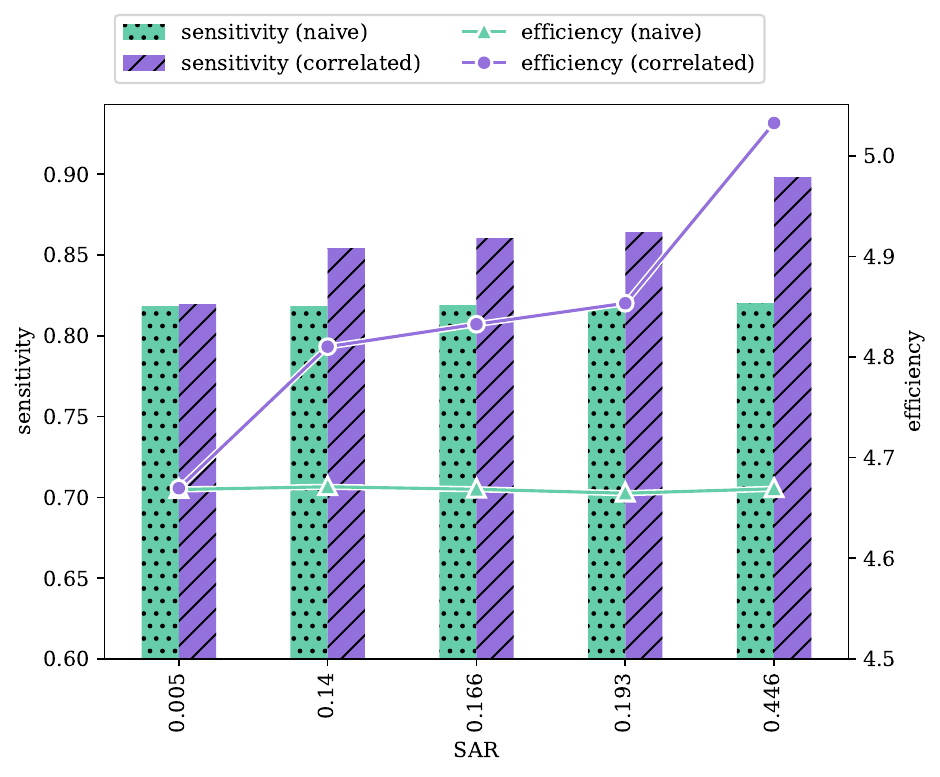}
        \subcaption{SAR} 
        \label{fig:sens_sar}
    \end{minipage}
    \vspace{1em} 
        \begin{minipage}{0.42\textwidth}
        \centering
        \includegraphics[width=\textwidth]{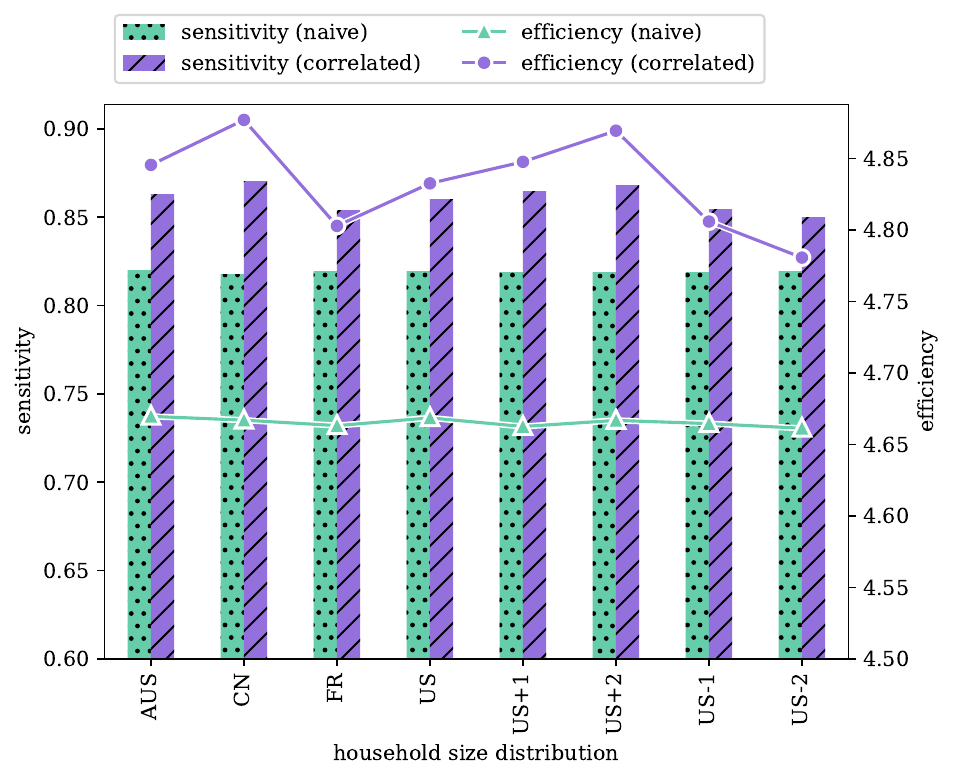}
        \subcaption{Household distribution} 
        \label{fig:sens_household}
    \end{minipage}
    \begin{flushleft}
        \footnotesize{\textit{Note.} Bars and points are obtained by taking the average outcome over 2000 replications.}
    \end{flushleft}
\end{figure}

Figure~\ref{fig:sens_prev} shows that smaller prevalence leads to lower sensitivity but higher efficiency. This is due to the existence of fewer positive samples in a positive pool, which results in larger FNR because of the dilution effect. Smaller prevalence also implies fewer positive pools, leading to fewer follow-up tests and therefore higher overall efficiency.

Figure~\ref{fig:sens_pool_size} shows that a larger pool size typically implies a stronger dilution effect, which causes sensitivity to decline. 
Efficiency increases with pool size initially because for smaller pools the number of pooled tests is the dominating factor in determining the efficiency. On the other hand, a larger pool (e.g., size of 24) is more likely to contain a positive, which requires more individual tests once the pool tests positive. This causes the efficiency to decline for larger pools.

In Figure~\ref{fig:sens_fnr}, sensitivity decreases and efficiency increases as the population-average individual test FNR, $\bar\beta$, rises. A higher $\bar\beta$ also implies a higher FNR of the pooled test, which explains the drop in sensitivity. Efficiency increases because a higher detection threshold causes more cases to be missed by the pooled tests and therefore fewer follow-up tests are required. 

Figure~\ref{fig:sens_sar} shows that the change in SAR does not affect the performance of \NP, as the protocol does not benefit from the correlation structure in the population. Meanwhile, correlated pooling achieves a better sensitivity and efficiency under larger SAR values. This is because a larger SAR creates a stronger correlation among household members, causing positive samples to be clustered in fewer pools. This in turn raises the probability of detecting positive pools and simultaneously lowers the number of follow-up tests needed.
This aligns with the advantage of \HCP over \CCP in the dynamic simulation.

In Figure~\ref{fig:sens_household}, the change in household size distribution does not affect the performance of \NP, but it does affect that of correlated pooling. Under household size distributions that have larger weights on larger household sizes (e.g., \texttt{CN}, \texttt{US+1}, \texttt{US+2}), positive pools under \CP tend to contain a larger number of positives, which implies improvement in both sensitivity and efficiency. 

While the results above are based on the baseline setting, we do expect the sensitivity analysis based on other parameter settings to show similar patterns. 

\subsection{Implication of Correlation for Decision-Making}\label{subsec:SIR}

In this section, we show that \CP enables more powerful epidemic control than \NP based on a deterministic SIR model \citepAP{kermack1927contribution}, which translates to important implications for policy-making similar to those derived from the dynamic simulation in Section~\ref{subsec:corr_as_modeling_choice}.

We let $S$, $I$, $R$ denote the fractions of susceptible, actively infected, removed (due to either natural recovery or being detected and isolated in screening followed by recovery) individuals in the population, respectively.\footnote{We assume, for simplicity, that an infected individual is infectious and a recovered individual does not become susceptible again.} We assume a constant fraction of the non-isolated population is screened every day. The disease dynamics can be represented by a set of three discrete-time equations, where a time step corresponds to a day:
\begin{align}\label{eq:SIR}
    S(t+1)-S(t) &=\ -b_I\cdot S(t)I(t)\nonumber\\
    I(t+1)-I(t) &=\ b_I\cdot S(t)I(t) - (b_R + f\cdot\text{sensitivity})\cdot I(t)\\
    R(t+1)-R(t) &=\ (b_R + f\cdot\text{sensitivity})\cdot I(t),\nonumber
\end{align}
where $b_I$, $b_R$ are the rates of transmission and recovery, respectively;\footnote{We assume $b_I>b_R$, since the epidemic dies out naturally even without intervention if $b_I\leq b_R$.} $f$ is the frequency of screening for non-isolated individuals, i.e., those in the $S$ and $I$ groups. 

We first derive the critical screening frequency required to control the epidemic, i.e., stabilize or reduce the number of active infections. To quantify the epidemic growth, we define the \textit{growth factor} $\lambda$ at time $t$ as the ratio of the number of new cases at time $t$ to the number of cases removed at time $t$: 
$\lambda(t)=\dfrac{b_I\cdot S(t)I(t)}{(b_R + f\cdot\text{sensitivity})\cdot I(t)}$.

According to Equation~\ref{eq:SIR}, the number of infected individuals grows when $\lambda(t)>1$ and declines when $\lambda(t)<1$. 
We further construct a time-invariant upper bound on $\lambda(t)$ by setting $S(t)=1$: $\lambda' = \dfrac{b_I}{b_R + f\cdot\text{sensitivity}}$.\footnote{Alternatively, $\lambda'$ can be interpreted as the growth factor in the early stage of the epidemic, where the majority of the population is susceptible, i.e., $S(t)\approx 1$.} Since $\lambda(t)\leq \lambda'$ for all $t$, any screening frequency $f$ that results in a $\lambda'$ less than 1 also implies $\lambda(t)<1$ for all $t$.
Therefore, we use $\lambda'=1$ as a threshold that characterizes whether the epidemic is brought under control.
At this threshold, the screening frequency has a critical value $f^*$ satisfying
$f^*\times\text{sensitivity}=b_I-b_R$, which implies that
\begin{equation}
    f^*\propto \text{sensitivity}^{-1}.\label{eq:critical_screening_freq}
\end{equation}

A larger value of $f$ would reduce $\lambda'$ even further, but it would increase test consumption, a key quantity of practical concern. Hence, we next use $f^*$ to derive the \textit{minimum} test consumption required for epidemic control. 
For a screening frequency $f$, test consumption per day satisfies:
\begin{align}
    \text{test consumption per day} &\propto\ \text{screening frequency}\ \times\ \text{\# tests consumed per person}\nonumber\\
    &=\ f\ \times \ \text{efficiency}^{-1}.\label{eq:daily_test_consumption}
\end{align}

By Equations~\ref{eq:critical_screening_freq} and~\ref{eq:daily_test_consumption}, 
\begin{align}
    \text{minimum test consumption per day} &\propto \ f^*\ \times \ \text{efficiency}^{-1}\nonumber\\
    &\propto\ \text{sensitivity}^{-1}\ \times\ \text{efficiency}^{-1}\label{eq:test_consumption}.
\end{align}

That is, the minimum test consumption per day is directly proportional to $\text{sensitivity}\times\text{efficiency}$, which manifests the significance of having both higher sensitivity and efficiency in group testing.
In fact, this product is precisely the effective efficiency metric $\gamma$ studied in Section~\ref{subsec:metric_of_interest}.

Recall that both sensitivity and efficiency depend on the pool size, prevalence level, and pooling choice. 
Therefore, one should maximize sensitivity $\times$ efficiency when optimizing the pool size for a group testing protocol in real-world decision-making.

Table~\ref{tab:policy_cp} compares the optimal \NP and \CP policies (by choosing a pool size that maximizes sensitivity $\times$ efficiency) under different prevalence levels. The last column of Table~\ref{tab:policy_cp} illustrates the reduction in minimum test consumption required for epidemic control using the optimal \CP policy relative to the optimal \NP policy. 

\begin{table}[h]
\caption{Comparison of the optimal \CP and \NP policies in terms of sensitivity $\times$ efficiency under different prevalence levels. \label{tab:policy_cp}}
\begin{tabular}{cccccc}
\toprule
\multirow{2}{*}{\footnotesize Prevalence} & \multicolumn{2}{c}{\footnotesize Optimal naive pooling}   & \multicolumn{2}{c}{\footnotesize Optimal correlated pooling} & \multirow{2}{*}{\begin{tabular}[c]{@{}c@{}}\footnotesize Reduction in test consumption\\\small when using \CP \end{tabular}} \\ \cmidrule(r){2-3}\cmidrule(l){4-5}
& \footnotesize Pool size &\footnotesize Sensitivity $\times$ Efficiency &\footnotesize Pool size   & \footnotesize Sensitivity $\times$ Efficiency  &     \\ \midrule
0.1\%   & 40  & 13.52  & 40   & 15.86 & 14.8\%  \\
0.5\%   & 15  & 6.29  & 20  & 7.26  & 13.4\%   \\
1\%    & 12   & 4.56   & 12  & 5.23  & 12.9\% \\
5\%  & 6 & 2.17   & 6  & 2.44   & 10.9\%    \\
10\%  & 4 & 1.59    & 4  & 1.72  & 7.4\%\\ \bottomrule
\end{tabular}
\end{table}

For example, when prevalence is 1\%, a pool size of 12 is optimal for both \NP and \CP in terms of maximizing sensitivity $\times$ efficiency. 
Using Equation~\ref{eq:test_consumption}, we derive the optimal \NP policy uses $\frac{1/4.56-1/5.23}{1/5.23} = 14.7\%$ more tests than the optimal \CP policy. 

Such a difference has a substantial impact on real-world policy-making.
As correlated pooling accounts for the naturally arising within-pool correlation, it is a more accurate model for reality than naive pooling. 
Hence, policies informed by models ignoring the correlation tends to overestimate the test consumption necessary for controlling the epidemic. 
This leads to two insights similar to those derived in Section~\ref{subsec:corr_as_modeling_choice}: 
\begin{itemize}
    \item If the available testing capacity meets the minimum test consumption required by the optimal \CP policy but not the optimal \NP policy, a correlation-oblivious policy-maker would decide that no screening policy can permit safe reopening and thus issue a lockdown. However, a correlation-aware policy maker would keep the economy open with a feasible screening policy.
    \item If the available testing capacity of the city meets the minimum test consumption required by the optimal \NP policy, the correlation-oblivious policy-maker would decide to conduct screening. However, they would choose a lower screening frequency than allowed in reality because \NP underestimates the actual efficiency. On the other hand, a correlation-aware policy-maker would choose a higher screening frequency and achieve better epidemic mitigation.
\end{itemize}

Furthermore, if the naturally-induced within-pool correlation is weak, explicit measures can be taken to facilitate \CP. For example, one can mandate that individuals from the same household get tested together so that their samples can be placed in the same pool without many logistical difficulties. For a city with limited resources, such measures could enable a safe reopen with population-wide screening, while it may not be feasible otherwise.

\section{Quantifying the $(1+\delta)$ Bound in Theorem~\ref{thm:eff}}\label{appdx:delta}

In Appendix~\ref{appdx:delta}, we numerically investigate the bound $1+\delta$ derived in Theorem~\ref{thm:eff} and show that it is consistently close to one under various conditions. 
We first derive an upper bound $\delta'$ for $\delta$ and then provide 95\% confidence interval for $\delta'$ under various pool sizes and detection thresholds. Appendix~\ref{appdx:cond_indep} lays out the conditional independence relations necessary for the upper bound derivation. Appendix~\ref{appdx:delta_upper_bound} derives the upper bound $\delta'$ for $\delta$. Appendix~\ref{appdx:CI_for_delta} presents the point estimate and 95\% confidence interval for $\delta'$ under various pool sizes and detection thresholds. Appendix~\ref{appdx:thm_2_implications} discusses the implications of Theorem~\ref{thm:eff} for test Consumption in practice.

For succinctness, we abbreviate the probability operator $\PP_{\cp,\alpha}(\cdot)$ and the expectation operator $\EE_{\cp,\alpha}[\cdot]$ as $\PP(\cdot)$ and $\EE[\cdot]$ in Appendix~\ref{appdx:delta}.

\subsection{Conditional Independence Relations}\label{appdx:cond_indep}

We rely on two conditional independence assumptions discussed previously in Section~\ref{sec:theory} to derive an upper bound $\delta'$ for $\delta$, which we formulate again below. 

\begin{assumption}\label{assu:W_given_V}
For all $i=1,\cdots, n$, $W_i$ is independent of $\{V_j\}_{j\neq i}$ and $\{W_j\}_{j\neq i}$ given $V_i$. 
\end{assumption}

\begin{assumption}\label{assu:V_given_E}
For all $i=1,\cdots, n$, $V_i$ is independent of $\{V_{j}\}_{j\neq i}$ given $E_i$ where $E_i = \mathbbm{1}\{V_i > 0\}$. 
\end{assumption}

Assumptions~\ref{assu:W_given_V}~and~\ref{assu:V_given_E} also imply a sequence of conditional independence results, which we use in the derivation of an upper bound for $\delta$ in Appendix~\ref{appdx:delta_upper_bound}.

First, we show that Assumption~\ref{assu:W_given_V}
implies a weaker conditional independence relation, namely $\{W_i\}_{i=1}^{n}$ are independent given all $\{V_i\}_{i=1}^{n}$. 
\begin{lemma}\label{lem:Ws_given_Vs}
$\{W_i\}_{i=1}^{n}$ are conditionally independent given $\{V_i\}_{i=1}^{n}$.
\end{lemma}
\proof{Proof of Lemma~\ref{lem:Ws_given_Vs}.} 
Starting from the joint conditional density, we have that
\begin{align*}
    f(w_{1:n}\mid v_{1:n}) &= \frac{f(w_{1:n}, v_{2:n}\mid v_1)}{f(v_{2:n}\mid v_1)}\\
    &= \frac{f(w_1\mid v_1)f(w_{2:n}, v_{2:n}\mid v_1)}{f(v_{2:n}\mid v_1)}\quad\text{by Assumption~\ref{assu:W_given_V}}\\
    &=f(w_1\mid v_1)f(w_{2:n}\mid v_{1:n})\\
    &= ...\quad\text{repeat the above calculations for $n-1$ times}\\
    &= \prod_{i=1}^{n-1}f(w_i\mid v_i) \cdot f(w_n\mid v_{1:n})\\
    &= \prod_{i=1}^{n-1}f(w_i\mid v_{1:n}) \cdot f(w_n\mid v_{1:n})\quad\text{by Assumption~\ref{assu:W_given_V}}\\
    &=\prod_{i=1}^{n}f(w_i\mid v_{1:n}).
\end{align*}
\Halmos
\endproof

Then, we derive a similar conditional independence relation that $\{V_i\}_{i=1}^{n}$ are independent given $\{E_i\}_{i=1}^{n}$. To see this, we first note that by the definition of independence, it immediately follows from Assumption~\ref{assu:V_given_E} that given $E_i$, $V_i$ is also independent of the indicators $E_j$ where $j\neq i$.

\begin{lemma}\label{lem:V_i_E_j_given_E_i}
For all $i=1,\cdots, n$, $V_i$ is conditionally independent of $\{E_{j}\}_{j\neq i}$ given $E_i$.
\end{lemma}

Lemma~\ref{lem:V_i_E_j_given_E_i}, together with Assumption~\ref{assu:V_given_E}, implies that given all indicator variables $\{E_{i}\}_{i=1}^{n}$, $\{V_i\}_{i=1}^{n}$ are independent. 
\begin{lemma}\label{lem:Vs_given_Es}
$\{V_i\}_{i=1}^{n}$ are conditionally independent given $\{E_i\}_{i=1}^{n}$.
\end{lemma}

\proof{Proof of Lemma~\ref{lem:Vs_given_Es}.} The proof technique is the same as that of Lemma~\ref{lem:Ws_given_Vs}. Starting from the joint conditional density, we have that 
\begin{align*}
    f(v_{1:n}\mid e_{1:n}) &= \frac{f(v_{1:n}, e_{2:n}\mid e_1)}{f(e_{2:n}\mid e_1)}\\
    &= \frac{f(v_1\mid e_1)f(v_{2:n}, e_{2:n}\mid e_1)}{f(e_{2:n}\mid e_1)}\quad\text{by Assumption~\ref{assu:V_given_E} and Lemma~\ref{lem:V_i_E_j_given_E_i}}\\
    &=f(v_1\mid e_1)f(v_{2:n}\mid e_{1:n})\\
    &= ...\quad\text{repeat the above calculations for $n-1$ times}\\
    &= \prod_{i=1}^{n-1}f(v_i\mid e_i) \cdot f(v_n\mid e_{1:n})\\
    &= \prod_{i=1}^{n-1}f(v_i\mid e_{1:n}) \cdot f(v_n\mid e_{1:n})\quad\text{by Lemma~\ref{lem:V_i_E_j_given_E_i}}\\
    &=\prod_{i=1}^{n}f(v_i\mid e_{1:n}).
\end{align*}

Hence, given $E_{1:n}$, $V_1,\cdots, V_n$ are independent. 
\Halmos
\endproof

It follows from Lemmas~\ref{lem:Ws_given_Vs}~and~\ref{lem:Vs_given_Es} that $(V_i, W_i)$, $i=1,\cdots, n$ are also conditionally independent, given the indicators $\{E_i\}_{i=1}^n$. 
\begin{lemma}\label{lem:W_V_given_Es}
$\{ V_i,W_i \}_{i=1}^{n}$ are conditionally independent given $\{E_i\}_{i=1}^n$.
\end{lemma}
\proof{Proof of Lemma~\ref{lem:W_V_given_Es}.}
We consider the joint conditional density of $(V_i, W_i)_{i=1}^{n}$ given $\{E_i\}_{i=1}^n$:
\begin{align*}
    f\left((v_i,w_i)_{i=1}^{n}\mid e_{1:n}\right) &= f(w_{1:n}\mid v_{1:n}, e_{1:n})f(v_{1:n}\mid e_{1:n})\\
    &=f(w_{1:n}\mid v_{1:n})f(v_{1:n}\mid e_{1:n})\\
    &= \prod_{i=1}^{n}f(w_i\mid v_{1:n})\prod_{i=1}^{n}f(v_i\mid e_{1:n})\quad\text{by Lemma~\ref{lem:Ws_given_Vs} and \ref{lem:Vs_given_Es}}\\
    &= \prod_{i=1}^{n}f(w_i\mid v_i)\prod_{i=1}^{n}f(v_i\mid e_i)\quad\text{by Assumptions~\ref{assu:W_given_V} and \ref{assu:V_given_E}}\\
    &=\prod_{i=1}^{n}f(w_i, v_i\mid e_{i})\\
    &= \prod_{i=1}^{n}f(w_i, v_i\mid e_{1:n})\quad\text{by Lemma~\ref{lem:Ws_given_Vs} and \ref{lem:Vs_given_Es}}.
\end{align*}

We are done.
\Halmos
\endproof

\subsection{Deriving an Upper Bound for $\delta$}\label{appdx:delta_upper_bound}

Now we are equipped with the tools needed to provide an upper bound for $\delta$. Recall that
\begin{equation}\label{eq:delta}
   \delta = \frac{\PP(Y=1, S_D=0\mid S>0)}{\PP( Y=1, S_D > 0\mid S>0)} = \frac{\PP(Y=1\mid S_D=0, S > 0)\PP(S_D = 0\mid S>0)}{\PP(Y=1\mid S_D > 0)\PP(S_D > 0\mid S>0)}.
\end{equation}

To bound $\delta$ from above, we provide upper and lower bounds for the terms in the numerator and denominator in Equation~\ref{eq:delta}, respectively. We start by proving an upper bound for the second term in the numerator. It also implies that $\PP(S_D > 0\mid S > 0)\geq \PP(S_D > 0\mid S = 1)$ for the second term in the denominator.
\begin{proposition} \label{prop:upper_right_in_bound}
$\PP(S_D = 0\mid S > 0) \leq \PP(S_D = 0\mid S = 1)$.
\end{proposition}

\proof{Proof of Proposition~\ref{prop:upper_right_in_bound}.}
We consider $\PP(S_D = 0\mid S = k)$ for any $k\in\{1,2,\cdots, n\}$. Since $S_D = \sum_{i=1}^{n}W_i$, we have that
\begin{align}
    \PP(S_D = 0\mid S = k) &= \PP\left(\bigcap_{i=1}^{n}\{W_i = 0\}\mid S = k\right)\nonumber\\
    &= \EE\left[\PP\left(\bigcap_{i=1}^{n}\{W_i = 0\}\mid E_{1:n}, S = k\right)\mid S = k\right]\nonumber\\
    &= \EE\left[\prod_{i=1}^{n}\PP(W_i = 0\mid E_{1:n})\mid S = k\right]\quad\text{by Lemma~\ref{lem:W_V_given_Es}}\nonumber\\
    &= \EE\left[\prod_{i=1}^{n}\EE[1 - p(V_i)\mid E_{1:n}]\mid S = k\right]\nonumber\\
    &= \EE\left[\prod_{i=1}^{n}\EE[1 - p(V_i)\mid E_{i}]\mid S= k \right]\quad\text{by Lemma~\ref{lem:V_i_E_j_given_E_i}}.\label{eq:SD_0_S_k_intermediate}
\end{align}

Note that for $i=1,2,\cdots, n$, we have
\begin{align}
    \EE[1 - p(V_i)\mid E_i] &= \PP(W_i = 0\mid E_i)\nonumber\\
    &=\begin{cases}
    1 & E_i = 0\nonumber\\
    \bar{\beta} & E_i = 1
    \end{cases}\nonumber\\
    &=\bar{\beta}^{E_i}.\label{eq:pop_avg_FNR}
\end{align}
where $\bar\beta$ is the \textit{population-average individual test FNR}, i.e., $\bar\beta = \EE[1 - p(V)\mid V > 0]$.\footnote{
Note that $\bar\beta$ is not to be confused with $\beta_{\pool,\alpha}$ ($\pool\in\{\np,\cp\}$) introduced in Section~\ref{sec:theory} which represents the overall FNR of a specific group testing protocol.}

Recall that $S = \sum_{i=1}^{n}\mathbbm{1}\{V_i > 0\} = \sum_{i=1}^{n}E_i$. Combining Equations~\ref{eq:SD_0_S_k_intermediate}~and~\ref{eq:pop_avg_FNR}, we find that
\begin{align*}
    \PP(S_D=0\mid S=k) &= \EE\left[\prod_{i=1}^{n}\bar{\beta}^{E_i}\mid S = k\right]\\
    &=\EE\left[\bar\beta^{\sum_{i=1}^{n}E_i}\mid S = k\right]\\
    & = \bar\beta^{k}.
\end{align*}

Since $\bar\beta\in [0,1]$, we find $\PP(S_D = 0\mid S = k)\leq \PP(S_D = 0\mid S= 1)$ for all $k\in \{1,2,\cdots, n\}$. By the law of iterated expectations, it follows that $\PP(S_D = 0\mid S > 0 ) \leq \PP(S_D = 0\mid S= 1)$. 
\Halmos
\endproof

Second, we provide a lower bound for the first term in the denominator in 
Equation~\ref{eq:delta}. To achieve this, we characterize a first-order stochastic dominance relation, given in Lemmas~\ref{lem:V_given_W_dominance}.

\begin{lemma}\label{lem:V_given_W_dominance}
$\PP(V_i\geq v\mid W_i=1)\geq \PP(V_i\geq v\mid W_i=0)$ for all $i\in\{1,2,\cdots, n\}$.
\end{lemma}
\proof{Proof of Lemma~\ref{lem:V_given_W_dominance}.}
Recall that $W_i = \text{Ber}(p(V_i))$ where $p(\cdot):\mathbb{R}_{\geq 0}\to [0,1]$ is monotone increasing.

By Bayes rule, we have that 
\begin{align*}
    \PP(V_i\geq v\mid W_i = 1) &= \frac{\PP(W_i = 1\mid V_i\geq v)\PP(V_i\geq v)}{\PP(W_i = 1)}\\
    \PP(V_i\geq v\mid W_i = 0) &=\frac{\PP(W_i = 0\mid V_i\geq v)\PP(V_i\geq v)}{\PP(W_i = 0)} = \frac{(1 -\PP(W_i = 1\mid V_i\geq v)\PP(V_i\geq v)}{1 - \PP(W_i = 1)}.
\end{align*}

Then,
\begin{align*}
    &\PP(V_i\geq v\mid W_i = 1)\geq \PP(V_i\geq v\mid W_i = 0)\\
    \Longleftrightarrow\quad & \frac{\PP(W_i=1\mid V_i\geq v)}{\PP(W_i = 1)}\geq\frac{1 - \PP(W_i = 1\mid V_i\geq v)}{1 - \PP(W_i=1)}\\
    \Longleftrightarrow\quad & \PP(W_i=1\mid V_i\geq v)\geq \PP(W_i = 1)\\
    \Longleftrightarrow\quad & \PP(W_i=1\mid V_i\geq v)(1 - \PP(V_i\geq v))\geq \PP(W_i = 1\mid V_i<v) (1 - \PP(V_i\geq v)).
\end{align*}

If $\PP(V_i \geq v) = 1$, then the inequality holds; otherwise, by monotonicity of $p(v)$ we have $$
\PP(W_i =1 \mid V_i \geq v)\geq p(v)\geq \PP(W_i = 1\mid V_i < v).
$$

We are done.
\Halmos
\endproof

\begin{proposition}\label{prop:lower_left_in_bound}
$\PP(Y = 1\mid S_D > 0)\geq \PP(Y= 1\mid S_D > 0, S = 1)$.
\end{proposition}

\proof{Proof of Proposition~\ref{prop:lower_left_in_bound}.}
We consider $\PP(Y = 1\mid S_D = k, S = s)$ for any $0\leq k\leq s\leq n$ and show that it is increasing in both $k$ and $s$. For $h(\BFv) = \frac{1}{n}\sum_{i=1}^{n}v_i$, we have that 
\begin{align*}
    \PP(Y=1\mid S_D = k, S = s) &= \EE[\PP(Y = 1\mid W_{1:n}, E_{1:n})\mid S_D = k, S = s]\nonumber\\
    &= \EE\left[\EE\left[p\left(\frac{1}{n}\sum_{i=1}^{n}V_i\right)\mid W_{1:n}, E_{1:n}\right]\mid S_D = k, S  =s\right].
\end{align*}

To derive the inner expectation, we study the joint conditional density of $V_1,\cdots, V_n$ given $W_{1:n}$ and $E_{1:n}$. We have that 
\begin{align*}
    f(v_{1:n}\mid w_{1:n}, e_{1:n}) &= \frac{f(v_{1:n}, w_{1:n}\mid e_{1:n})}{f(w_{1:n}\mid e_{1:n})}\\
    &= \prod_{i=1}^{n}\frac{f(v_i, w_i\mid e_{1:n})}{f(w_i\mid e_{1:n})}\quad\text{by Lemma~\ref{lem:W_V_given_Es}}\\
    &= \prod_{i=1}^{n}f(v_i\mid w_i, e_{1:n})\\
    &= \prod_{i=1}^{n}\frac{f(w_i\mid v_i)f(v_i\mid e_{1:n})}{\int f(w_i\mid v_i)f(v_i\mid e_{1:n})dv_i}\\
    &= \prod_{i=1}^{n}\frac{f(w_i\mid v_i)f(v_i\mid e_i)}{\int f(w_i\mid v_i)f(v_i\mid e_i)dv_i}\quad\text{by Assumption~\ref{assu:V_given_E}}\\
    &= \prod_{i=1}^{n}f(v_i\mid w_i, e_i).
\end{align*}

Hence, given $W_{1:n}$ and $E_{1:n}$, $\{V_i\}_{i=1}^{n}$ are independent, with the distribution of $V_i$ given by $V_i\mid W_i, E_i$.
Since $V_{1},\cdots, V_n$ are identically distributed, we have that $\{V_i\mid W_i = 1, E_i = 1\}_{i=1}^{n}$ and $\{V_i\mid W_i = 0, E_i = 1\}_{i=1}^{n}$ are also identically distributed, respectively. Denote the distributions for $V_i\mid W_i = 1, E_i = 1$ and $V_i\mid W_i = 0, E_i = 1$ by $F_{V\mid W = 1}$ and $F_{V\mid W = 0}$, respectively. Then, $\sum_{i=1}^{n}V_i$ is the sum of $S_D$ i.i.d random variables with distribution $F_{V\mid W = 1}$ and $S - S_D$ i.i.d random variables with distribution $F_{V\mid W = 0}$. That is, the distribution of $\sum_{i=1}^{n}V_i$ only depends on $\{E_i\}_{i=1}^{n}$ and $\{W_i\}_{i=1}^{n}$ through their respective sums, $S$ and $S_D$. Hence, since $p(v)$ is monotone increasing, $\PP(Y = 1\mid S_D = k, S = s)$ is increasing in $s$. Moreover, since $F_{V\mid W = 1}$ first-order stochastic dominates $F_{V\mid W = 0}$ by Lemma~\ref{lem:V_given_W_dominance}, $\PP(Y = 1\mid S_D = k, S = s)$ is also increasing in $k$. Therefore, we have
\begin{align*}
    \PP(Y= 1\mid S_D > 0) &= \EE[\PP(Y = 1\mid S_D, S)\mid S_D > 0]\\
    &\geq \EE[\PP(Y = 1\mid S_D = 1, S  =1)\mid S_D > 0]\\
    &= \PP(Y = 1\mid S_D = 1, S  =1).
\end{align*}

We are done.
\Halmos
\endproof

\begin{proposition}\label{prop:upper_left_in_bound}
$\PP(Y = 1 \mid S_D = 0, S > 0)\leq \PP(Y = 1\mid S_D = 0, S = n)$.
\end{proposition}
\proof{Proof of Proposition~\ref{prop:upper_left_in_bound}.}
As shown in the proof of Proposition~\ref{prop:lower_left_in_bound}, we have that $\PP(Y = 1\mid S_D = k, S = s)$ is increasing in $s$. Hence,
\begin{align*}
    \PP(Y = 1\mid S_D = 0, S > 0) &= \EE[\PP(Y =1\mid S_D, S)\mid S_D = 0, S > 0]\\
    &\leq \EE[\PP(Y =1\mid S_D, S = n)\mid S_D = 0, S > 0]\\
    &= \PP(Y =1\mid S_D = 0, S = n),
\end{align*}
which concludes the proof.
\Halmos
\endproof

Combining Propositions~\ref{prop:upper_right_in_bound}, \ref{prop:lower_left_in_bound} and \ref{prop:upper_left_in_bound}, we find that 
\begin{align}
    \delta' &= \frac{\PP(Y = 1\mid S_D = 0, S =n)\PP(S_D = 0\mid S = 1)}{\PP(Y = 1\mid S_D =S=1 )\PP(S_D = 1\mid S = 1)} \nonumber\\
    &= \frac{\PP(Y = 1\mid S_D = 0, S =n)}{\PP(Y = 1\mid S_D =S=1 )}\cdot \frac{\bar\beta}{1 - \bar\beta},\label{eq:delta_prime}
\end{align}
is an upper bound for $\delta$.

\subsection{Confidence Interval for $\delta'$}\label{appdx:CI_for_delta}
In this section, we provide a point estimate and 95\% confidence interval for $\delta'$ under different pool sizes and detection thresholds. We show that $\delta'$ is consistently small under various conditions. Below we describe the methodology in detail. We assume that $h(\BFv) = \frac{1}{n}\sum_{i=1}^{n}v_i$ throughout this subsection. 

We use Monte Carlo simulation to estimate $\PP(Y = 1\mid S_D = 0, S=n)$ and $\PP(Y = 1\mid S_D = S = 1)$ separately. 
Let $V_1,\cdots, V_n\overset{i.i.d}{\sim}F_{V\mid W=0}$ where $F_{V\mid W = 0}$ is the distribution for $V_i\mid W_i= 0,E_i= 1$. Then, as shown in the proof of Proposition~\ref{prop:lower_left_in_bound}, $X = \PP(Y = 1\mid V_{1:n}) = p\left(\frac{1}{n}\sum_{i=1}^{n}V_i\right)$ is an unbiased estimator for $\PP(Y=1\mid S_D = 0,S=n)$, i.e. $\PP(Y=1\mid S_D = 0,S=n) = \EE[X]$. To sample from $F_{V\mid W = 0}$, we first sample $V$ from $V\mid V > 0$, the viral load distribution described in Table~\ref{tab: GMM_VL}, then we sample $W \sim \text{Ber}(p(V))$. We keep the sampled $V$ if the sampled $W$ is equal to zero and discard $V$ otherwise. We generate $B = 10^6$ samples $X_1,\cdots, X_B$ for estimating $\PP(Y=1\mid S_D = 0,S=n)$. 

Similarly, let $V\sim F_{V\mid W = 1}$ where $F_{V\mid W = 1}$ is the distribution for $V_i\mid W_i= 1,E_i= 1$. Then, $Z = \PP(Y = 1\mid V,0,\cdots, 0) = p(V/n)$ is an unbiased estimator for $\PP(Y=1\mid S_D = S = 1)$, i.e. $\PP(Y=1\mid S_D = S = 1) = \EE[Z]$. Sampling from $F_{V\mid W = 1}$ follows a similar procedure as sampling from $F_{V\mid W = 0}$. We generate $B = 10^6$ samples $Z_1,\cdots, Z_B$ for estimating $\PP(Y=1\mid S_D = S = 1)$. 

Hence, the point estimate for $\delta'$ is given by $$
\hat{\delta}' = \frac{\bar{X}}{\bar{Z}}\cdot \frac{\bar\beta}{1- \bar\beta}. 
$$

To provide a confidence interval for $\delta'$, we first find confidence intervals for the $\EE[X]$ and $\EE[Z]$ separately.
We derive the confidence interval for $\EE[Z]$ based on central limit theorem. Using normal approximation, the $q=99.99\%$ confidence interval for $\EE[Z]$ is given by $[L_Z, U_Z] = [\bar{Z}- 3.891\cdot \sigma_{\bar{Z}}, \bar{Z} + 3.891\cdot \sigma_{\bar{Z}}]$.
On the other hand, $\EE[X]$ is close to zero in the regime we consider, and the samples $X_i$ can differ by several orders of magnitude. Thus, instead of using the normal approximation, we employ bootstrapping \citepAP{EfroTibs93} with $10^4$ replications to construct the $\frac{95}{q}\%$ confidence interval for $\EE[X]$, denoted by $[L_X, U_X]$. 

Because the samples $X_i$'s and $Z_i$'s are independent, the Cartesian product $[L_X, U_X]\times [L_Z, U_Z]$ is a $\left(\frac{95}{q}\cdot q\right)\% = 95\%$ confidence interval for $(\EE_{1,\alpha}[X], \EE_{1,\alpha}[Z])$. It follows that $\left[\frac{L_X}{U_Z}, \frac{U_X}{L_Z}\right]$ (assuming that $0 < L_Z \leq U_Z$ and $0\leq L_X\leq U_X$) is a 95\% confidence interval for $\delta'$.

Table~\ref{tab:CI_for_delta} summarizes the point estimate and 95\% confidence interval for $\delta'$ under different pool sizes and detection thresholds. We see that under all conditions, $\hat\delta'$ is consistently small, with the maximum $\hat\delta'$ achieved at $n=2$ and $\bar\beta=2.5\%$. 
\begin{table}[h]
\TABLE
{Point estimate and 95\% confidence interval for $\delta'$ under different pool sizes $n$ and population-average individual test FNR $\bar\beta$.\label{tab:CI_for_delta}}
{\addtolength{\tabcolsep}{6pt}\begin{tabular}{ccccccc}
\toprule
$n$ & $\bar\beta$ & $\bar X$ & $\bar Z$ & $\hat\delta'$ & 95\% CI for $\delta'$ (lb) & 95\% CI for $\delta'$ (ub) \\ \midrule
2 & 0.025       & 3.35E-02 & 0.960    & 8.96E-04      & 8.90E-04                   & 9.02E-04                   \\
2  & 0.05        & 1.35E-02 & 0.946    & 7.51E-04      & 7.44E-04                   & 7.59E-04                   \\
2  & 0.1         & 2.94E-03 & 0.938    & 3.48E-04      & 3.41E-04                   & 3.55E-04                   \\
2  & 0.2         & 1.73E-04 & 0.932    & 4.64E-05      & 4.26E-05                   & 5.03E-05                   \\
4   & 0.025       & 1.00E-02 & 0.903    & 2.84E-04      & 2.81E-04                   & 2.86E-04                   \\
4   & 0.05        & 1.94E-03 & 0.888    & 1.15E-04      & 1.13E-04                   & 1.17E-04                   \\
4   & 0.1         & 1.06E-04 & 0.881    & 1.33E-05      & 1.25E-05                   & 1.42E-05                   \\
4  & 0.2         & 6.49E-07 & 0.853    & 1.90E-07      & 3.70E-08                   & 4.34E-07                   \\
6  & 0.025       & 4.48E-03 & 0.871    & 1.32E-04      & 1.31E-04                   & 1.33E-04                   \\
6  & 0.05        & 4.82E-04 & 0.856    & 2.96E-05      & 2.89E-05                   & 3.03E-05                   \\
6  & 0.1         & 7.98E-06 & 0.846    & 1.05E-06      & 8.84E-07                   & 1.23E-06                   \\
6  & 0.2         & 2.53E-11 & 0.802    & 7.89E-12      & 2.87E-14                   & 2.32E-11                   \\
12 & 0.025       & 1.12E-03 & 0.817    & 3.51E-05      & 3.47E-05                   & 3.55E-05                   \\
12  & 0.05        & 3.34E-05 & 0.801    & 2.20E-06      & 2.12E-06                   & 2.28E-06                   \\
12   & 0.1         & 1.57E-08 & 0.779    & 2.24E-09      & 1.48E-09                   & 3.13E-09                   \\
12  & 0.2         & 1.73E-26 & 0.710    & 6.08E-27      & 7.34E-35                   & 1.83E-26                   \\ \bottomrule
\end{tabular}\addtolength{\tabcolsep}{-6pt}}{}
\end{table}

\subsection{Implications of Theorem~\ref{thm:eff} for Test Consumption in Practice}\label{appdx:thm_2_implications}

We show that in this setting, correlated pooling consumes no more follow-up tests per positive identified than naive pooling for a wide range of pool sizes and PCR test sensitivities ($80\% - 97.5\%$). 

Using Monte Carlo simulation with $10^6$ replications, we find that across a wide range of $\bar\beta$ and pool sizes, $\delta'$ is consistently close to zero (Table~\ref{tab:CI_for_delta}). The maximum value of $\delta'$ is $8.96\times 10^{-4}$ (95\% CI: $(8.90\times 10^{-4}, 9.02\times 10^{-4})$, obtained when $n=2$ and $\bar\beta = 2.5\%$. As $n$ increases, the relaxed bound converges to 1.

Now we provide intuition for why $\delta'$ is small. We first examine a representative curve of PCR test sensitivity versus sample viral load under $\bar\beta = 5\%$.  
Based on the viral load distribution among infected individuals given in Table~\ref{tab: GMM_VL}, when $\bar\beta=5\%$, the PCR test sensitivity grows rapidly from 0 to 1 over a narrow range of log viral load in the sample (as shown in Figure~\ref{fig:pcr_sensitivity_174}).

Specifically, a $\log_{10}$ viral load of 3.45 gives a PCR test sensitivity of 0.3\%, while a $\log_{10}$ viral load of 3.65 gives a PCR test sensitivity of 99.8\%. The fraction of infected individuals that have $\log_{10}$ viral load between 3.45 and 3.65 is only 2.8\%, indicating that the majority of positive samples either test positive with high probability (if the $\log_{10}$ viral load is above 3.65) or test positive with low probability (if the $\log_{10}$ viral load is below 3.45). 
Though not depicted here, the $p(v)$ curves corresponding to different $\bar\beta$ follow the same pattern. 

Based on the above observations, we argue that correlated pooling's test consumption per positive identified nearly meets or exceeds that of naive pooling in practice. We first observe that  $\PP_{1,\alpha}(Y=1 \mid S_D=0, S=n)$, which is in the numerator of $\delta'$, is small. If a pool contains only $n$ positives that would all test negative individually, i.e., $S_D=0$, then they likely all have viral loads below the narrow region where an individual test's sensitivity rises. 
Thus, the viral load in the pool, which is the average of the viral loads of these positive samples, is likely also below the narrow region, making it likely to test negative, i.e., $Y=0$. 

On the other hand, we argue that $\PP(Y = 1\mid S_D =S=1)$, which is in the denominator of $\delta'$, is reasonably large. In other words, if a pool contains only one positive sample and it would test positive individually, then the pool is likely to test positive. With its viral load drawn from the distribution described in Table~\ref{tab: GMM_VL}, a positive sample that would test positive individually has its viral load way above the narrow region with a reasonably large probability. Hence, even when such a sample is diluted by a factor equal to the pool size, the pooled sample likely still has its viral load above the narrow region and is likely to test positive, i.e., $Y = 1$.

\clearpage

\bibliographystyleAP{informs2014} 
\bibliographyAP{references} 

\end{APPENDICES}

%
%
%

\end{document}